\documentclass[twocolumn]{aastex6}
\usepackage{graphicx}
\usepackage[flushleft]{threeparttable}

\shorttitle{ASPECS: Survey Description}
\shortauthors{Walter et al.\ }

\def\Msun{M$_\odot$}

\def\Ci{[C\,{\sc i}]}

\def\Cii{[C\,{\sc ii}]}
\def\Ci{[C\,{\sc i}]}

\def\kms{km\,s$^{-1}$}

\def\Kkmspc{K~km\,s$^{-1}$\,pc$^2$}

\def\lsim{\mathrel{\rlap{\lower 3pt \hbox{$\sim$}} \raise 2.0pt \hbox{$<$}}}
\def\gsim{\mathrel{\rlap{\lower 3pt \hbox{$\sim$}} \raise 2.0pt \hbox{$>$}}}

\begin{document}

\title{
ALMA Spectroscopic Survey in the Hubble Ultra Deep Field: Survey Description
}
\author{
Fabian Walter\altaffilmark{1,2,3}, 
Roberto Decarli\altaffilmark{1}, 
Manuel Aravena\altaffilmark{4}, 
Chris Carilli\altaffilmark{3,5}, 
Rychard Bouwens\altaffilmark{6}, 
Elisabete da Cunha\altaffilmark{7,8}, 
Emanuele Daddi\altaffilmark{9}, 
R.\,J.~Ivison\altaffilmark{10,11}, 
Dominik Riechers\altaffilmark{12}, 
Ian Smail\altaffilmark{13}, 
Mark Swinbank\altaffilmark{13}, 
Axel Weiss\altaffilmark{14}, 
Timo Anguita\altaffilmark{15,16}, 
Roberto Assef\altaffilmark{4}, 
Roland Bacon\altaffilmark{17},
Franz Bauer\altaffilmark{18,19,20}, 
Eric F.~Bell\altaffilmark{21}, 
Frank Bertoldi\altaffilmark{22}, 
Scott Chapman\altaffilmark{23}, 
Luis Colina\altaffilmark{24}, 
Paulo C.~Cortes\altaffilmark{25,26}, 
Pierre Cox\altaffilmark{27}, 
Mark Dickinson\altaffilmark{27}, 
David Elbaz\altaffilmark{9}, 
Jorge G\'onzalez-L\'opez\altaffilmark{28}, 
Edo Ibar\altaffilmark{29}, 
Hanae Inami\altaffilmark{17},
Leopoldo Infante\altaffilmark{28}, 
Jacqueline Hodge\altaffilmark{6}, 
Alex Karim\altaffilmark{22}, 
Olivier Le Fevre\altaffilmark{30}, 
Benjamin Magnelli\altaffilmark{22}, 
Roberto Neri\altaffilmark{31}, 
Pascal Oesch\altaffilmark{32}, 
Kazuaki Ota\altaffilmark{33,5}, 
Gerg\"{o} Popping\altaffilmark{10}, 
Hans--Walter Rix\altaffilmark{1}, 
Mark Sargent\altaffilmark{34}, 
Kartik Sheth\altaffilmark{35}, 
Arjen van der Wel\altaffilmark{1},
Paul van der Werf\altaffilmark{6}, 
Jeff Wagg\altaffilmark{36}
}
\altaffiltext{1}{Max-Planck Institut f\"{u}r Astronomie, K\"{o}nigstuhl 17, D-69117, Heidelberg, Germany. E-mail: {\sf walter@mpia.de}}
\altaffiltext{2}{Astronomy Department, California Institute of Technology, MC105-24, Pasadena, California 91125, USA}
\altaffiltext{3}{National Radio Astronomy Observatory, Pete V.\,Domenici Array Science Center, P.O.\, Box O, Socorro, NM, 87801, USA}
\altaffiltext{4}{N\'{u}cleo de Astronom\'{\i}a, Facultad de Ingenier\'{\i}a, Universidad Diego Portales, Av. Ej\'{e}rcito 441, Santiago, Chile}
\altaffiltext{5}{Cavendish Laboratory, University of Cambridge, 19 J J Thomson Avenue, Cambridge CB3 0HE, UK}
\altaffiltext{6}{Leiden Observatory, Leiden University, PO Box 9513, NL2300 RA Leiden, The Netherland}
\altaffiltext{7}{Centre for Astrophysics and Supercomputing, Swinburne University of Technology, Hawthorn, Victoria 3122, Australia}
\altaffiltext{8}{Research School of Astronomy and Astrophysics, Australian National University, Canberra, ACT 2611, Australia}
\altaffiltext{9}{Laboratoire AIM, CEA/DSM-CNRS-Universite Paris Diderot, Irfu/Service d'Astrophysique, CEA Saclay, Orme des Merisiers, 91191 Gif-sur-Yvette cedex, France}
\altaffiltext{10}{European Southern Observatory, Karl-Schwarzschild-Strasse 2, 85748, Garching, Germany}
\altaffiltext{11}{Institute for Astronomy, University of Edinburgh, Royal Observatory, Blackford Hill, Edinburgh EH9 3HJ}
\altaffiltext{12}{Cornell University, 220 Space Sciences Building, Ithaca, NY 14853, USA}
\altaffiltext{13}{6 Centre for Extragalactic Astronomy, Department of Physics, Durham University, South Road, Durham, DH1 3LE, UK}
\altaffiltext{14}{Max-Planck-Institut f\"ur Radioastronomie, Auf dem H\"ugel 69, 53121 Bonn, Germany}
\altaffiltext{15}{Departamento de Ciencias F\'{\i}sicas, Universidad Andres Bello, Fernandez Concha 700, Las Condes, Santiago, Chile}
\altaffiltext{16}{Millennium Institute of Astrophysics, Chile}
\altaffiltext{17}{Universit\'{e} Lyon 1, 9 Avenue Charles Andr\'{e}, 69561 Saint Genis Laval, France}
\altaffiltext{18}{Instituto de Astrof\'{\i}sica, Facultad de F\'{\i}sica, Pontificia Universidad Cat\'olica de Chile Av. Vicu\~na Mackenna 4860, 782-0436 Macul, Santiago, Chile}
\altaffiltext{19}{Millennium Institute of Astrophysics (MAS), Nuncio Monse{\~{n}}or S{\'{o}}tero Sanz 100, Providencia, Santiago, Chile}
\altaffiltext{20}{Space Science Institute, 4750 Walnut Street, Suite 205, Boulder, CO 80301, USA}
\altaffiltext{21}{Department of Astronomy, University of Michigan, 1085 South University Ave., Ann Arbor, MI 48109, USA}
\altaffiltext{22}{Argelander Institute for Astronomy, University of Bonn, Auf dem H\"{u}gel 71, 53121 Bonn, Germany}
\altaffiltext{23}{Dalhousie University, Halifax, Nova Scotia, Canada}
\altaffiltext{24}{ASTRO-UAM, UAM, Unidad Asociada CSIC, Spain}
\altaffiltext{25}{Joint ALMA Observatory - ESO, Av. Alonso de C\'ordova, 3104, Santiago, Chile}
\altaffiltext{26}{National Radio Astronomy Observatory, 520 Edgemont Rd, Charlottesville, VA, 22903, USA}
\altaffiltext{27}{Steward Observatory, University of Arizona, 933 N. Cherry St., Tucson, AZ  85721, USA}
\altaffiltext{28}{Instituto de Astrof\'{\i}sica, Facultad de F\'{\i}sica, Pontificia Universidad Cat\'olica de Chile Av. Vicu\~na Mackenna 4860, 782-0436 Macul, Santiago, Chile}
\altaffiltext{29}{Instituto de F\'{\i}sica y Astronom\'{\i}a, Universidad de Valpara\'{\i}so, Avda. Gran Breta\~na 1111, Valparaiso, Chile}
\altaffiltext{30}{Aix Marseille Universite, CNRS, LAM (Laboratoire d'Astrophysique de Marseille), UMR 7326, F-13388 Marseille, France}
\altaffiltext{31}{IRAM, 300 rue de la piscine, F-38406 Saint-Martin d'H\`eres, France}
\altaffiltext{32}{Astronomy Department, Yale University, New Haven, CT 06511, USA}
\altaffiltext{33}{Kavli Institute for Cosmology, University of Cambridge, Madingley Road, Cambridge CB3 0HA, UK}
\altaffiltext{34}{Astronomy Centre, Department of Physics and Astronomy, University of Sussex, Brighton, BN1 9QH, UK}
\altaffiltext{35}{NASA Headquarters, Washington DC, 20546-0001, USA}
\altaffiltext{36}{SKA Organization, Lower Withington Macclesfield, Cheshire SK11 9DL, UK}

\begin{abstract}

We present the rationale for and the observational description of
ASPECS: The ALMA SPECtroscopic Survey in the {\it Hubble} Ultra--Deep
Field (UDF), the cosmological deep field that has the deepest
multi--wavelength data available. Our overarching goal is to obtain an
unbiased census of molecular gas and dust continuum emission in
high--redshift (z$>$0.5) galaxies. The $\sim$1$'$ region covered
within the UDF was chosen to overlap with the deepest available
imaging from {\it HST}. Our ALMA observations consist of full
frequency scans in band~3 (84--115\,GHz) and band~6 (212--272\,GHz) at
approximately uniform line sensitivity ($L'_{\rm
CO}\sim$2\,$\times$10$^{9}$\,\Kkmspc), and continuum noise levels of
3.8\,$\mu$Jy\,beam$^{-1}$ and 12.7\,$\mu$Jy\,beam$^{-1}$,
respectively. The molecular surveys cover the different rotational
transitions of the CO molecule, leading to essentially full redshift
coverage. The \Cii{} emission line is also covered at redshifts
$6.0<z<8.0$. We present a customized algorithm to identify line
candidates in the molecular line scans, and quantify our ability to
recover artificial sources from our data. Based on whether multiple CO
lines are detected, and whether optical spectroscopic redshifts as
well as optical counterparts exist, we constrain the most likely line
identification.  We report 10 (11) CO line candidates in the 3\,mm
(1\,mm) band, and our statistical analysis shows that $<$4 of these (in
each band) are likely spurious. Less than 1/3 of the total CO flux in
the low--J CO line candidates are from sources that are not associated
with an optical/NIR counterpart. We also present continuum maps of
both the band~3 and band~6 observations. The data presented here form
the basis of a number of dedicated studies that are presented in
subsequent papers.

\end{abstract}

\section{Introduction}

Characterizing the molecular gas content of distant galaxies is
essential in order to understand the evolution of the cosmic star
formation rate density \citep{madau14}, and the build--up of stellar
mass \citep{bell03} throughout cosmic time \citep{carilli13}.  A
unique way to fully characterize the molecular gas content in galaxies
in the early universe is through spectral line scans in well--studied
cosmological deep fields. In comparison to targeted observations of
individual galaxies, spectral scans have the advantage that molecular
gas reservoirs can be characterized without pre--selection through
other information (e.g., stellar mass, star--formation rate). Such
spectral line scans can also potentially reveal the presence of
gas--rich `dark' galaxies, i.e., galaxies that are invisible in the
optical wavebands, and that would not be selected as targets to search
for molecular gas emission \citep[e.g.,][]{walter12}. In a sense,
spectral line scans follow the spirit of the original {\it HST} deep
fields \citep[e.g.,][]{williams96,beckwith06}, as essentially no prior
knowledge/selection based on galaxy properties enters the choice of
field.

\begin{figure}
\includegraphics[width=0.99\columnwidth]{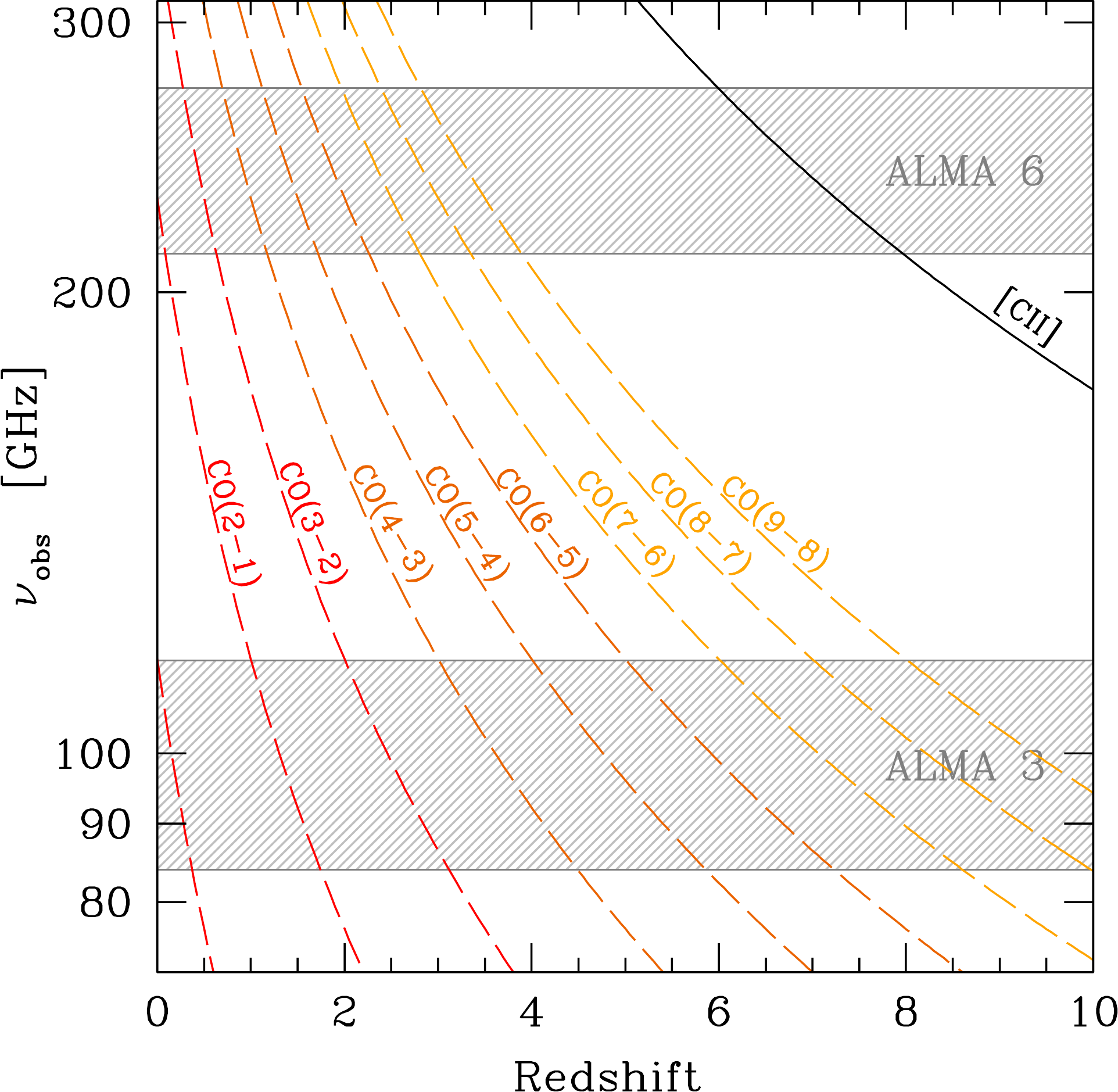}\\
\caption{CO and \Cii{} redshift coverage of our molecular line scans at 1mm and 3mm. See Table \ref{tab_z_range} for the exact redshift ranges of each transition. The 1mm+3mm synergy provides continuous CO redshift coverage at virtually any redshift, with only a tiny gap at $0.6309<z<0.6950$. The \Cii{} emission line is covered in the redshift range 6$<$z$<$8 and is discussed in {\it Paper~V}}
\label{fig_z_range}
\end{figure}

As the main constituent of the molecular gas in galaxies, molecular
hydrogen (H$_2$), is too weak to be detected, the next most abundant
tracer is typically used to measure the molecular gas content:
$^{12}$CO (hereafter: CO). Although this molecule is 10$^4$ times less
abundant, the line can be detected in various environments. As a
consequence, this molecule has been used at low and high redshift to
measure gas masses and kinematics. The CO line emission is observed in
various rotational transitions in galaxies \citep[e.g.,][]{carilli13}.
The rotational ground--state (J=1--0) of CO is at 115.271\,GHz, and
the higher rotational states (J$>$1) are approximately equally spaced
by that frequency\footnote{In reality, the spacing changes slightly as
the dipole moment changes for the higher transitions as a result of
centrifugal forces.}. The amount of high--J emission depends on the
{\em a priori} unknown excitation of the molecular gas. Nevertheless,
full frequency scans in the lowest frequency ALMA bands cover CO
emission at essentially all redshifts (see Fig.~\ref{fig_z_range}).

We here present the rationale for and the observational description of
ASPECS: The ALMA SPECtroscopic Survey in the {\it Hubble} Ultra--Deep
Field (UDF).  This paper is structured as follows: Sec.~2 summarizes
our field choices, as well as the observations and data reduction. In
Sec.~3 we describe our methodology to identify line candidates in our
data cubes, and present the continuum maps of both the band~3 and
band~6 observations. In Sec.~4 we compare our findings to simple
expectations based on previous multi--wavelength analysis of the
galaxies in the field.  We present our summary in Sec.~5.

A number of accompanying papers build on the data presented in this
paper (hereafter: {\it Paper~I}). In {\it Paper~II} (Aravena et
al.~2016a) we analyse the continuum information (mostly based on the
band~6 observations); in {\it Paper III} (Decarli et al.~2016a) we
discuss the implications for CO luminosity functions and the redshift
evolution of the cosmic molecular gas density; in {\it Paper~IV}
(Decarli et al.~2016b) we examine the properties of those galaxies in
the UDF that show bright CO emission; in {\it Paper~V} (Aravena et
al.~2016b) we search for \Cii{} emitters; in {\it Paper~VI} (Bouwens
et al.~2016) we investigate where high--redshift galaxies from ASPECS
lie in relation to known IRX--$\beta$ and IRX--stellar mass
relationships, and finally, in {\it Paper~VII} (Carilli et al.\ 2016)
we describe implications on intensity mapping experiments. Throughout
the paper we assume a standard cosmology with $H_0=70$ km s$^{-1}$
Mpc$^{-1}$, $\Omega_{\rm m}=0.3$ and $\Omega_{\Lambda}=0.7$, broadly
in agreement with the most recent {\em Planck} measurements
\citep{planck15}. Where required, we refer to the AB photometric
system (Oke \& Gunn 1973) for the magnitude definitions and to
Chabrier (2003) for the stellar initial mass function.

\begin{figure*}
\includegraphics[width=0.99\columnwidth]{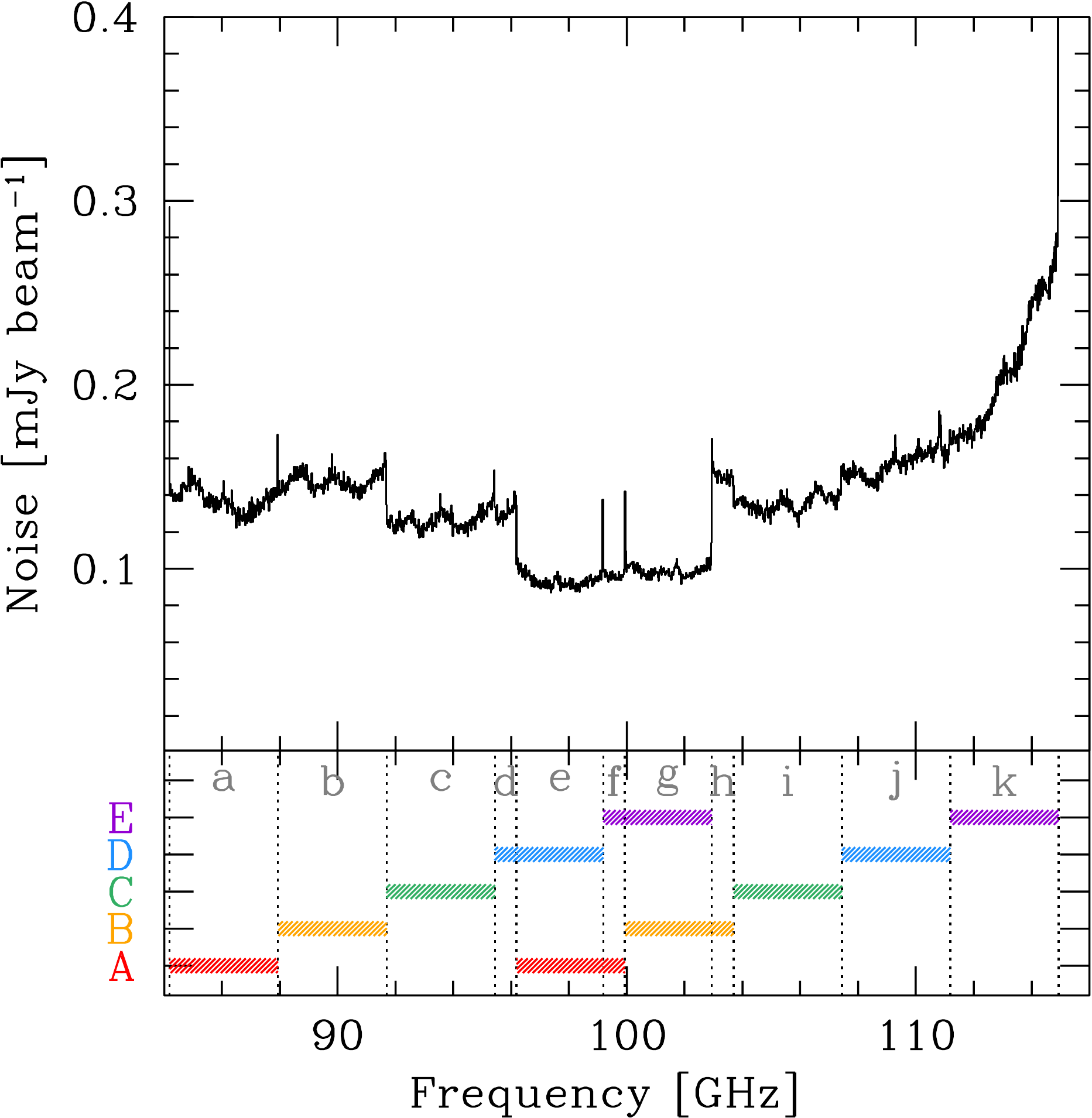}
\includegraphics[width=0.99\columnwidth]{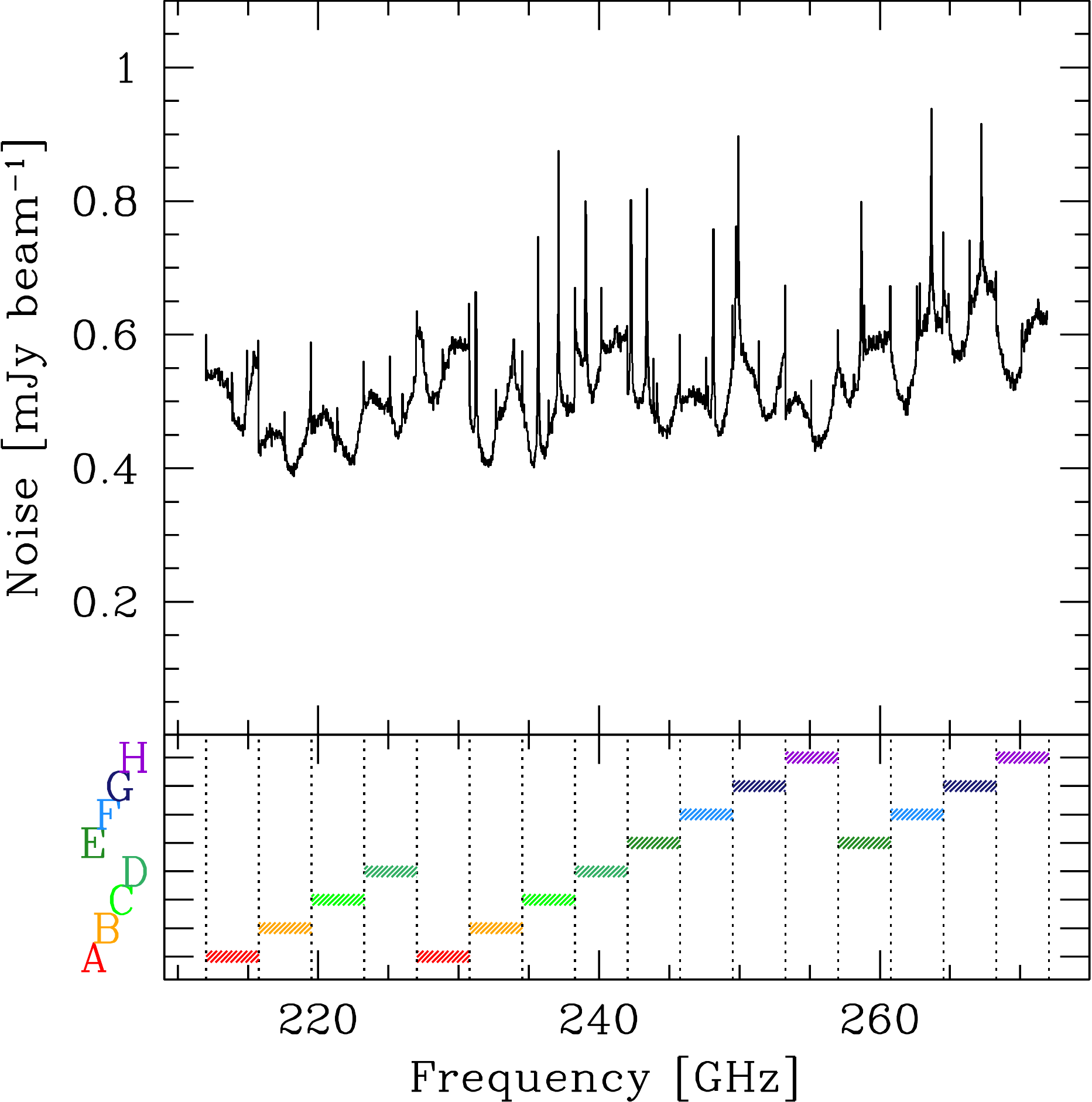}\\
\caption{RMS noise as a function of frequency in the 3mm ({\em left}) and 1mm ({\em right}) scans. At 3mm, each channel is 19.5\,MHz wide (five of the native channels), corresponding to 70\,\kms{} at 84 GHz, and 51\,\kms{} at 115 GHz. The original frequency settings (A-H) are labeled in the bottom panel, together with the frequency blocks (a-k) used in the data reduction. At 1mm, the channels are 31.3\.MHz wide (four of the native channels), corresponding to 44\,\kms{} at 212\,GHz, and to 34\,\kms{} at 272\,GHz. To first order, we reach uniform sensitivity as a function of frequency in both bands. The increase in noise towards high frequencies ($>$113\,GHz) in band 3 is due to the atmosphere (O$_2$).}
\label{fig_noise}
\end{figure*}

\begin{table}
\caption{\rm Lines and corresponding redshift ranges covered in the molecular line scans. For the 3mm data, comoving volume and volume--weighted average redshifts are computed within the primary beam, accounting for its frequency dependence. For the 1mm data, the area is fixed ($3700$ arcsec$^{2}$, as set by the size of the final mosaic).} \label{tab_z_range}
\begin{center}
\begin{tabular}{cccccc}
\hline
Transition  & $\nu_0$ & $z_{\rm min}$ & $z_{\rm max}$ & $\langle z \rangle$ & Volume \\
            & [GHz]   &     &        &       & [Mpc$^3$] \\
    (1)     & (2)     & (3) & (4)    & (5)   & (6) \\
\hline
\multicolumn{6}{c}{band 3: 3mm (84.176--114.928 GHz)}\\
CO(1-0) 	     & 115.271 & 0.0030  & 0.3694 & 0.2801 &   89 \\
CO(2-1) 	     & 230.538 & 1.0059  & 1.7387 & 1.4277 & 1920 \\
CO(3-2) 	     & 345.796 & 2.0088  & 3.1080 & 2.6129 & 3363 \\
CO(4-3) 	     & 461.041 & 3.0115  & 4.4771 & 3.8030 & 4149 \\
CO(5-4) 	     & 576.268 & 4.0142  & 5.8460 & 4.9933 & 4571 \\
CO(6-5) 	     & 691.473 & 5.0166  & 7.2146 & 6.1843 & 4809 \\
CO(7-6) 	     & 806.652 & 6.0188  & 8.5829 & 7.3750 & 4935 \\
\hline
\Ci{}$_{1-0}$        & 492.161 & 3.2823  & 4.8468 & 4.1242 & 4287 \\
\Ci{}$_{2-1}$        & 809.342 & 6.0422  & 8.6148 & 7.4031 & 4936 \\
\hline
\multicolumn{6}{c}{band 6: 1mm (212.032--272.001 GHz)}\\
CO(2-1) 	     & 230.538 & 0.0000  & 0.0873 & 0.0656 &  1.4 \\
CO(3-2) 	     & 345.796 & 0.2713  & 0.6309 & 0.4858 &  314 \\
CO(4-3) 	     & 461.041 & 0.6950  & 1.1744 & 0.9543 & 1028 \\
CO(5-4) 	     & 576.268 & 1.1186  & 1.7178 & 1.4297 & 1759 \\
CO(6-5) 	     & 691.473 & 1.5422  & 2.2612 & 1.9078 & 2376 \\
CO(7-6) 	     & 806.652 & 1.9656  & 2.8044 & 2.3859 & 2864 \\
\hline
\Ci{}$_{1-0}$        & 492.161 & 0.8094  & 1.3212 & 1.0828 & 1233 \\
\Ci{}$_{2-1}$        & 809.342 & 1.9755  & 2.8171 & 2.3973 & 2875 \\
\Cii{}$_{3/2-1/2}$   &1900.548 & 5.9873  & 7.9635 & 6.9408 & 4431 \\
\hline
\end{tabular}
\end{center}
\end{table}

\section{Observations and data reduction}\label{sec_observations}

\subsection{Choice of Frequencies}

Given the unknown excitation of the molecular gas in a given
high--redshift galaxy, when inferring H$_2$ masses, it is advantageous
to observe the CO emission in the lowest rotational state possible to
minimize excitation corrections, modulo the impact of the Cosmic
Microwave Background (da Cunha et al.\ 2013b).  With ALMA, the lowest
frequencies are accessible in band~3, which therefore is the primary
band of choice.  An important complement are line scan observations in
band~6, as the combination of both bands results in the following: (i)
other than one small gap in redshift, there is essentially complete
redshift coverage at all redshifts (see Fig.~\ref{fig_z_range} and
Tab.~\ref{tab_z_range}), (ii) the CO excitation (or limits on it) can
be immediately constrained through the detection of multiple
rotational transitions, (iii) deep continuum maps in the respective
observing bands are available `for free', and (iv), the
highest--redshift sources at $6<z<8$ can be probed through \Cii{}
emission.

Band~7 (275--373\,GHz) observations may be preferred when one is
interested only in the continuum flux densities of the galaxies but
such observations would only recover very high J (J$>$6) transitions
at $z>2$, which may not be highly excited in main sequence galaxies
\citep{daddi15}. Also, the field of view is smaller than in band~6,
necessitating more extensive mosaicing. The bandwidth of band~7
($\sim100$\,GHz) requires more than 13 frequency tunings (each with a
bandwidth of 8\,GHz). For all of these reasons, band~6 is preferred
over band~7 to complement the band~3 observations.

We obtained full frequency scans in both ALMA band~3 and band~6. In
band~3 this implied 5 frequency setups, labelled A--E in
Fig.~\ref{fig_noise}. Both the upper and lower sideband cover
3.75\,GHz, with a gap of $\sim$8\,GHz. For that reason, the central
range in band 3 was covered twice, resulting in observations with
lower noise in that frequency window. Such an overlap region did not
result from the setup of the band~6 frequency scan, as the gap between
the upper and lower sideband in band~6 is 12\,GHz (see right panel in
Fig.~\ref{fig_noise}).  
panels of Fig.~\ref{fig_noise} shows the resulting noise as a function
of frequency.  The noise increase in band~3 towards the higher
frequencies is due the atmospheric oxygen line significantly
increasing the system temperatures above $>$113\,GHz. As a consequence
of the higher frequency, the noise in band~6 observations was
significantly higher (and less well--behaved due to skylines etc) than
in band~3.

\subsection{Choice of field}

In principle such molecular line scan observations could be obtained
at (almost) any position in the sky that is not affected by foreground
emission (either our Galaxy, or other nearby galaxies). However, the
analysis and interpretation of the detected galaxies is greatly
facilitated if a field is chosen for which multi--wavelength
observations already exist. It also should be a field that is easily
accessible to ALMA. The {\it Hubble} Ultra Deep Field
\citep[UDF,][]{beckwith06} is the cosmological field with the deepest
observations in all important wavebands, with 18,000 catalogued
galaxies \citep{coe06}.  The UDF is situated in the 30$'$ Extended
{\it Chandra} Deep Field South \citep[ECDFS][]{lehmer05} / GOODS-South
\citep{giavalisco04} / CANDELS \citep{grogin11,koekemoer11} region, so
the large--scale structure around this field is well quantified.

\begin{figure*}
\begin{center}
\includegraphics[width=0.99\columnwidth]{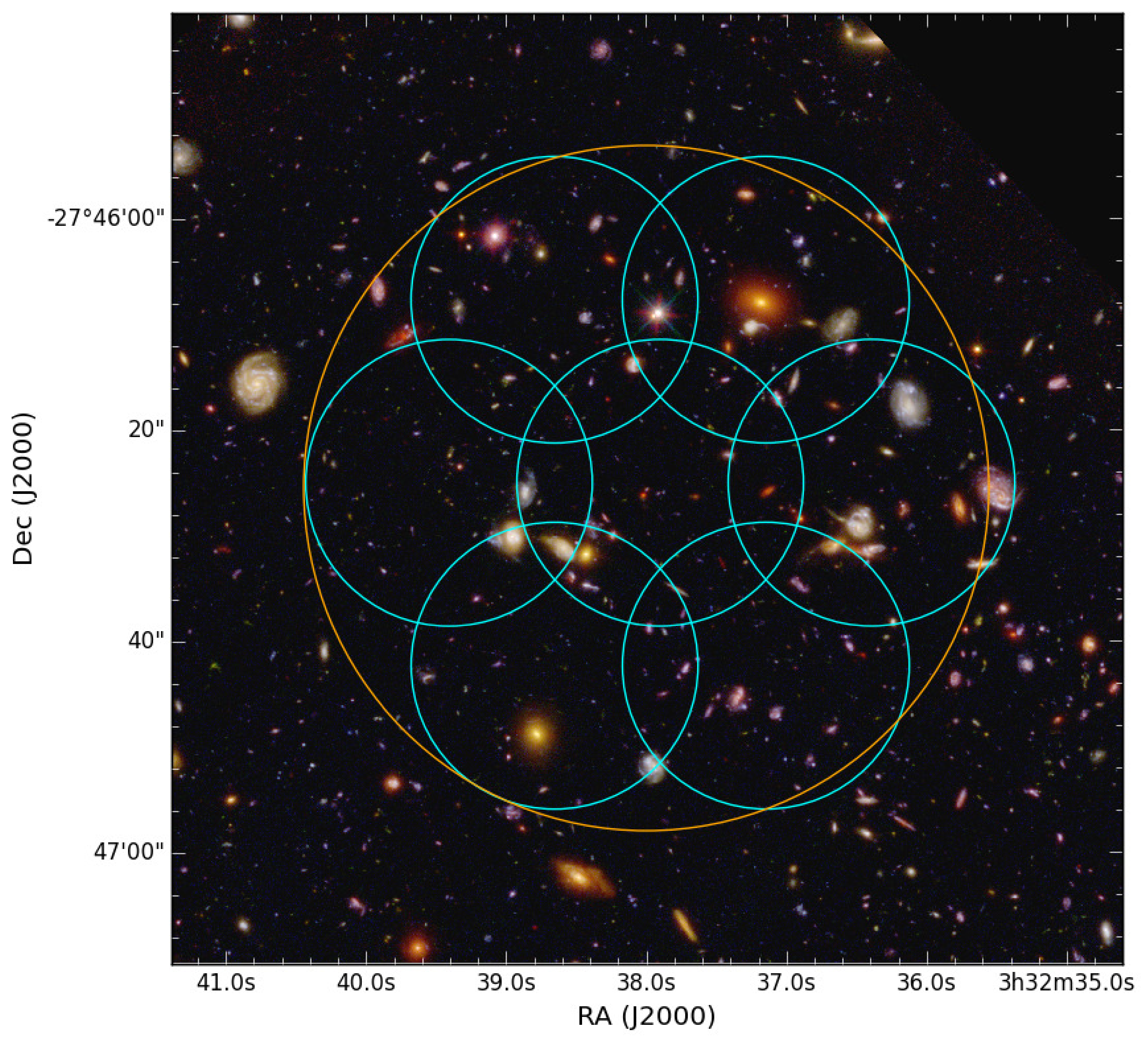}
\includegraphics[width=0.99\columnwidth]{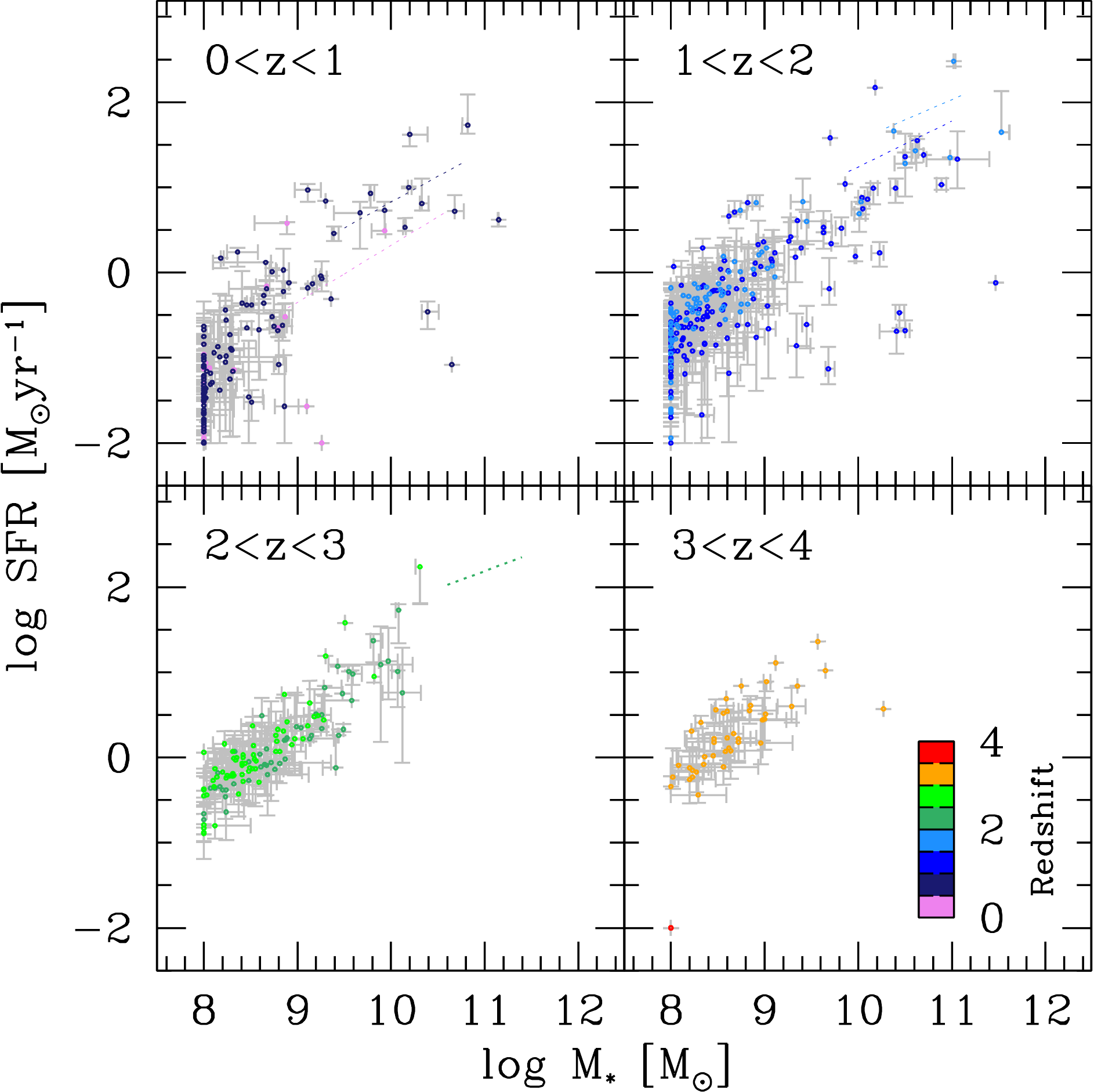}\\
\end{center}
\caption{{\it Left:} The FWHM of the primary beam (i.e. areal coverage) of our 3mm (orange) and 1mm (cyan) observations, overlaid on a three--color {\it HST} F435W / F775W / F105W image of the field from the XDF survey \citep{illingworth13}. The circles show the primary beam of each pointing at the central frequencies of the two scans. {\it Right:} Star formation rates vs. stellar masses for the galaxies in the target field, derived from MAGPHYS fitting (described in Sec.~\ref{sec_magphys}). The four panels show galaxies in different redshift ranges.}
\label{fig_pointings}
\end{figure*}

The goal of the ALMA frequency scan was to reach a sensitivity such
that the predicted `knee' of the CO luminosity function could be
reached at $z\sim2$ \citep[e.g.,][]{sargent14}. Given that multiple
frequency settings were needed to cover both band~3 and band~6, and
given the limited amount of time available in ALMA cycle~2, this
implied that only the area corresponding to one pointing in band~3
could be covered by our observations. This $\sim1'$ region was covered
with a 7--point mosaic in band 6 (see Fig.~\ref{fig_pointings}).  Our
pointing was chosen to lie in the deepest part of the UDF, the
so--called UDF12 (Ellis et al.\ 2013) or eXtremely Deep Field
\citep[XDF,][]{illingworth13} (hereafter: XDF), and included the
highest number of $z$--drop galaxy candidates, i.e. galaxies at $z>6$,
that could be detectable in \Cii{} emission. The field also comprises
a significant overlap with the deepest MUSE observations of the UDF
(Bacon et al., in prep.). The region covered by our observations
comprise $\sim$10\% of the total area of the UDF (corresponding to a
co-moving survey volume of ~18,000 Mpc$^3$ out to $z\sim8$) and
harbors roughly $\sim$1500 optical/NIR--selected galaxies. In
Fig.~\ref{fig_pointings} we also present the star formation rates and
stellar masses of all galaxies covered by our observations, based on
fitting of the galaxies' spectral energy distribution
(Sec.~\ref{sec_magphys}).

\subsection{Choice of Array Configuration}

ALMA has been designed to reach high (sub--arcsec) angular resolution.
However, to be sensitive to the full molecular gas reservoir in a
galaxy, observations in a compact array configuration are essential to
ensure that no extended CO emission is missed by the
interferometer. Note that this is not related to the `missing short
spacing' problem\footnote{The missing short spacing problem means that
the interferometer is `blind' to spatial scales above a certain
size. Given the likely clumpiness of high--redshift galaxies, missing
short spacings should typically not be a concern in high--redshift
galaxy observations.}. For instance, in observations with extended
ALMA array configurations the synthesized beam will end up being
smaller than the typical size of a high--redshift galaxy. As a result,
the amount of emission per beam is only a fraction of the total
emission of the galaxy, while the noise does not change. In the case
of low S/N detections, this will result in the non-detection of a
source, whereas the emission would be detected by a compact
configuration. Our observations were taken in the C34-2 and C34-1
configurations, resulting in beam sizes of $3.6''\times 2.1''$
(band~3) and $1.7''\times 0.9''$ (band~6), i.e. well matched to the
expected sizes of the galaxies under consideration.

\subsection{Observations}

The project consists of two spectral scans, one at 3mm (band~3) covering the
frequency range 84--115 GHz and one at 1mm (band~6) covering the
frequency range 212--272 GHz (see Figs.~\ref{fig_z_range} and~\ref{fig_noise}). 
The  time allocated for both projects amounts to a total of $\sim$40\,hours including overheads
(split approximately 50--50 between band~3 and band~6).

The 3mm observations (ALMA Project ID: 2013.1.00146.S) were carried
out between July $1^{\rm st}$, 2014 and January $6^{\rm th}$,
2015. The 3mm scan consisted of a single pointing (RA=03:32:37.90
Dec=--27:46:25.0, J2000.0) and 5 frequency settings (see
Fig.~\ref{fig_noise}). Each setting had $4\times1.875$ GHz spectral
windows (two in the upper side band, and two in the lower side band),
and was observed in three execution blocks. The native channel width
is $3.9025$\,MHz, or $\sim$12\,km\,s$^{-1}$ at $\sim$100\,GHz.
Observations were carried out in a relatively compact (C34-2) array
configuration with 29--41 antennas, with baselines ranging between 12
and 612\,m. The quasar J0348--2749 was observed in the majority of the
execution blocks as phase and amplitude calibrator, while Uranus and
the quasars J0334--4010 and J0334--4008 were used as flux and bandpass
calibrators.  Data were calibrated and imaged with the {\it Common
Astronomy Software Applications} package (CASA) version 4.2.2 of the
ALMA pipeline.

To combine the different setups we adopted the following approach: 
1) For each execution block, we split out
cubes in frequency ranges as shown in Fig.~\ref{fig_noise} (a-k). 2)
The frequency ranges marked with the letters e-g, i.e., where upper
and lower side band observations from different frequency settings
overlap, were re-sampled using the CASA task \textsf{ms.cvel}.  3) We
then combined all the available data for each individual frequency range (a-k)
using the CASA task \textsf{concat}. 4) Upper and lower side band data
come with different weighting scales, although the data quality is
comparable. We therefore ran \textsf{statwt} in order to homogenize
the weighting system in the concatenated data. 5) We combined all the
frequency ranges using the task \textsf{concat} again.

We imaged the 3mm cube after averaging over two and five native
channels (7.8\,MHz and 19.5\,MHz respectively) using natural
weighting. The 19.5\,MHz channels correspond to 70\,\kms{} at 84 GHz,
and 51\,\kms{} at 115 GHz. We created a band~3 continuum map as well
(see discussion in Sec.~\ref{sec_continuum}). The corresponding
primary beams of the ALMA antennas are $75''$ at 84 GHz and $55''$ at
115 GHz. The restored synthesized beam size is $3.5''\times 2.0''$
(FWHM) with PA=84$^\circ$. We thus adopted a pixel scale of
$0.5''$\,pixel$^{-1}$, and an image size of $90''\times90''$. A
primary beam correction has been applied for all quantitative
analysis.  The final data set covers the frequency range
84.176--114.928 GHz, and reaches an rms of
$0.1-0.25$\,mJy\,beam$^{-1}$ per 19.5\,MHz channel (see
Fig.~\ref{fig_noise}). For comparison, the PdBI spectral scan at 3mm
in the {\it Hubble} Deep Field North (Decarli et al.\ 2014, Walter et
al.\ 2012, 2014) reached a sensitivity of $\sim0.3$\,mJy\,beam$^{-1}$
per 90\,\kms{} channel, or $\sim0.4$\,mJy\,beam$^{-1}$ at the sampling
adopted here. Therefore these ALMA observations are a factor 3--4
deeper at $\nu<113$\,GHz than the previous 100\,hour (on--source)
effort with PdBI (Decarli et al.\ 2014).

The 1mm observations (ALMA Project ID: 2013.1.00718.S) were carried
out between December $12^{\rm th}$, 2014 and April $21^{\rm st}$,
2015. In order to cover a similar area as the 3mm pointing, a 7--point
mosaic was observed, centred on the same coordinates as for the 3mm
observations (see Fig.~\ref{fig_pointings}). For each pointing
position, eight frequency settings were needed to cover the entire
band (see Fig.~\ref{fig_noise}), resulting in continuous coverage from
212--272\,GHz. In this case, there was no overlap between different
spectral windows of various frequency tunings (see
Fig.~\ref{fig_noise}).  Observations were carried out in the most
compact available array configuration (C34-1) with 30--34
antennas. Baselines ranged between 12 and 350\,k$\lambda$. The quasar
J0348--2749 was adopted as phase and amplitude calibrator, while
Uranus and the quasar J0334--4008 acted as flux and bandpass
calibrators.  The cube was imaged in spectral samplings of 4, 8 and 12
native channels, corresponding to $15.6$\,MHz, $31.2$\,MHz, and
$46.8$\,MHz respectively, as well as in a continuum image. The
$31.2$\,MHz sampling corresponds to 44\,\kms{} at 212\,GHz and to
34\,\kms{} at 272\,GHz. We adopted natural weighting, yielding a
synthesized beam of $1.5''\times 1.0''$ with PA=-79$^\circ$. We
adopted a pixel scale of $0.3''$ per pixel. The final mosaic covers a
region of approximately $75''\times 70''$ to the half--sensitivity
point.

\section{Line search}\label{sec_search}

The data reduction resulted in two data cubes, one in band~3 and
one in band~6, as well as continuum maps, which we discuss later (Sec.~\ref{sec_continuum}). 
We here describe our methodology to search for line
emitting sources in these  cubes.

\subsection{The blind line search}

For our blind search of line candidates, we developed an IRAF--based
routine, \textsf{findclumps}, which operates directly on the imaged
data cubes. The script performs floating averages of a number of
channels, computes the rms of the averaged maps, and searches for
peaks exceeding a certain S/N threshold using the IRAF task
\textsf{daofind}. The position, frequency, and S/N of the recovered
candidates is saved.  As input, we used the $7.8$\,MHz and $31.2$\,MHz
sampling for the 3\,mm and 1\,mm cubes respectively. Since the
significance of a line detection is maximized when averaging over a
frequency range comparable with the actual width of the line, we ran
our search over 3, 5, 7, and 9--channel windows, i.e., kernel line
widths of $\sim$50--300\,\kms{} (an inspection using larger
line--widths did not result in additional detections -- this is also
supported by our completeness test, see below).

The list of line candidates identified by this procedure is then
trimmed in order to keep only candidates that lie within $\sqrt{2}
\times$ the primary beam radius at 3mm (= $53''$ at 84\,GHz, $39''$ at
115\,GHz), equivalent to a response of $\sim$30\% and within a fixed
radius of $30.9''$ at 1mm (given that the latter is a mosaic).

The floating-average approach and the use of different windows of
spectral sampling allow us to avoid missing candidates because of a
priori choices in terms of spectral bins. However, our candidate lists
are subject to multiplicity both spatially and spectrally. Moreover,
the 1mm search is bound to pick up bright continuum sources as
potential line candidates. We therefore masked a posteriori the line
candidates associated with the two brightest 1mm continuum sources
(see Sec.~\ref{sec_continuum} and {\it Paper II}). We consider as
duplicates line candidates that are offset by less than one
synthesized beam ($\sim\!2.5''$ at 3mm, $\sim\!1.5''$ at 1mm) and that
appear in consecutive channels in the floating average.

When assessing the reliability of our line candidates, we need to keep two
separate issues in mind (`fidelity' and `completeness', which we discuss 
in Secs.~\ref{sec_fidelity} and~\ref{sec_completeness}).

\subsubsection{Fidelity}\label{sec_fidelity}

First, is a given line detection significant? This question is harder
to address in practice than one would naively think: the S/N of a
single detection will be a function of the width of the line, and the
noise in the cubes is not Gaussian. The best way to address this
question is to perform two independent searches: (a) for positive
emission; these candidates would correspond to both real astrophysical
sources and noise peaks, (b) for negative emission; these candidates
would only correspond to non-astrophysical sources\footnote{An
interesting hypothesis is that at least some of the negative sources
are in fact real absorption systems due to absorption against the
CMB. However, our checks revealed that none of the significant
negative sources are either associated with a galaxy visible in the
UDF, nor with a strong continuum emission. We conclude that the
negative sources revealed by our search are physically
implausible.}. These latter sources can be used to define a term that
we refer to as {\em fidelity}, i.e. we can statistically subtract the
unphysical `negative' lines from the physical `positive' ones.

\begin{figure}
\includegraphics[width=0.99\columnwidth]{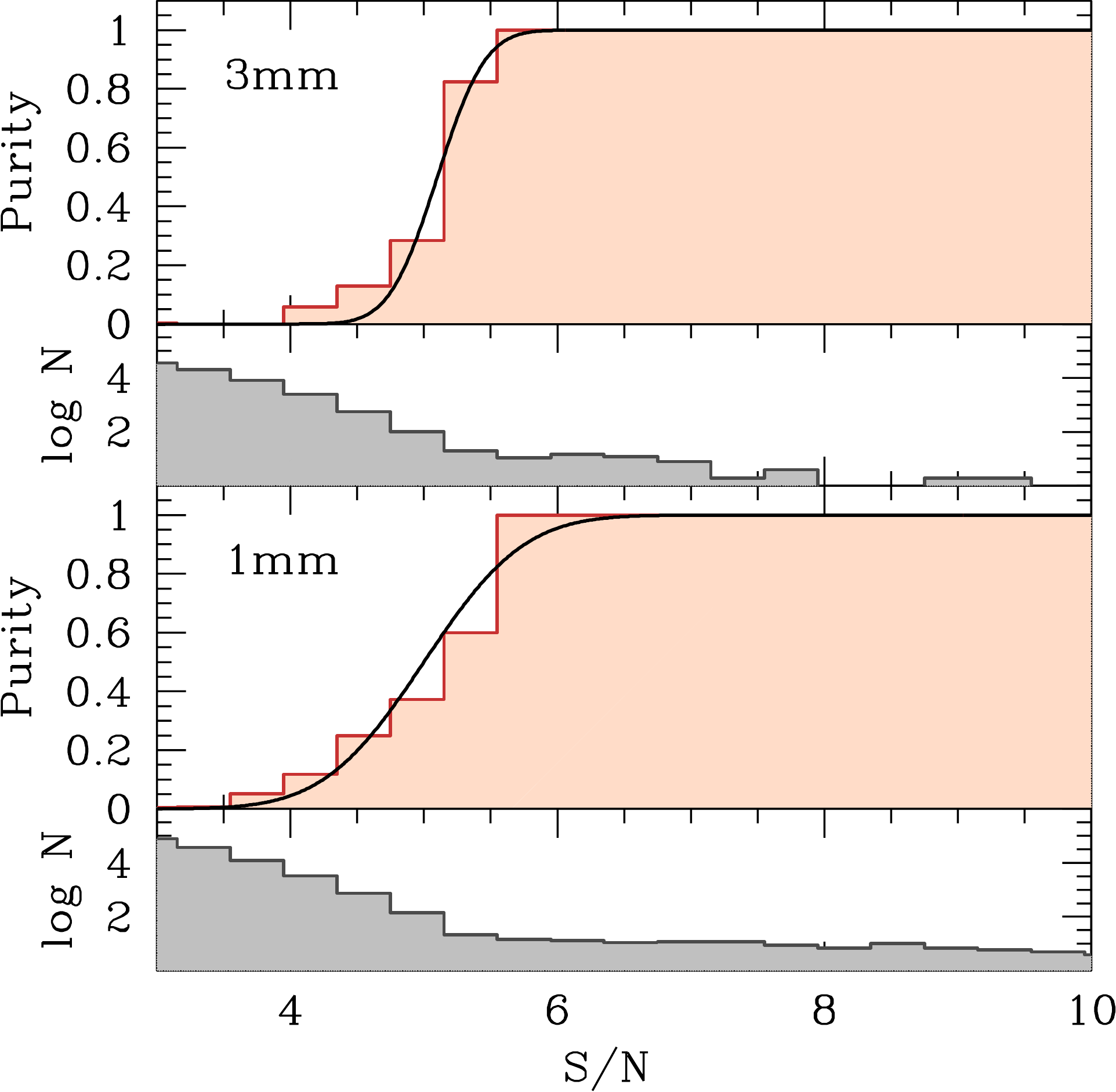}\\
\caption{The {\em fidelity} in our line search, plotted as a red histogram as a function of the line $S/N$ of the individually detected candidates. The fidelity is defined as in eq.~\ref{eq_fidelity}. The number of candidates as a function of $S/N$ is also shown. We model the fidelity dependence on $S/N$ as an error function (solid black line). The search reaches 60\% fidelity at $S/N\sim5.2$ both at 1mm and 3mm, although the latter shows a sharper increase of fidelity with $S/N$. We choose a fidelity level of $>$60\% for the sources that enter our analysis, implying that, at the lowest significance, out of a sample of 10 candidates, 6 are likely real, and 4 sources at similar $S/N$ were also detected with negative signal.}
\label{fig_fidelity}
\end{figure}

We thus assess the degree of fidelity in our line search by running
the same search over the positive and negative peaks. The basic
assumption is that, given the interferometric nature of our data set,
and that we do not expect to detect absorption features against very
high-S/N continuum emission, all the `negative' line candidates will
be noise peaks, while the `positive' line candidates will be a mixture
of noise peaks and genuine lines. The search for negative peaks is
performed in the exact same way as the one for positive emission. By
comparing the results of these two searches, we can quantify the
fidelity of our search at a given line candidate significance as
follows: \begin{equation}\label{eq_fidelity} {\rm fidelity}(S/N) = 1 -
\frac{N_{\rm neg}(S/N)}{N_{\rm pos}(S/N)} \end{equation} where $N_{\rm
pos}(S/N)$ and $N_{\rm neg}(S/N)$ are the number of positive and
negative line candidates with a given $S/N$, respectively. This
definition is such that, if the number of negative candidates at a
given $S/N$ is comparable to the number of positive candidates, then
the fidelity is null; if it is negligible, then the fidelity is close
to 100\%. For the analysis of our blind search, we request a fidelity
level of 60\% or higher. This threshold was chosen so that at the
lowest accepted significance, more than half of the `positive' line
candidates are real. We determine the signal-to-noise ($S/N$) ratio
computed by \textsf{findclumps} as follows: For each floating-averaged
channel, we compute the map rms (which will constitute the `noise'
term) and we take the peak pixel value at the position of a line
candidate as `signal'.  We emphasize that, since the averaging window
is not optimized to match the actual width of a line candidate (also
this approach assumes spatially-unresolved line emission), this
definition of $S/N$ is by construction conservative. The $S/N$ values
of each line candidate are reported in Tab.~\ref{tab_lines}. In
Fig.~\ref{fig_fidelity} we show how the fidelity of our line search
changes as a function of the line $S/N$. It is convenient to have an
analytical description of the fidelity dependence on $S/N$. While not
physically motivated, the following error function provides a good
description of the observed trend, with the following
parameterization: \begin{equation}\label{eq_fidelity2} {\rm
fidelity}(S/N) = \frac{1}{2}\ {\rm
erf}\left(\frac{S/N-C}{\sigma}\right)+0.5 \end{equation} where $C_{\rm
3mm}$=5.1, $C_{\rm 1mm}$=5.0, $\sigma_{\rm 3mm}$=0.4, $\sigma_{\rm
1mm}$=0.8. This implies that we reach 60, 80, and 95\% fidelity levels
at $S/N$=$5.17$, $5.34$, and $5.57$ at 3mm, and at $S/N$=$5.15$,
$5.50$, and $5.97$ at 1mm. We will use this equation to assess the
fidelity for our individual line detections.

\subsubsection{Completeness}\label{sec_completeness}

The second question concerns our ability to extract faint sources from
our data cubes ({\em completeness}). We address this by inserting
artificial line sources of various strengths and widths in our data
cubes, then calculating our ability to recover them in our line
search. This is a standard way of deriving the completeness of sources
in the low S/N regime.

\begin{figure*}
\includegraphics[width=0.99\columnwidth]{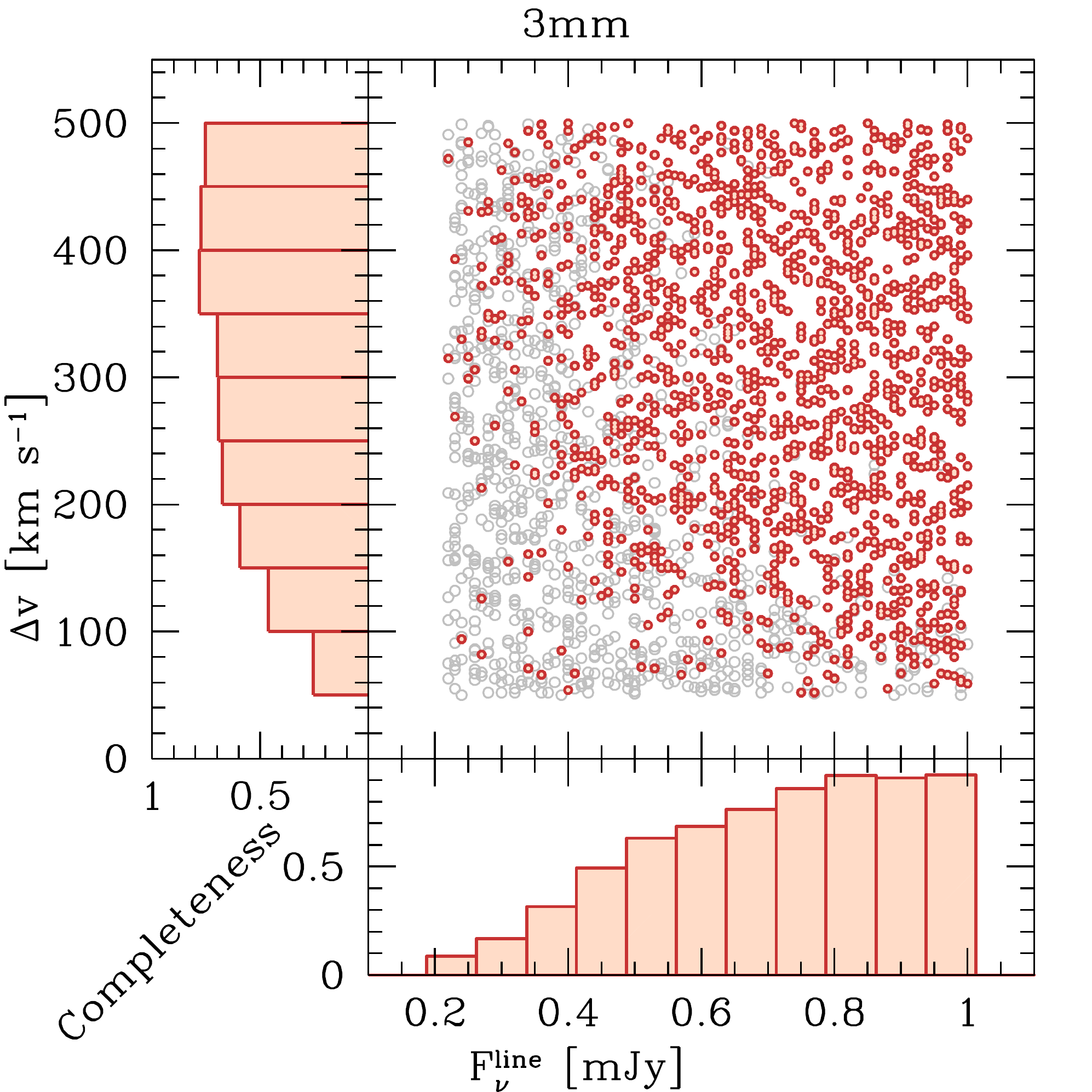}
\includegraphics[width=0.99\columnwidth]{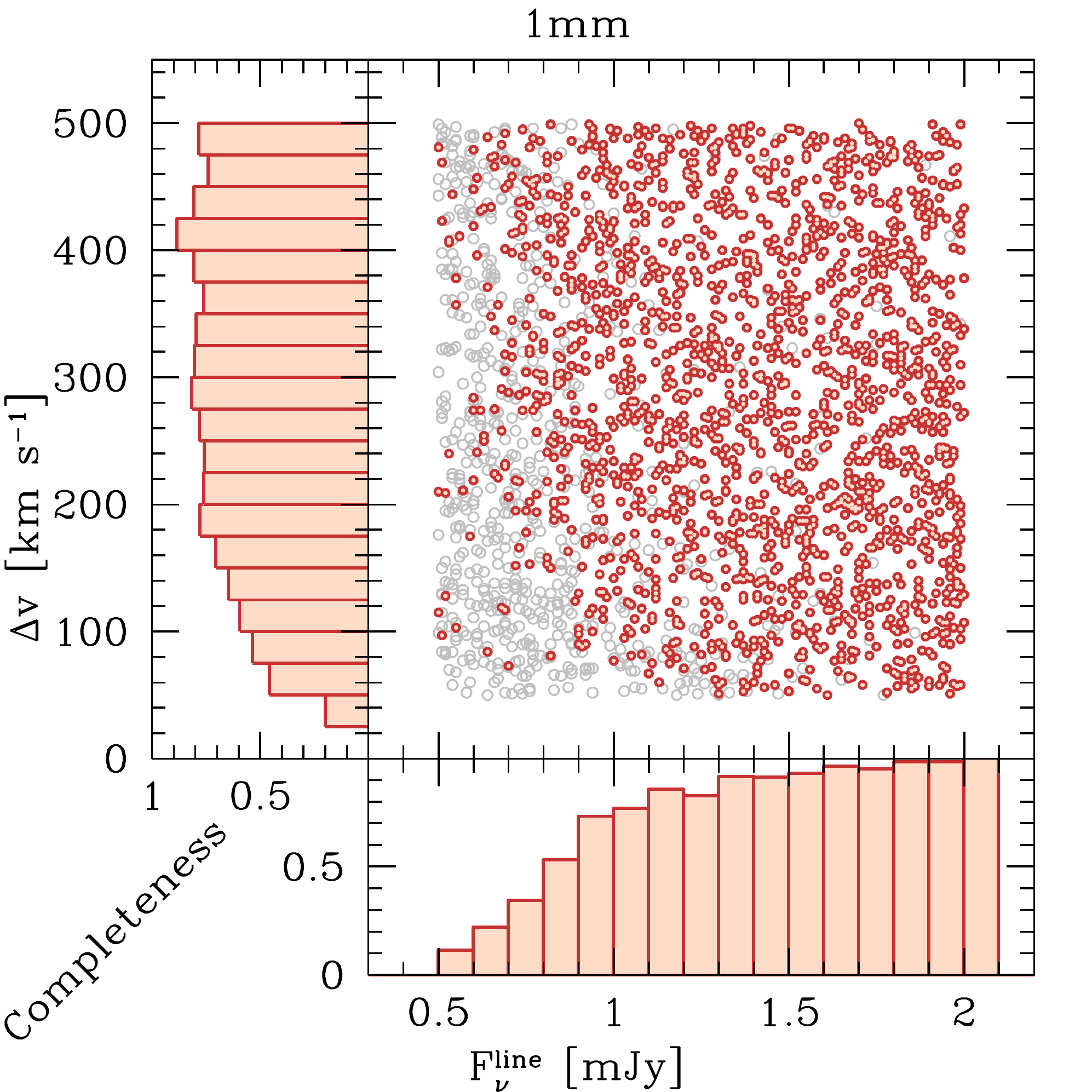}\\
\caption{Completeness assessment of our line search. In each diagram, each circle represents an artificially injected line candidate. Filled symbols highlight the candidates that we recover in our analysis. The histograms show the marginalization along the y- and x-axis respectively, showing the level of completeness (i.e., the fraction of input line candidates that our script successfully identifies) as a function of the line width ($\Delta v$) and peak flux density ($F_\nu^{\rm line}$), respectively. The 3mm case is shown on the left, the 1mm one is on the right.}
\label{fig_completeness}
\end{figure*}

We assess the level of {\em completeness} in our blind line search by
adding 2500 artificial line sources to the data cube,
re--running our line searching algorithm, and comparing the number of
recovered sources with the input catalog. The line candidates are
assumed to have a gaussian profile along the spectral axis, and the
shape of the synthetic beam in the maps. The line spatial positions,
peak frequencies, peak flux densities and widths are randomly
generated with uniform distributions as follows: RA, Dec and the line
peak frequency are required to be homogeneously distributed within the
cubes. The line peak flux density range between 0.22--1.00\,mJy (at
3mm) and between 0.5--2.0\,mJy (at 1mm), where the fainter side is set
to roughly match the 1-$\sigma$ typical limit of each channel. The
line widths span the range 50--500\,\kms{}. In
Fig.~\ref{fig_completeness} we show how the completeness of our line
search is a function of the input width and peak flux density of the
lines. At 3mm, the completeness is $>50$\%  for peak flux
densities $F_\nu^{\rm line}>0.45$\,mJy, and for line widths $\Delta
v>100$\,\kms{}. We also observe a minor dependence of the completeness
on the frequency due to the decreasing sensitivity towards the high
frequency end of the scan (see Fig.~\ref{fig_noise}). The line search
in the 1mm mosaic shows a completeness $>50$\%  for peak flux
densities $>$0.8\,mJy and widths $>100$\,\kms{}.  These
completeness corrections will be used extensively in {\it Paper~III}.

\begin{table*}
\caption{{\rm Catalogue of the line candidates identified in our analysis. (1) Line ID. (2-3) Right ascension and declination (J2000).  (4) Central frequency and uncertainty, based on Gaussian fit. (5) Velocity integrated flux and uncertainty. (6) Line Full Width at Half Maximum, as derived from a Gaussian fit. (7) signal-to-noise as measured by the line searching algorithm. (8) Spatially coincident optical/NIR counterpart? (9) Comments on line identification.}} \label{tab_lines}
\begin{center}
\begin{tabular}{ccccccccl}
\hline
ID  & RA	  & Dec 	& Frequency	  & Flux	     & FWHM     & $S/N$ & Opt/NIR & Comments \\  
ASPECS...    & (J2000.0)   & (J2000.0)	& [GHz] 	  & [Jy\,km\,s$^{-1}$] & [\kms] &		   & c.part? &          \\  
 (1) & (2)	  & (3) 	& (4)		  & (5) 	     & (6)	& (7)		   &  (8)    & (9)      \\  
\hline
\multicolumn{9}{l}{\bf ~3mm (band 3)} \\
 3mm.1  & 03:32:38.52 & -27:46:34.5 & $ 97.567_{-0.003}^{+0.003}$ & $0.72\pm0.03$ & $500_{-30}^{+30}$  & 19.91     & Y & J=3; J=7,8 also detected \\
 3mm.2  & 03:32:39.81 & -27:46:11.6 & $ 90.443_{-0.003}^{+0.003}$ & $0.44\pm0.08$ & $540_{-30}^{+30}$  & 12.80     & Y & J=2; J=5 tentatively detected. Confirmed\\
\multicolumn{8}{c}{ }                                                	     	                                    & by opt. spectroscopy \\
 3mm.3  & 03:32:35.55 & -27:46:25.7 & $ 96.772_{-0.003}^{+0.003}$ & $0.13\pm0.01$ & $ 57_{-30}^{+30}$  &     9.48  & Y & J=2 is ruled out by optical spectroscopy \\
 3mm.4  & 03:32:40.64 & -27:46:02.5 & $ 91.453_{-0.003}^{+0.003}$ & $0.23\pm0.03$ & $ 73_{-30}^{+30}$  &     5.86  & N & lack of counterpart suggests J$>$2 \\
 3mm.5  & 03:32:35.48 & -27:46:26.5 & $110.431_{-0.003}^{+0.003}$ & $0.18\pm0.02$ & $ 82_{-25}^{+25}$  &     5.42  & Y & J=2 confirmed by optical spectroscopy \\
 3mm.6  & 03:32:35.64 & -27:45:57.6 & $ 99.265_{-0.003}^{+0.003}$ & $0.23\pm0.02$ & $160_{-30}^{+30}$  &     5.40  & N & lack of counterpart suggests J$>$2  \\
 3mm.7  & 03:32:39.26 & -27:45:58.8 & $100.699_{-0.003}^{+0.003}$ & $0.08\pm0.01$ & $ 60_{-30}^{+25}$  &     5.40  & N & lack of counterpart suggests J$>$2  \\
 3mm.8  & 03:32:40.68 & -27:46:12.1 & $101.130_{-0.003}^{+0.003}$ & $0.19\pm0.01$ & $100_{-30}^{+25}$  &     5.30  & N & no match with nearby galaxy; J$>$2  \\
 3mm.9  & 03:32:36.01 & -27:46:47.9 & $ 98.082_{-0.003}^{+0.003}$ & $0.09\pm0.01$ & $ 64_{-30}^{+30}$  &     5.28  & N & lack of counterpart suggests J$>$2  \\
 3mm.10 & 03:32:35.66 & -27:45:56.8 & $102.587_{-0.003}^{+0.003}$ & $0.24\pm0.02$ & $120_{-25}^{+25}$  &     5.18  & Y & J=3 ($z$=$2.37$) would match $z_{\rm grism}=2.33$\\
\hline
\multicolumn{9}{l}{\bf ~1mm (band 6)} \\
 1mm.1 & 03:32:38.54  & -27:46:34.5 & $227.617_{-0.003}^{+0.003}$ & $0.79\pm0.04$ & $463_{-10}^{+80}$ & 18.28     & Y & J=7 \\
 1mm.2 & 03:32:38.54  & -27:46:34.5 & $260.027_{-0.059}^{+0.003}$ & $1.10\pm0.05$ & $478_{-70}^{+11}$ & 16.46     & Y & J=8 \\
 1mm.3 & 03:32:38.54  & -27:46:31.3 & $225.181_{-0.003}^{+0.003}$ & $0.22\pm0.02$ & $101_{-18}^{+18}$ &     5.87  & Y & J=3 would imply $z$=0.54, and $z_{\rm grism}=0.59$ \\
 1mm.4 & 03:32:37.36  & -27:46:10.0 & $258.333_{-0.003}^{+0.016}$ & $0.27\pm0.02$ & $150_{-20}^{+20}$ &     5.62  & N & if \Cii{}, tentative CO(6-5) detection is reported. \\
 \multicolumn{8}{c}{ }                                               	     	                                   & Possibly lensed by foreground Elliptical?\\
 1mm.5 & 03:32:38.59  & -27:46:55.0 & $265.320_{-0.031}^{+0.003}$ & $0.72\pm0.03$ & $211_{-10}^{+37}$ &     5.47  & N & lack of other lines suggests J=4\\
 1mm.6 & 03:32:36.58  & -27:46:50.1 & $222.553_{-0.003}^{+0.003}$ & $0.56\pm0.02$ & $302_{-40}^{+12}$ &     5.45  & Y & J=4 yields $z$=$1.07$, J=5 yields $z$=$1.59$, \\
 \multicolumn{8}{c}{ }                                               	     	                                   & J=6 yields $z$=$2.11$, tentative second line for J=4 or J=6 \\
 1mm.7 & 03:32:37.91  & -27:46:57.0 & $257.042_{-0.003}^{+0.003}$ & $1.78\pm0.03$ & $179_{-11}^{+11}$ &     5.43  & N & lack of other lines suggests J=4 \\
 1mm.8 & 03:32:37.68  & -27:46:52.6 & $222.224_{-0.003}^{+0.022}$ & $0.39\pm0.02$ & $210_{-12}^{+30}$ &     5.33  & N & lack of counterpart excludes J=2,3; lack of second \\
 \multicolumn{8}{c}{ }                                               	     	                                   & line exclude CO. \Cii{}? \\
 1mm.9 & 03:32:36.14  & -27:46:37.0 & $249.085_{-0.003}^{+0.016}$ & $0.34\pm0.02$ & $150_{-20}^{+20}$ &     5.19  & N & J=4; lack of counterparts excludes J$<$4, and lack\\
  \multicolumn{8}{c}{ }                                              	     	                                   & of other lines excludes J$>$4 \\
 1mm.10 & 03:32:37.08 & -27:46:19.9 & $237.133_{-0.003}^{+0.003}$ & $0.49\pm0.04$ & $281_{-12}^{+48}$ &     5.18  & N & J=4 or 6 due to lack of counterparts and other\\
 \multicolumn{8}{c}{ }                                               	     	                                   & lines. J=4 favoured because of excitation\\
 1mm.11 & 03:32:37.71 & -27:46:41.0 & $223.067_{-0.025}^{+0.003}$ & $0.27\pm0.02$ & $169_{-12}^{+35}$ &     5.16  & N & lack of other lines suggests J=3\\
\hline
\end{tabular}
\end{center}
\end{table*}
							 	
\subsection{Line candidates}

\subsubsection{Properties}

 For our subsequent analysis we consider only those sources that
have a {\em fidelity} of greater than 60\% and where the extracted
line is detected at $>$2$\sigma$ in consecutive channels (width:
$\sim$25\,\kms{} at 3\,mm, $\sim$40\,\kms{} at 1\,mm). Our blind
search resulted in 10 line candidates from the 3\,mm search, and 11
line candidates from the 1\,mm search (see Tab.~\ref{tab_lines} and
the figures in the Appendix). Given our requirement on the {\em
fidelity} in our search, we expect that $<$4 out of these line
candidates are spurious in each band.  We show the candidates, sorted
by S/N of the line emission, in Figs.~\ref{fig_ps_3mm_a} (band 3)
and~\ref{fig_ps_1mm_a} (band 6). In each case, the left panel shows an
{\it HST} color composite, and the middle panel shows the {\it HST}
image in greyscale, and the CO line candidates in contours. The right
panel shows the spectrum extracted at the position of the line
candidate. The basic parameters of the candidate lines (RA, Dec,
frequency, integrated flux, line width and $S/N$) are summarized in
Tab.~\ref{tab_lines}.

\subsubsection{Optical/NIR counterparts}

We have searched for optical/NIR counterparts by matching the
positions of the sources in the multi--wavelength catalogs (Sec.~2.2)
with our line candidates. Whether a specific CO line candidate has a
counterpart or not is summarized in column~8 of Tab.~\ref{tab_lines}
(see also Figs.~\ref{fig_ps_3mm_a} and~\ref{fig_ps_1mm_a}). The lines
that show an optical/NIR counterpart with matching redshift are
discussed in detail in {\it Paper~IV}.

\subsubsection{Redshift determination}

Given the (almost) equi--distant spacing of the rotational transitions
of CO, it is not straightforward to assign a unique redshift to each
candidate in a number of cases.

\paragraph{Multiple CO lines?}

For certain redshifts, more than one CO transition is covered by our
band~3 and band~6 scans. We use this information to constrain the
redshift of some of the candidate. Likewise, in other cases a certain
redshift solution can be ruled out if other detectable lines are not
detected. This information is given in the `comments' column of
Tab.~\ref{tab_lines}.

\paragraph{Optical/NIR spectroscopic redshifts:}

In some cases, spectroscopic redshifts are available for the
optical/NIR counterparts, either through longslit spectroscopy
\citep{lefevre05,kurk13,skelton14,morris15}, or {\it HST} grism
observations \citep{morris15,momcheva16}. We also record this
information in the `comments' column of Tab.~\ref{tab_lines}.

\paragraph{Lack of optical/NIR counterparts:}

In a number of cases, no optical/NIR counterpart of the line candidate
is visible in the {\it HST} image. This is can be due to the fact
that the source is spurious. But if the candidate was real, and
assuming that there is no signficant reddening by dust, then the
exquisite depth of the available optical/NIR observations (in
particular the {\it HST}/WFC3 IR images and the Spitzer/IRAC images)
can place constraints on the stellar mass of galaxies as a
function of redshift. Our MAGPHYS fits (see Sec.~\ref{sec_magphys}) of
the available photometry suggest that a galaxy securely detected in
H-band (1.6$\mu$m) at $>50$\,nJy (corresponding to a secure,
$>$10-$\sigma$ detection in a few bands) has a stellar mass of
$>$4$\times10^6$\,\Msun{}, $>$2$\times10^7$\,\Msun{}, and
$>$10$^8$\,\Msun{} at $z$=0.5, 1.0, and 2.0, respectively. Because of
the combination of low molecular gas content, and likely
elevated $\alpha_{\rm CO}$ values \citep{bolatto13}, we do not expect
to detect CO in galaxies with $M_*\ll 10^9$\,\Msun{}. Therefore
can use the lack of an optical counterpart to set constraints on the
redshift of the candidate. In particular, we assume that line
candidates selected in band~3 and lacking an optical/NIR
counterpart are at $z>2$ (i.e., the line is identified as CO(3-2) or a
higher-J transition). In the case of band~6 candidates, we give
priority to the constraints from the multiple line
(non--)detection. The `lack of counterpart' argument is chosen only to
rule out the lowest--$z$ scenarios (J$<$4, corresponding to
$z<0.695$). This additional constraint on the line candidates is also
given in the `comments' column of Tab.~\ref{tab_lines}.

The total CO flux of all line candidates is 2.55\,Jy\,km\,s$^{-1}$,
whereas the total flux of the candidates that have no optical/NIR
counterpart is 0.83\,Jy\,km\,s$^{-1}$ (from Tab.~\ref{tab_lines}),
i.e. $\sim$33\% of the total. As some of the line candidates that
do not show an optical/IR counterpart are likely spurious, and
considering that the brightest CO detections with optical/NIR
counterparts dominate the total emission, the flux fraction of real
objects without optical/NIR counterpart is likely lower.

\begin{figure*}
\includegraphics[width=0.99\columnwidth]{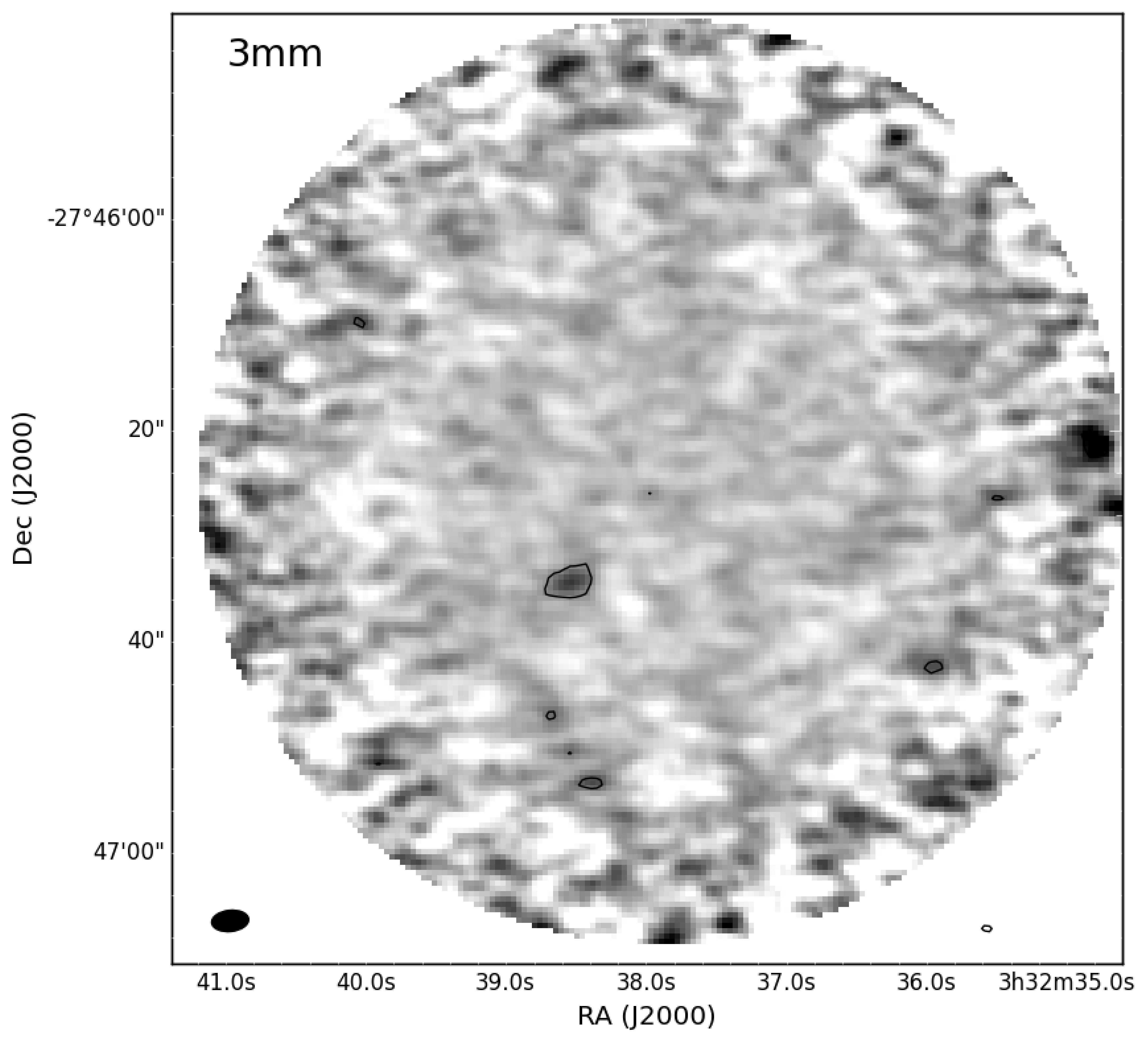}
\includegraphics[width=0.99\columnwidth]{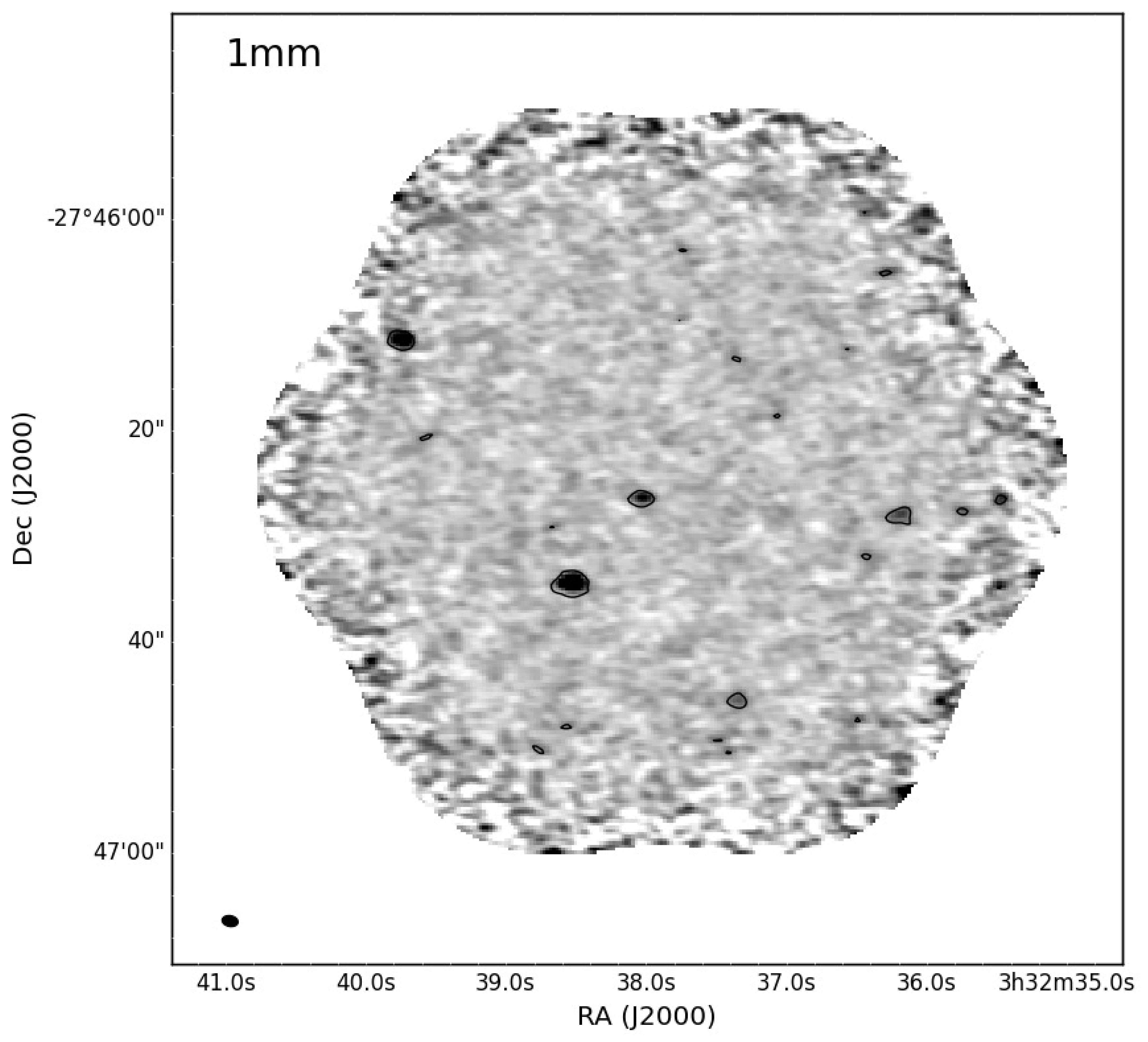}\\
\caption{Continuum images at 3mm ({\em left}) and 1mm ({\em right}). In both panels, we plot a contour at the 3$\sigma$ level, where 1-$\sigma$ is 3.8\,$\mu$Jy\,beam$^{-1}$ in the 3mm observations and 12.7\,$\mu$Jy\,beam$^{-1}$ in the 1mm observations. Both images have been primary-beam corrected. Note that at 3\,mm, only one source is clearly detected at $S/N>3$. The 1mm continuum map is extensively discussed in {\it Paper II}.}
\label{fig_continuum}
\end{figure*}

\subsection{Other CO-- and \Cii{}--detected galaxies}

This paper describes our blind search results. An alternative approach
to finding line emission in the galaxies covered by our observations
is to search the cubes at the position of optical galaxies that have
accurate spectroscopic redshifts. Such additional information
(position and redshift) could in principle help to identify plausible
CO or \Cii{} emission lines at lower significance than those revealed
by the automatic search.  We have performed such a search, which has
resulted in the detection of 3 additional galaxies that are
tentatively detected in CO emission. These detections are presented
and discussed in {\it Paper~IV}. We note that the inclusion of these 3
galaxies would not change the statistical analysis based on the much
larger sample presented here. Similarly, in {\it Paper~V} we
investigate the presence of \Cii{} emission in galaxies for which a
photometric redshift from SED fitting or the detection of a clear
drop-out in the $z$ band suggests redshifts $z>6$.

\subsection{Continuum emission}\label{sec_continuum}

The frequency scans can be used to obtain very high--sensitivity maps
of the continuum, by collapsing the two data cubes along the frequency
axis, after removing the few channels that contain significant line
emission. The resulting continuum maps with noise levels at their
center of 3.8\,$\mu$Jy\,beam$^{-1}$ (band~3) and
12.7\,$\mu$Jy\,beam$^{-1}$ (band 6) are shown in
Fig.~\ref{fig_continuum} and will be discussed in detail in {\it
Paper~II}.

\section{Comparison with expectations}\label{sec_magphys}

We present a detailed comparison of the evolution of the CO luminosity
functions, and the resulting cosmic density of molecular hydrogen in
{\it Paper~III}. As a sanity check, we here briefly compare the number
of CO--detected galaxies with previous expectations based on a
multi--wavelength analysis of the galaxies in the UDF.

For each galaxy in the UDF, \citet{dacunha13} estimated stellar
masses, SFRs, IR luminosities, and expected CO and \Cii{} fluxes and
luminosities by fitting the optical/NIR photometry provided by
\citet{coe06}, using the MAGPHYS spectral energy fitting code
\citep{dacunha08}. We show the resulting star formation rates, and
stellar masses, in four redshift bins in the right hand panel of
Fig.~\ref{fig_pointings}. Note that typical selections of main
sequence galaxies for CO follow--up usually target stellar masses
M$_{\rm star}>10^{10}$\,M$_\odot$ and star formation rates
SFR$>50$\,M$_\odot$\,yr$^{-1}$ (e.g. Daddi et al. 2008, Tacconi et
al. 2008, Genzel et al.\ 2008, Tacconi et al.\ 2012, Daddi et al.\
2015, Genzel et al.\ 2015).  I.e. this selection would target galaxies
in the top right part of each diagram, as the UDF contains many
galaxies that are much less massive / star forming.

\begin{figure}
\includegraphics[width=0.99\columnwidth]{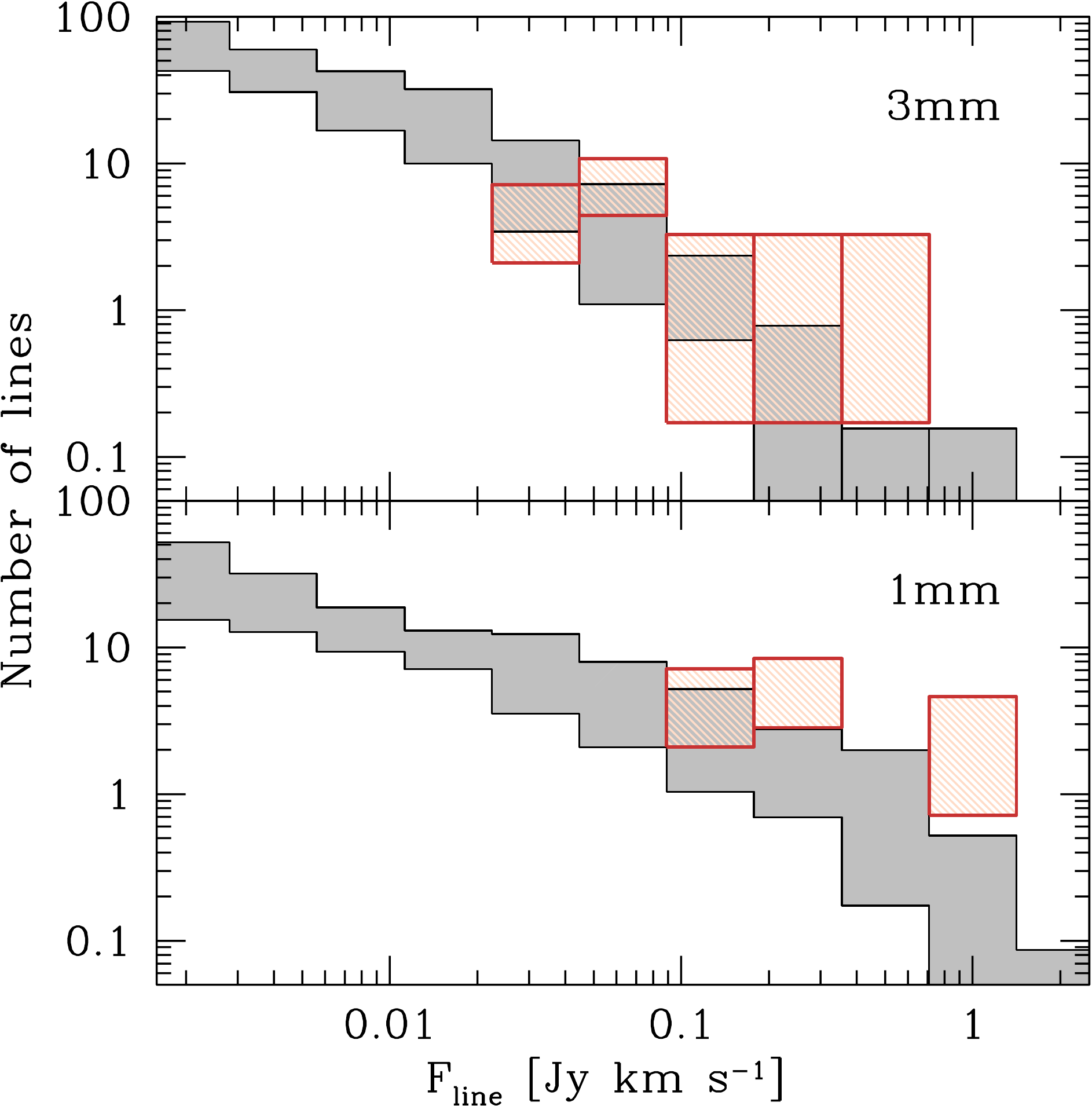}\\
\caption{Comparison between the MAGPHYS-based predictions of line fluxes from \citet{dacunha13}, in grey, and
the flux distribution of the line candidates actually observed in our survey (red boxes). The numbers from \citet{dacunha13}
are computed over the whole UDF, and scaled down to match the same area coverage of our survey. We consider here only the transitions
that we cover in our scan (see Tab.~\ref{tab_z_range}). The lower and upper sides of the shaded grey area refer the cases of 
Milky Way- and M82-like CO excitation. In the case of our ALMA constraints, the vertical size of the boxes show the Poissonian
uncertainties in the number of lines detected in a certain flux range. Our ALMA constraints are not corrected for the fidelity
and completeness of our line search. The number of detected lines is in general agreement with the expectations, in particular if one keeps in mind that ASPECS~1mm.1/2, whose high--J CO emission dominates the highest flux bin at 1mm, was not included in the \citet{dacunha13} study.}
\label{fig_magphys}
\end{figure}

In Fig.~\ref{fig_magphys} we show the expected numbers of line
detections in the 3\,mm and 1\,mm bands, respectively. In this plot,
the expected number of lines from \citet{dacunha13}, originally
computed for the entire $3'\times3'$UDF, has been scaled to the areal
coverage of our survey. In \citet{dacunha13}, two extreme CO
excitation cases were considered in order to transform predicated
CO(1--0) luminosities into higher--J line luminosities: the
low--excitation case of the global Milky Way disk, and the
high--excitation case of the nucleus of the local starburst galaxy M82
\citep{weiss07}. For each line flux plotted on the abscissa, this
range of excitation conditions is indicated by the grey region on the
ordinate.

In this figure, we compare to our observations, which are plotted as
red--shaded regions. For each flux bin on the abscissa, the number
counts with the Poissonian error bars are shown on the ordinate. For
this back--of--the envelope calculation, we do not correct our
measurements for completeness or fidelity (this is done in detail in
{\it Paper~III}). A number of things need to be kept in mind in this
comparison: the total number of detected sources is low, which results
in large uncertainties in the measurements on the ordinate. At 1\,mm,
the data in the highest flux bin (around 1\,mJy\,km\,s$^{-1}$) is
significantly higher than the predictions. Note however that
measurement includes the two high--J CO detections of ASPECS~1mm.1/2,
a galaxy that was not included in the UDF catalog on which the
predictions by \citet{dacunha13} were based.  Larger areas are
required to see if there is indeed an excess of high--J CO emission
present. Overall, we conclude that within the large uncertainties,
there is reasonable agreement between the observations and previous
expectations. This is discussed in detail in {\it Paper~III}.

\section{Summary}

We present the rationale for and the observational description of
ASPECS, our complete band~3 and band~6 spectral line scan with ALMA of
the {\it Hubble} Ultra--Deep Field (UDF). This field was chosen
because it has the deepest multi--wavelength data available, it will
remain a key cosmological deep field in the future (in particular in
the era of JWST) and is easily observable by ALMA.  We discuss our
survey design of the full frequency scans in band~3 (84--115\,GHz) and
band~6 (212--272\,GHz) and report the relevant parameters of our final
dataset. Critically, ALMA allows us to reach approximately uniform
depth (line sensitivity: $\sim L'_{\rm CO}\sim 2 \times
10^{9}$\,\Kkmspc) across a broad range of redshifts.

The spectral line scans cover the different rotational transitions of
the CO molecule at different redshifts, leading to essentially full
redshift coverage.  We present a customized algorithm to identify line
candidates in our data. This algorithm takes varying linewidths of the
possible emission lines into account. We assess the {\rm fidelity} of
our line search by comparing the number of positive candidates to the
respective number of negative candidates, the latter being unphysical.
We also calculate the {\em completeness} of our search, by quantifying
our ability to recover artificial sources in our data. We present CO
spectra and {\it HST} postage stamps of the most signficant
detections. Based on whether multiple CO lines are detected, and
whether optical spectroscopic (either slit or grism) redshifts as well
as optical/NIR counterparts exist, we give constraints on the most
likely line identification of our candidates.

Out of the 10 line candidates (3mm band) reported in our search
(Tab.~\ref{tab_lines}), we expect $<$4 candidates to be spurious,
given our statistical analysis.  There are a number of line candidates
at positions where no optical/NIR counterpart is present. The total CO
flux of these candidates is less than 33\% of the total flux of all
candidates, i.e. candidate sources without counterparts only
contribute a small fraction of the total measured flux in the targeted
field.  We also present continuum maps of both the band~3 and band~6
observations. The observed flux distribution of the line candidates is
in general agreement with the empirical expectations by
\citet{dacunha13} based on SED modeling of the optical/NIR emission of
galaxies in the UDF.

The data presented in this paper ({\it Paper~I}) form the basis of a
number of dedicated studies presented in subsequent papers:

\noindent $\bullet$ In {\it Paper~II} (Aravena et al.\ 2016a) we
present 1.2\,mm continuum number counts, dust properties of individual
galaxies, and demonstrate that our observations

recover the cosmic infrared background at the wavelengths considered. 

\noindent $\bullet$ In {\it Paper~III} (Decarli et al.\ 2016a) we
discuss the implications for CO luminosity functions and the resulting
constraints on the gas density history of the Universe. Based on our
data we show that there is a sharp decrease (by a factor of $\sim$5)
in the cosmic molecular gas density from redshift $\sim$ 3 to 0.

\noindent $\bullet$ In {\it Paper~IV} (Decarli et al.\ 2016b) we
examine the properties of those galaxies in the UDF that show bright
CO emission, and discuss these also in the context of the bright
optical galaxies that are not detected in CO.

\noindent $\bullet$ In {\it Paper~V} (Aravena et al.\ 2016b) we search
for \Cii{} emitters in previously reported Lyman--break galaxies at
6$<$z$<$8.

\noindent $\bullet$ In {\it Paper VI} (Bouwens et al.\ 2016) we
investigate where high--redshift galaxies from ASPECS lie in relation
to known IRX--$\beta$ and IRX--stellar mass relationships, concluding
that less dust continuum emission is detected in z$>$2.5 than expected
(unless high dust temperatures, T$\sim$50\,K, are assumed).

\noindent $\bullet$ Finally, in {\it Paper VII} (Carilli et al.\ 2016)
we discuss implications on CO intensity mapping experiments, and
contributions towards the emission from the cosmic microwave
background.

The data presented here demonstrate the unique power of ALMA spectral
scans in well--studied cosmological deep fields. The current size of
the survey is admittedly small, limited by the amount of time
available in ALMA `early science'. More substantial spectral scan
surveys with ALMA of the full UDF (and beyond) will become feasible
once ALMA is fully operational.

\section*{Acknowledgments}

We thank the referee for a constructive report that helped improve the
presentation of the data. FW, IRS, and RJI acknowledge support through
ERC grants COSMIC--DAWN, DUSTYGAL, and COSMICISM, respectively. MA
acknowledges partial support from FONDECYT through grant 1140099.  DR
acknowledges support from the National Science Foundation under grant
number AST-\#1614213 to Cornell University. FEB and LI acknowledge
Conicyt grants Basal-CATA PFB--06/2007 and Anilo ACT1417. FEB also
acknowledge support from FONDECYT Regular 1141218 (FEB), and the
Ministry of Economy, Development, and Tourism's Millennium Science
Initiative through grant IC120009, awarded to The Millennium Institute
of Astrophysics, MAS. IRS also acknowledges support from STFC
(ST/L00075X/1) and a Royal Society / Wolfson Merit award. Support for
RD and BM was provided by the DFG priority program 1573 `The physics
of the interstellar medium'.  AK and FB acknowledge support by the
Collaborative Research Council 956, sub-project A1, funded by the
Deutsche Forschungsgemeinschaft (DFG). PI acknowledges Conict grants
Basal-CATA PFB--06/2007 and Anilo ACT1417. RJA was supported by
FONDECYT grant number 1151408.  This paper makes use of the following
ALMA data: 2013.1.00146.S and 2013.1.00718.S. ALMA is a partnership of
ESO (representing its member states), NSF (USA) and NINS (Japan),
together with NRC (Canada), NSC and ASIAA (Taiwan), and KASI (Republic
of Korea), in cooperation with the Republic of Chile. The Joint ALMA
Observatory is operated by ESO, AUI/NRAO and NAOJ. The 3mm-part of
ALMA project had been supported by the German ARC.

\appendix

\section{Line candidates from the blind search}\label{sec_blind_cat}

In this appendix, we show postage stamps and extracted spectra for all the line candidates identified with the blind line search.

\begin{figure*}
\includegraphics[width=0.48\columnwidth]{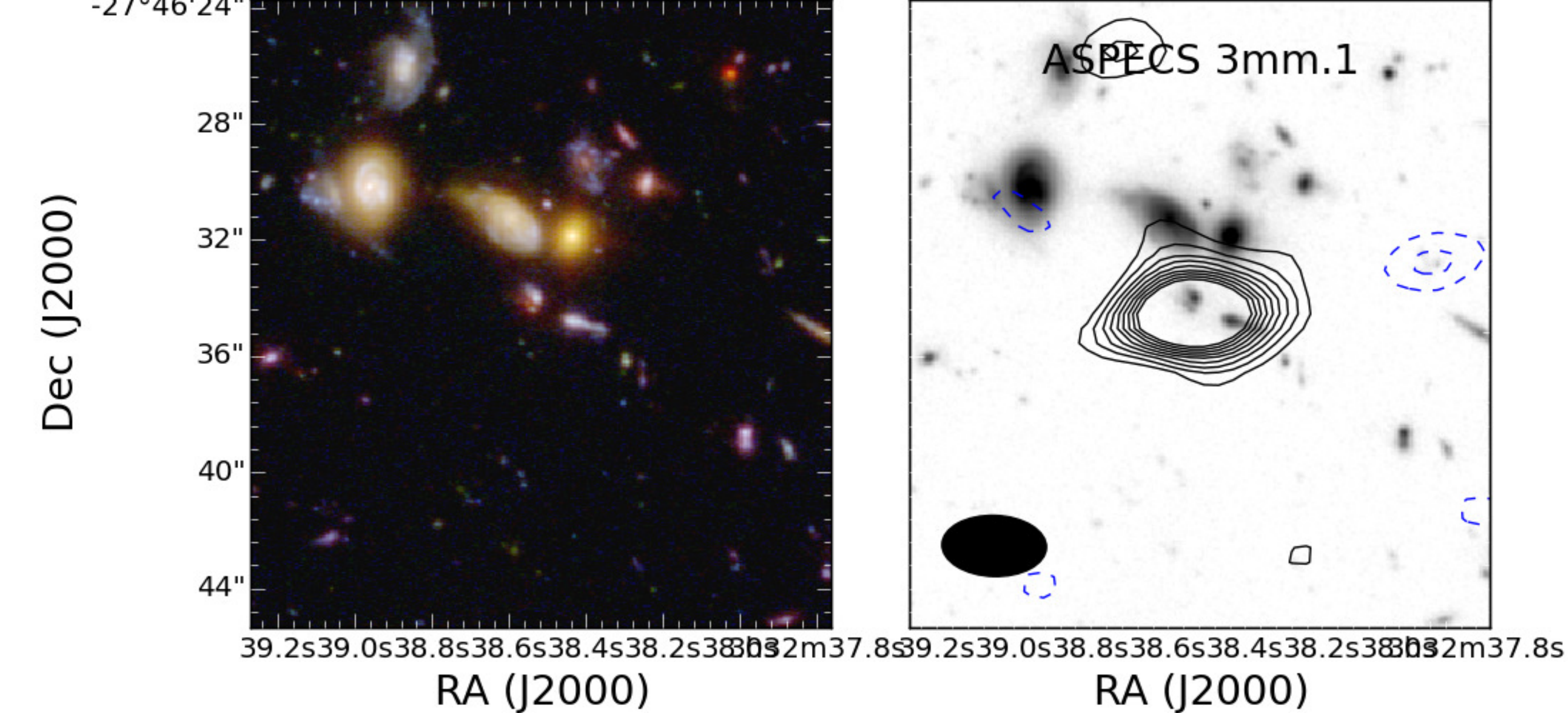}
\includegraphics[width=0.38\columnwidth]{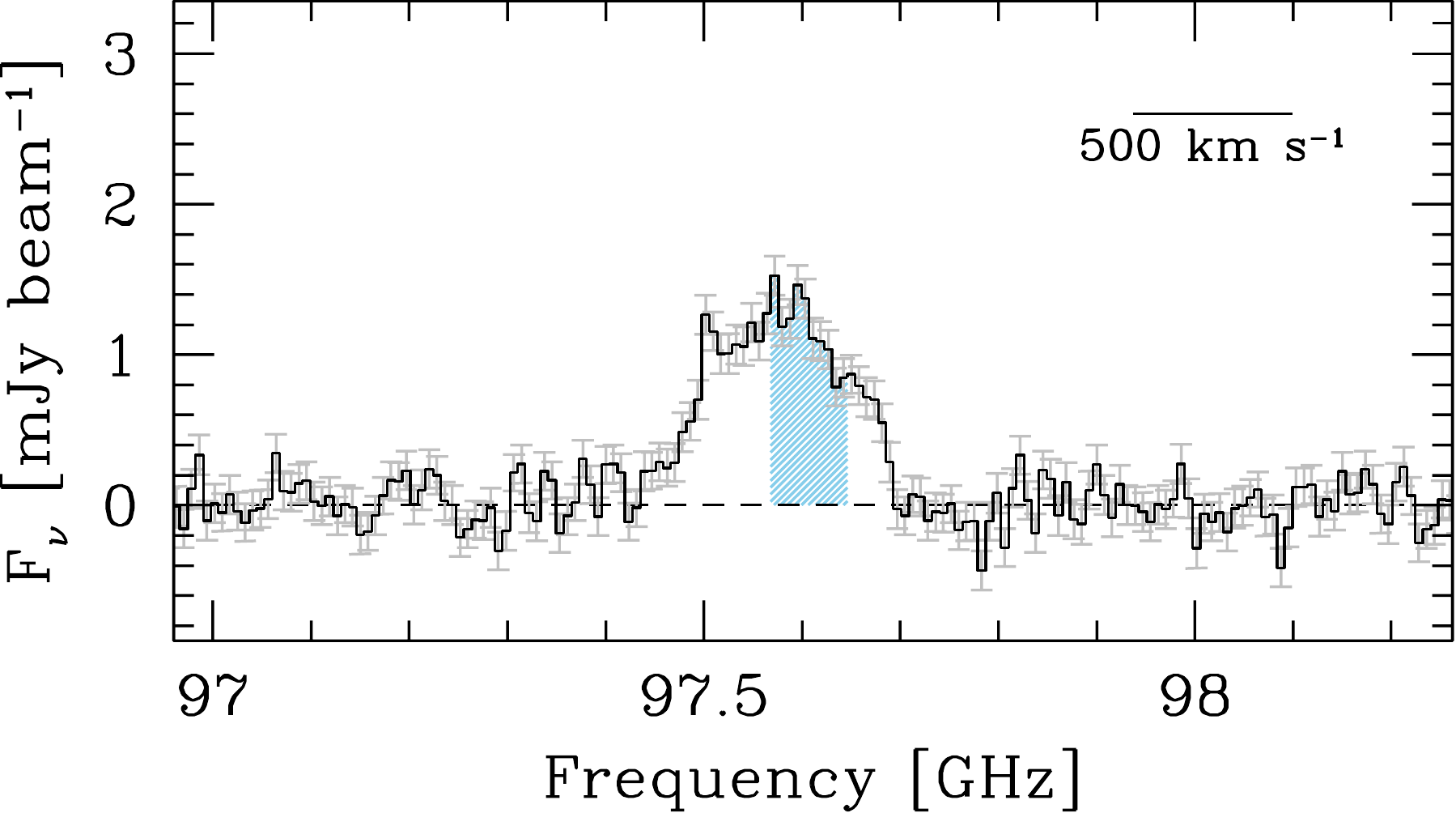}\\
\includegraphics[width=0.48\columnwidth]{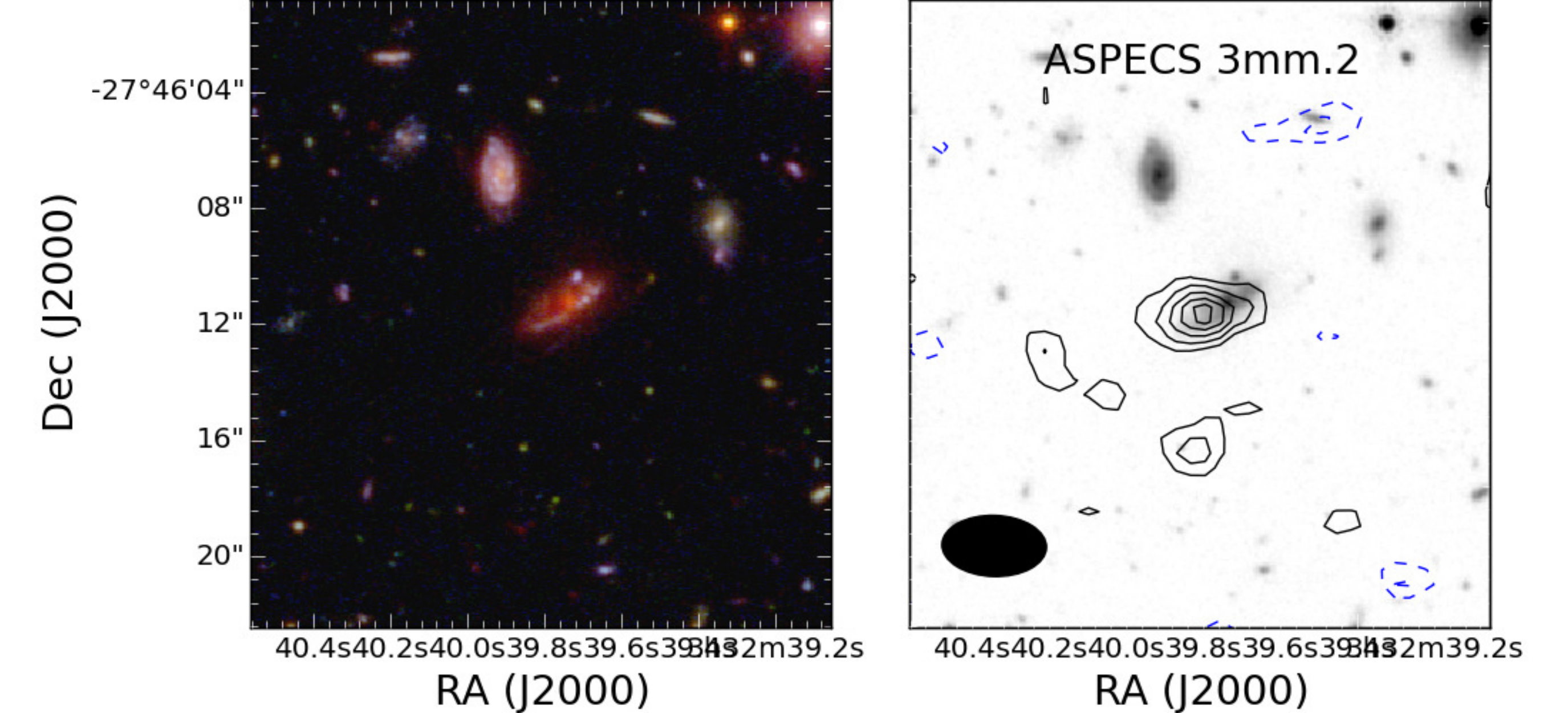}
\includegraphics[width=0.38\columnwidth]{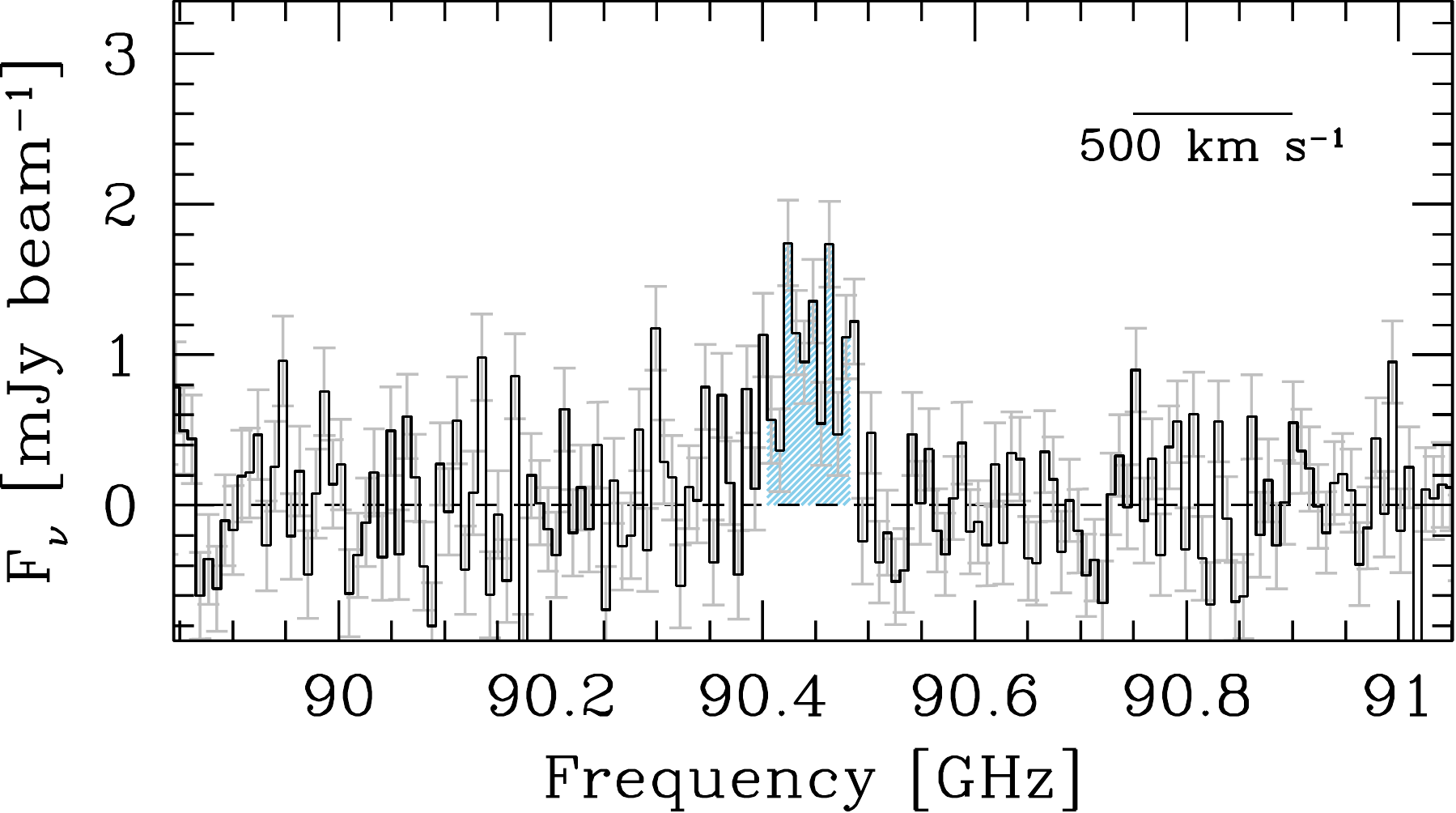}\\
\includegraphics[width=0.48\columnwidth]{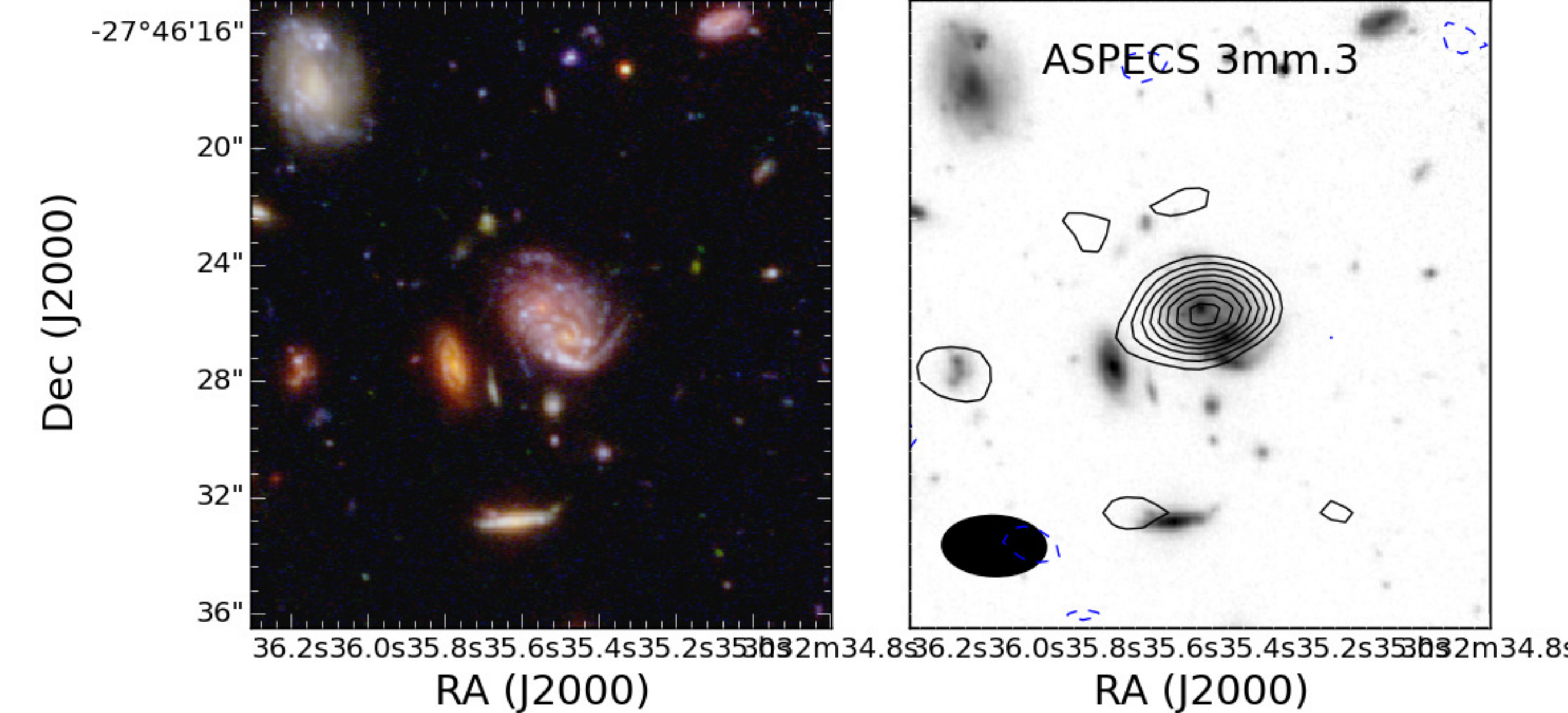}
\includegraphics[width=0.38\columnwidth]{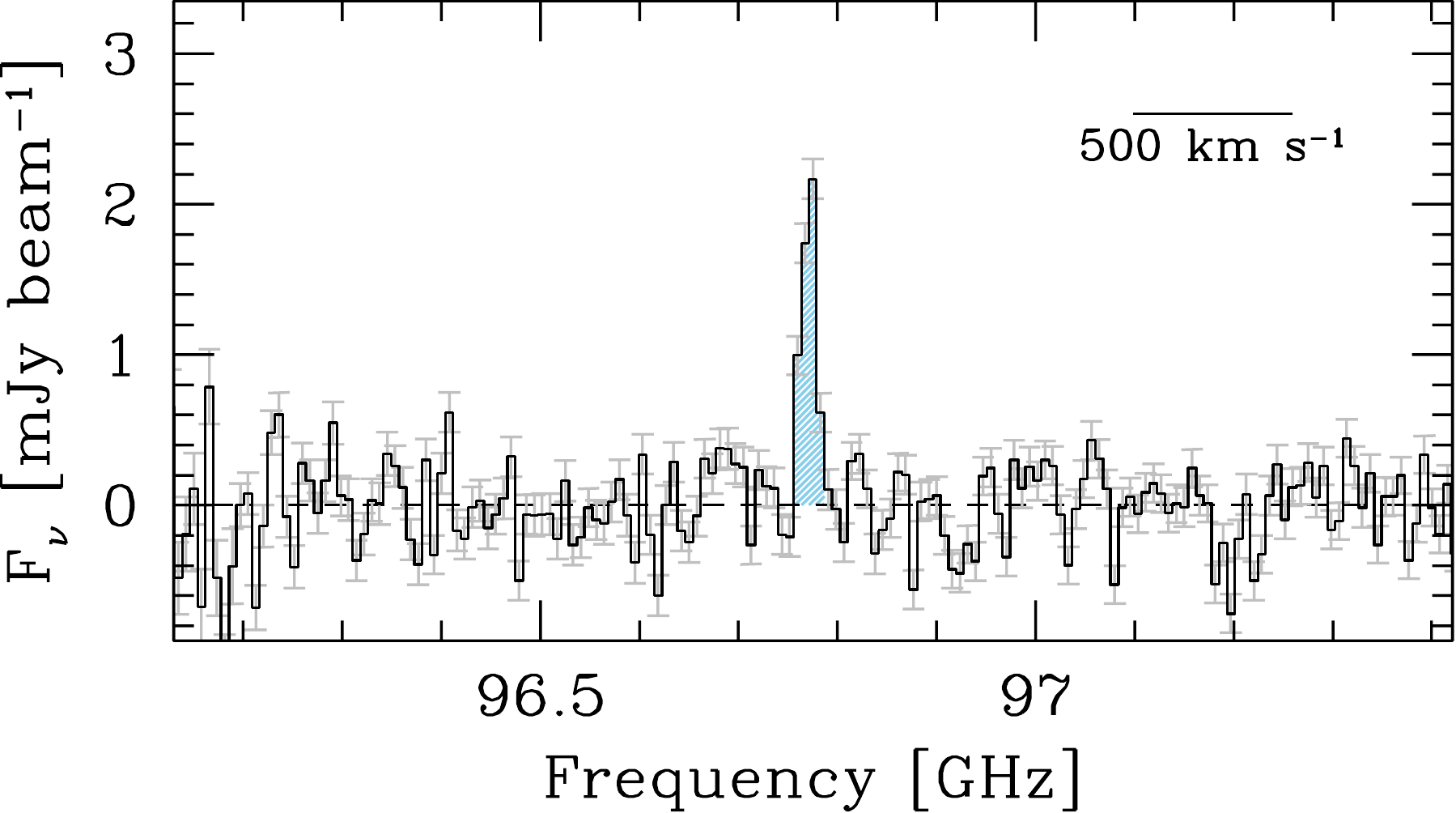}\\
\includegraphics[width=0.48\columnwidth]{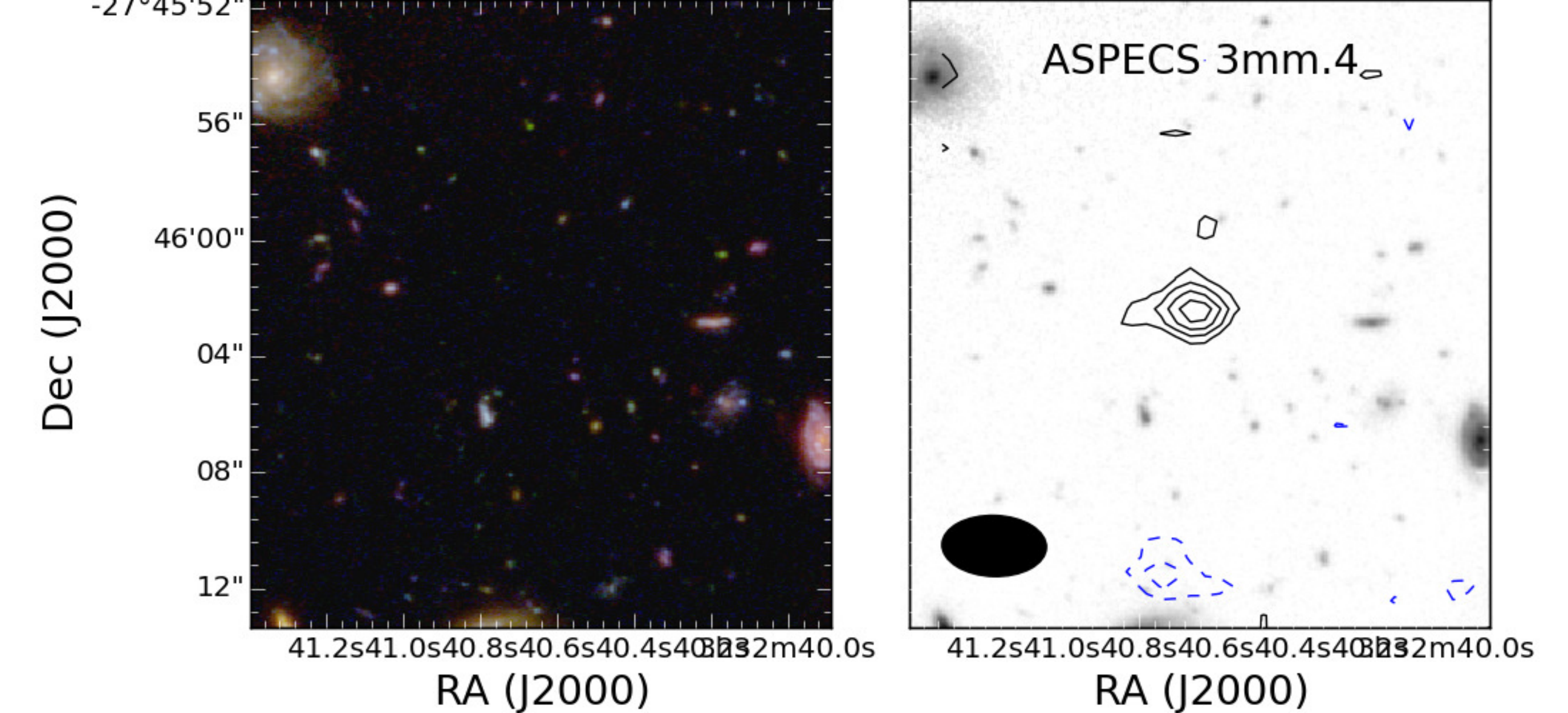}
\includegraphics[width=0.38\columnwidth]{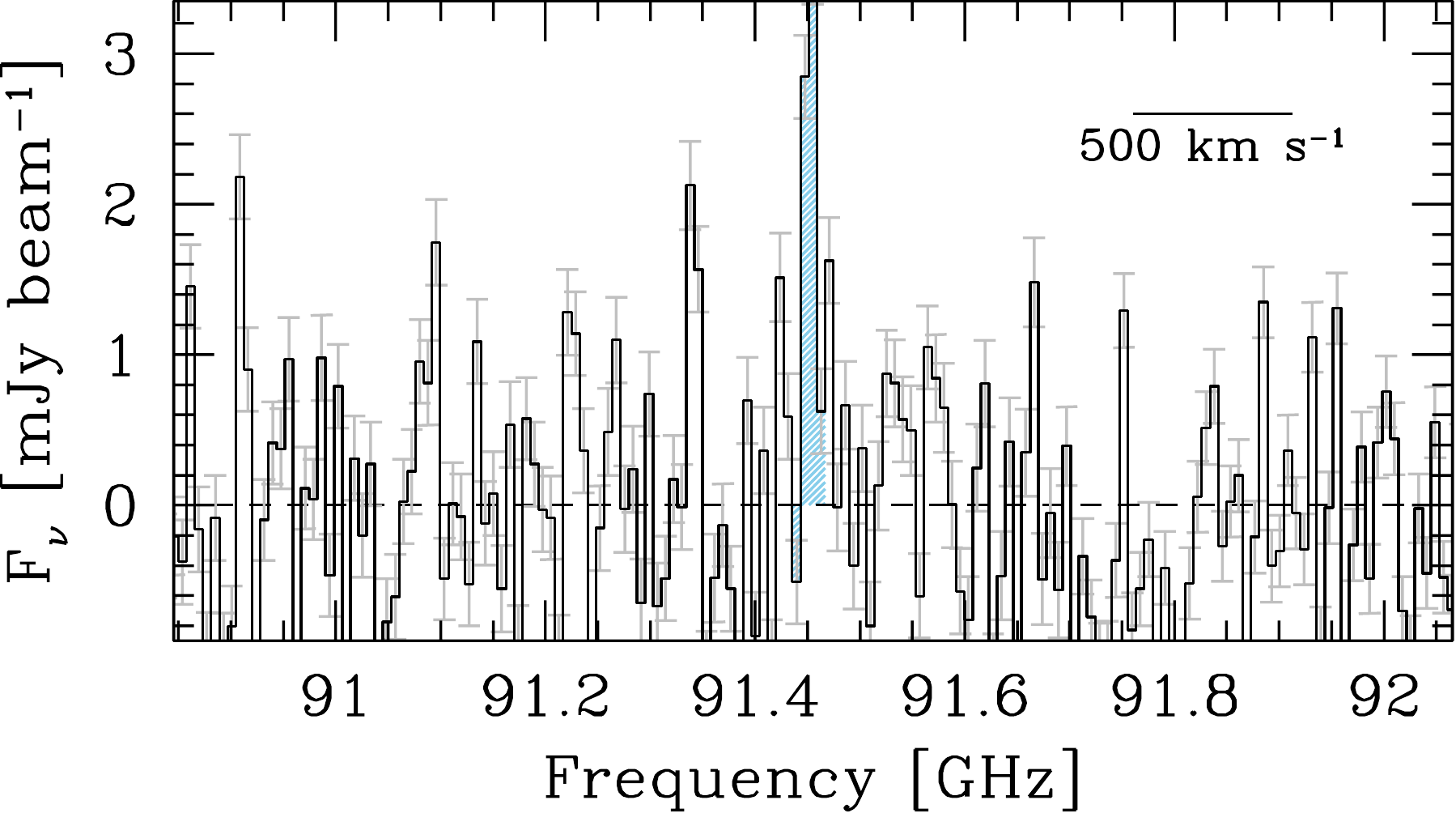}\\
\includegraphics[width=0.48\columnwidth]{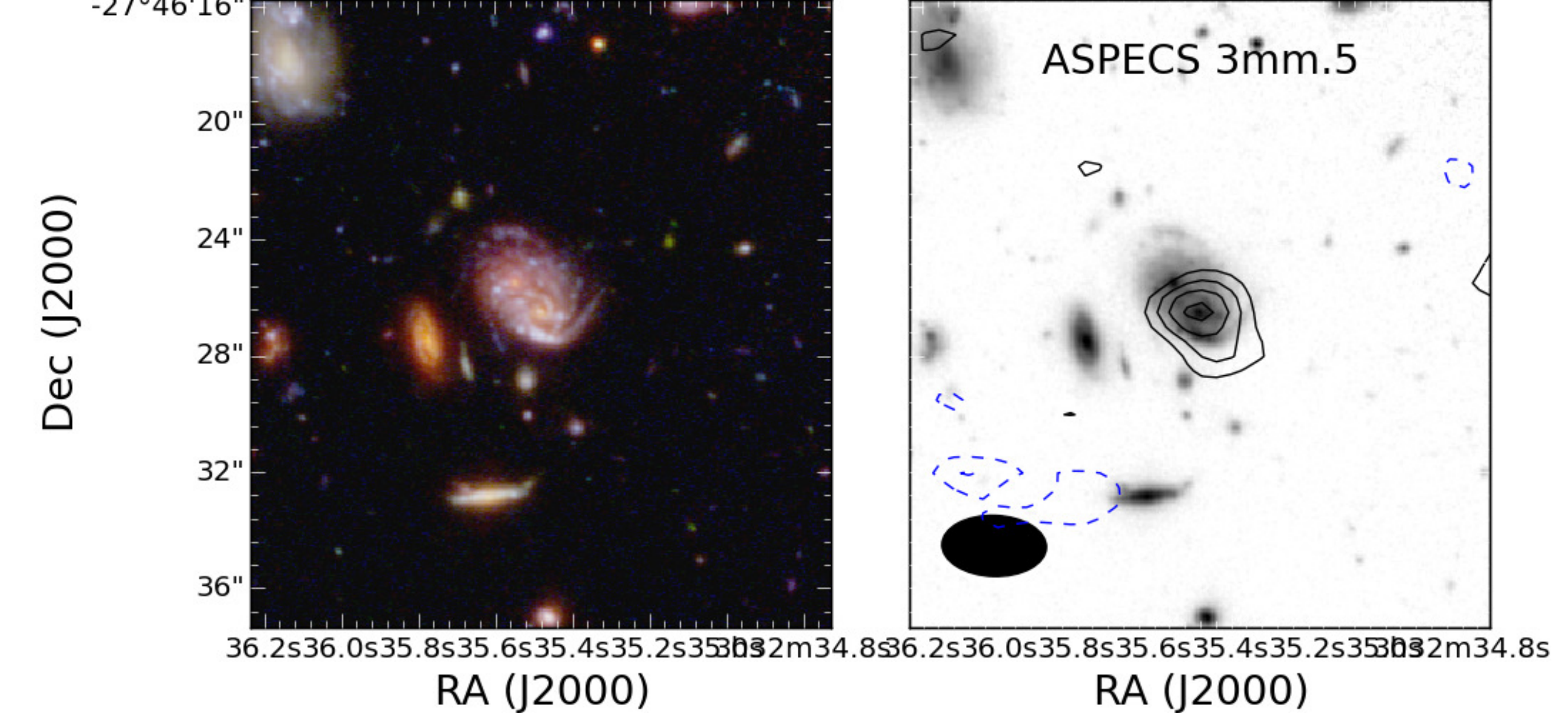}
\includegraphics[width=0.38\columnwidth]{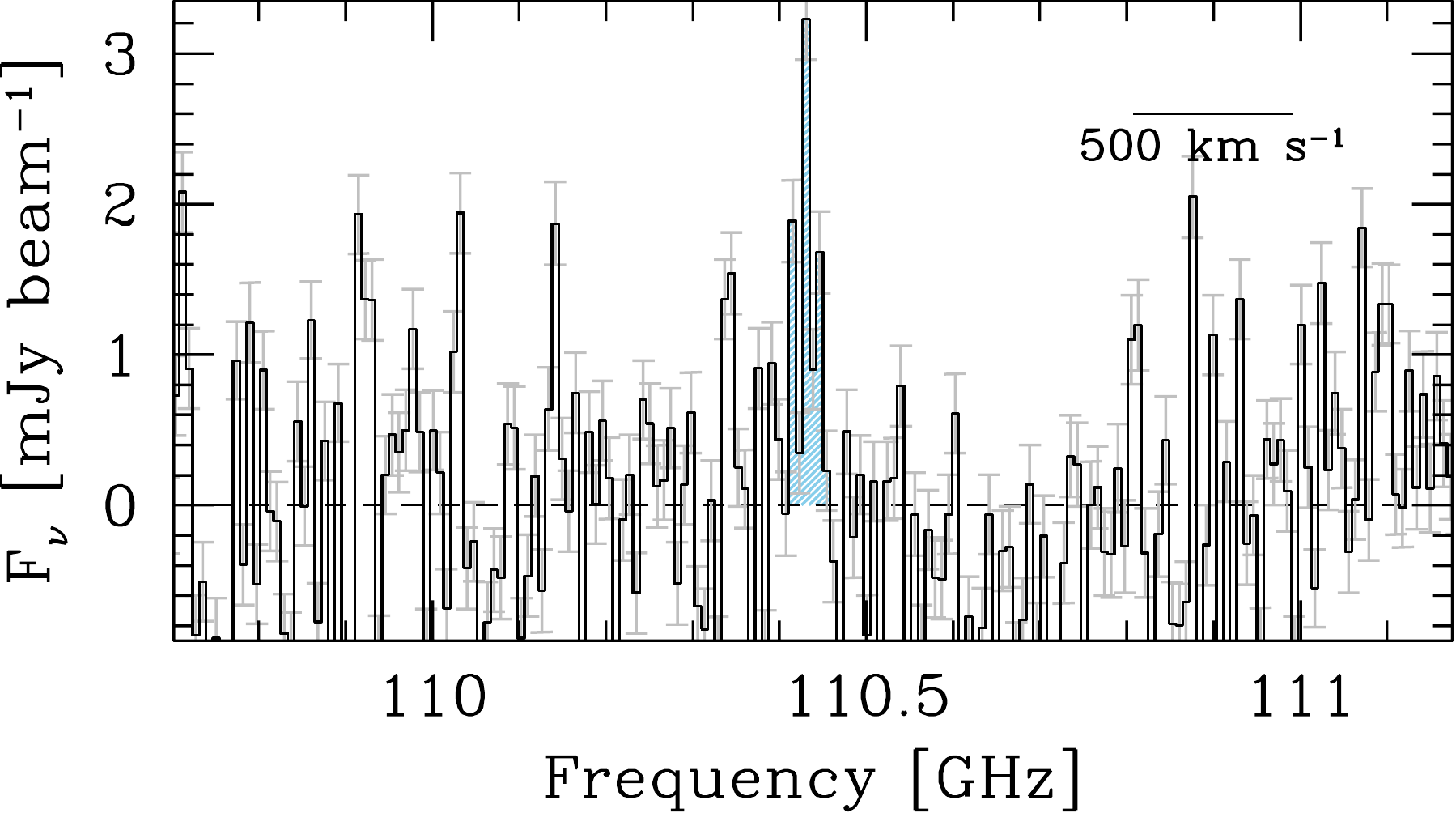}\\
\caption{{\em Left:} Optical/NIR {\it HST} multi--color image centred on the line candidates discovered in the blind search at 3mm (using the F125W, F775W and F435W filters, \citep{illingworth13}). {\em Middle:} CO contours of the candidate line maps resulting from our line search described in Sec.~\ref{sec_search}. Positive (negative) contours of the CO emission are plotted in solid black (dashed blue), where the contours mark the $\pm2$,3,4,\ldots-$\sigma$ isophotes ($\sigma$ is derived from the respective line map). Each postage stamp is $20''\times20''$  and the size of the synthesized beam is show in the lower left. {\em Right:} spectrum of the line candidate. The blue shading marks the channels that the line-searching algorithm used to compute the line S/N (this is why the shading does not cover the entire width of the brightest source). All line parameters are summarized in Tab.~\ref{tab_lines}.}
\label{fig_ps_3mm_a}
\end{figure*}

\begin{figure*}
\figurenum{\ref{fig_ps_3mm_a}}
\includegraphics[width=0.48\columnwidth]{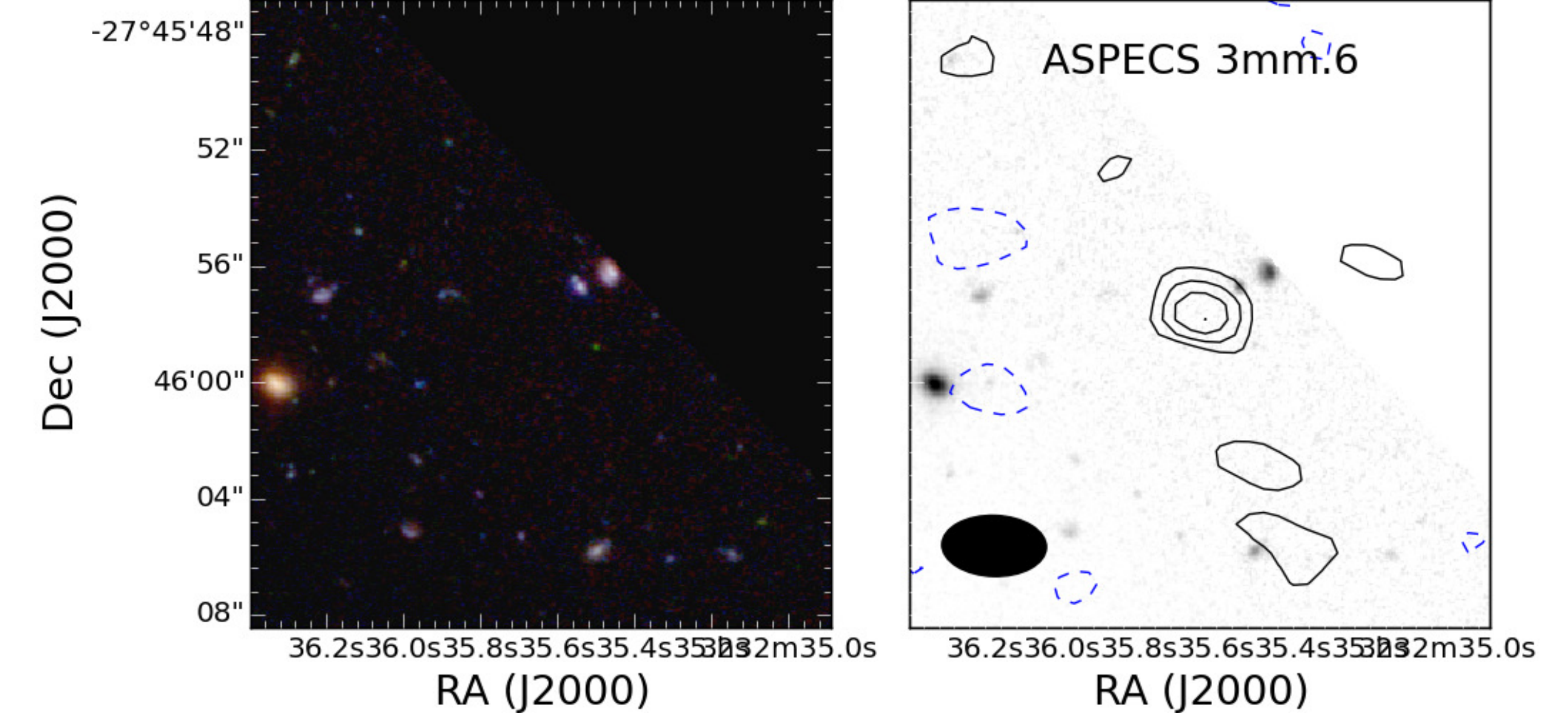}
\includegraphics[width=0.38\columnwidth]{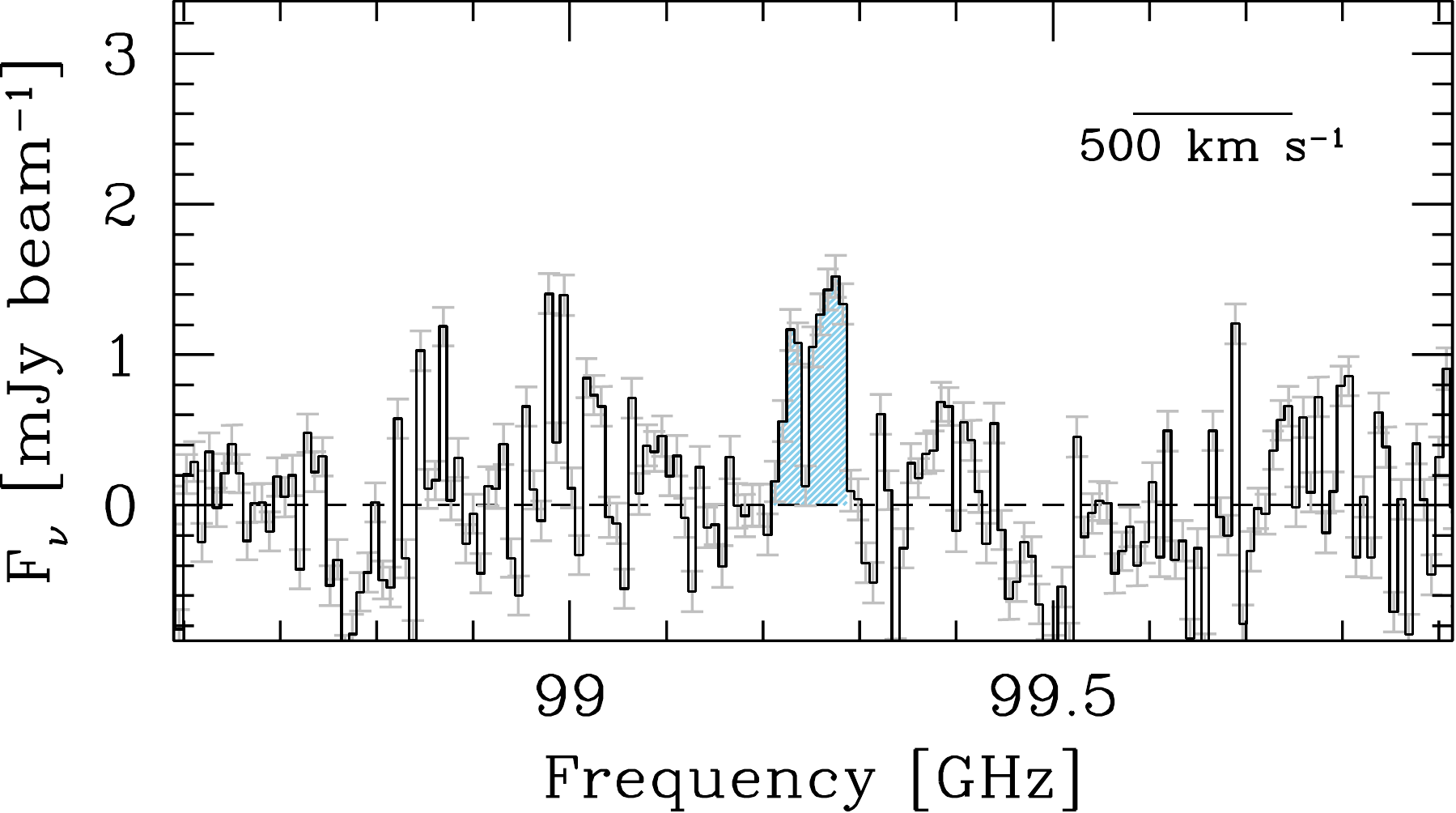}\\
\includegraphics[width=0.48\columnwidth]{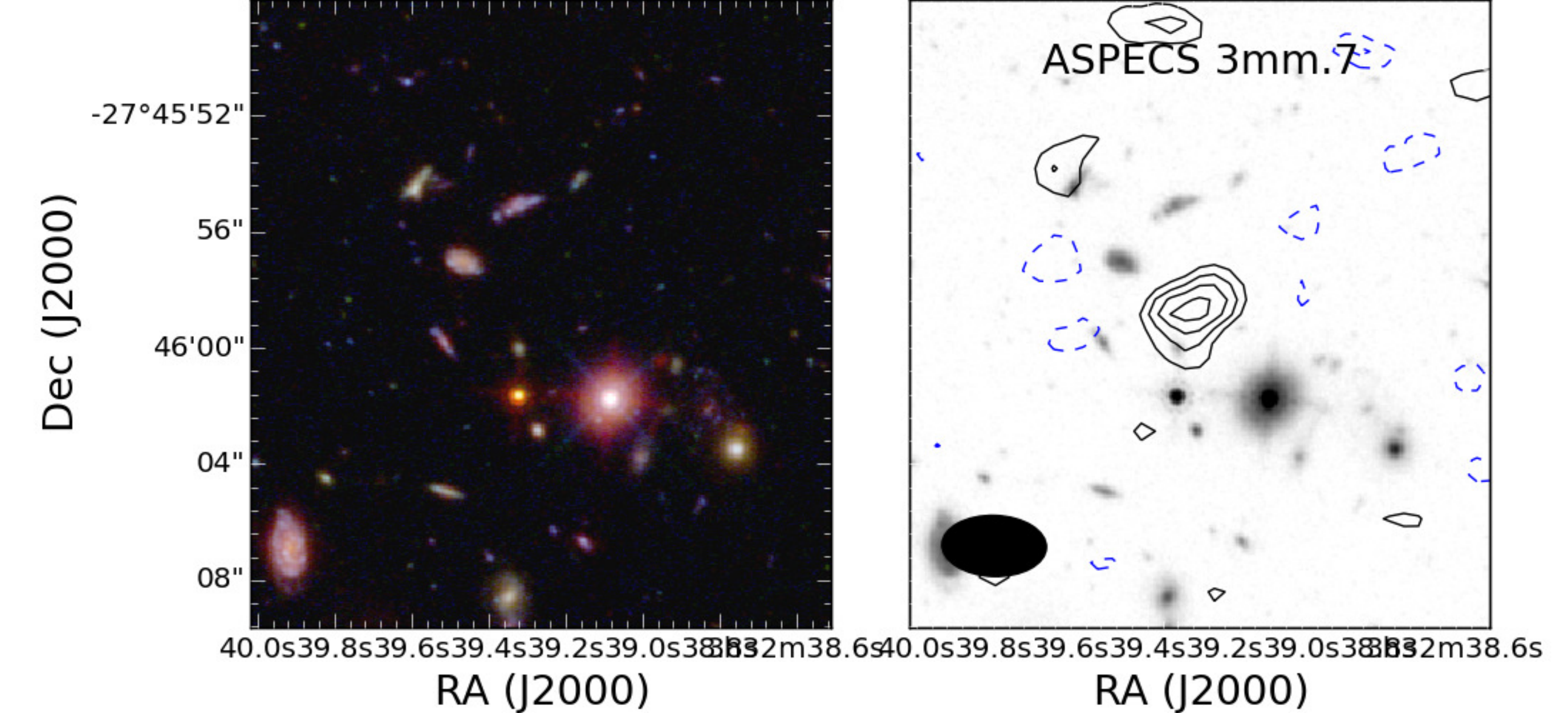}
\includegraphics[width=0.38\columnwidth]{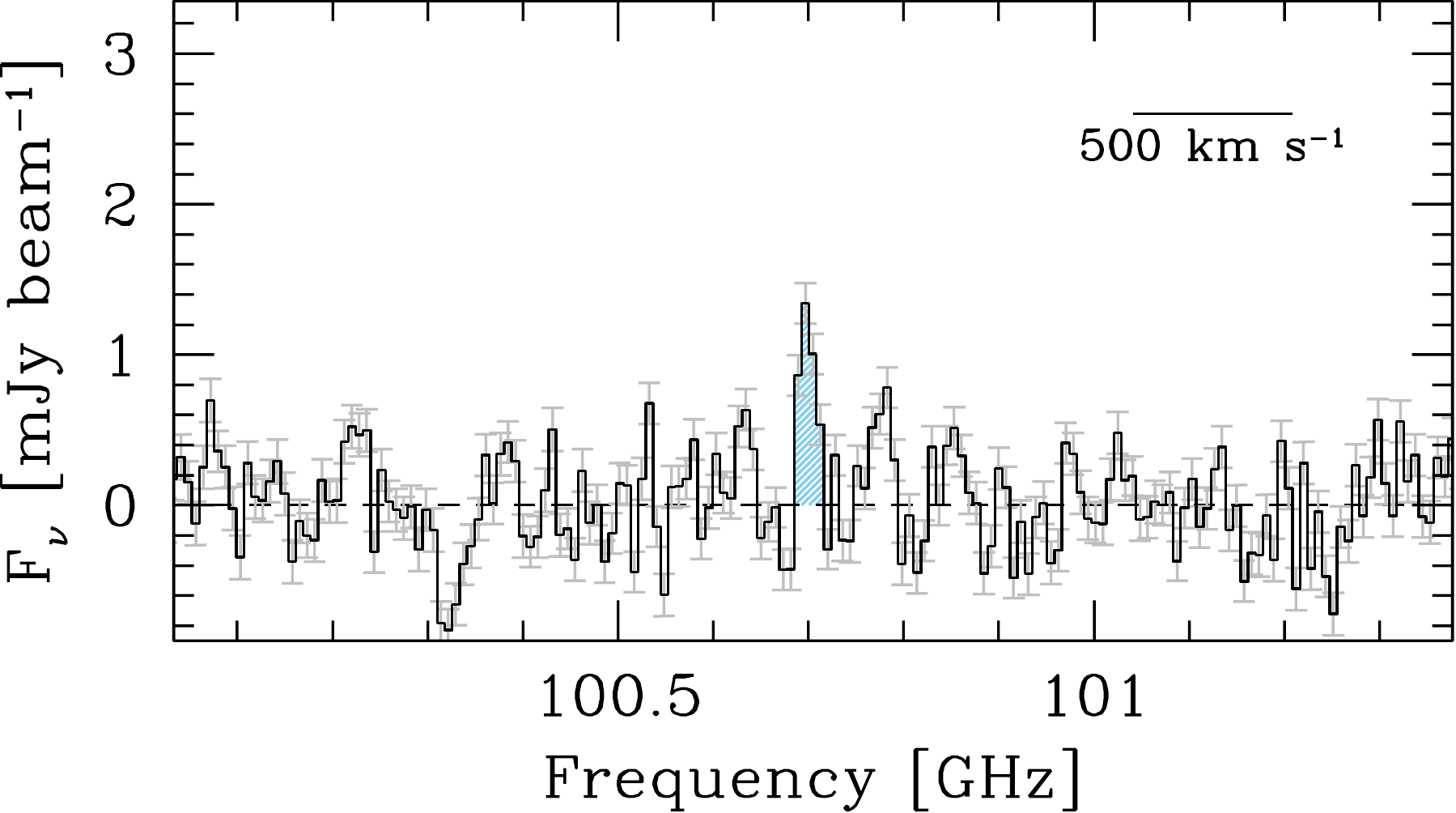}\\
\includegraphics[width=0.48\columnwidth]{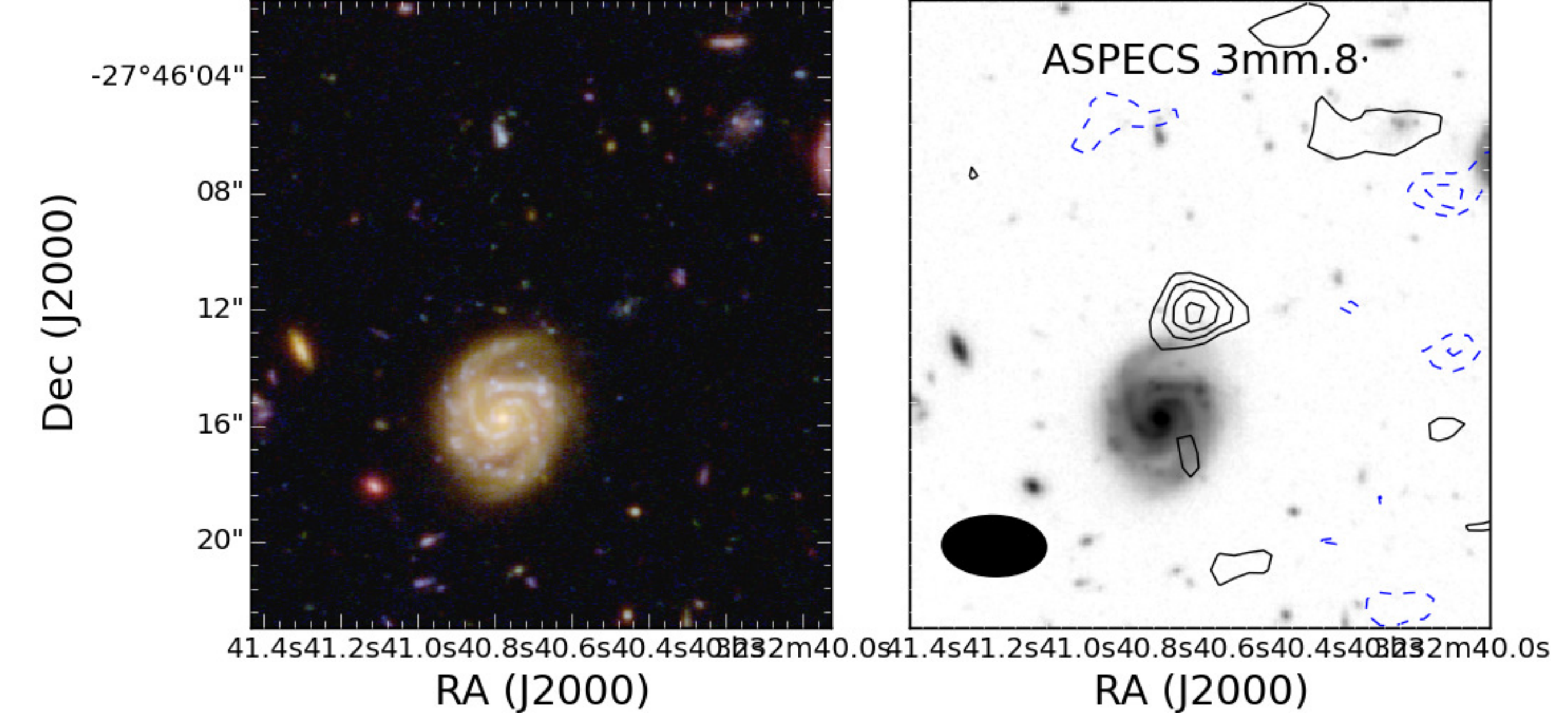}
\includegraphics[width=0.38\columnwidth]{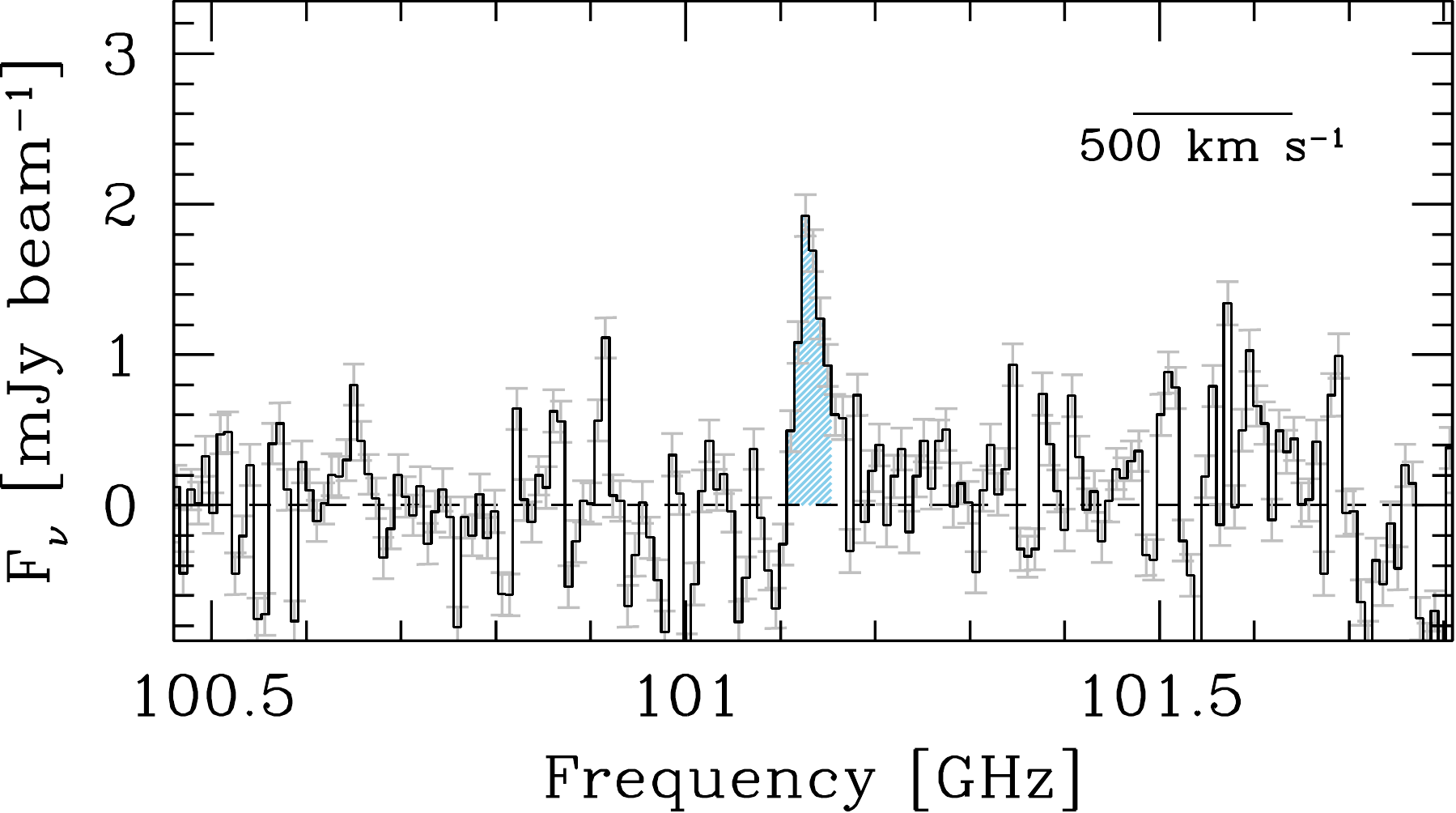}\\
\includegraphics[width=0.48\columnwidth]{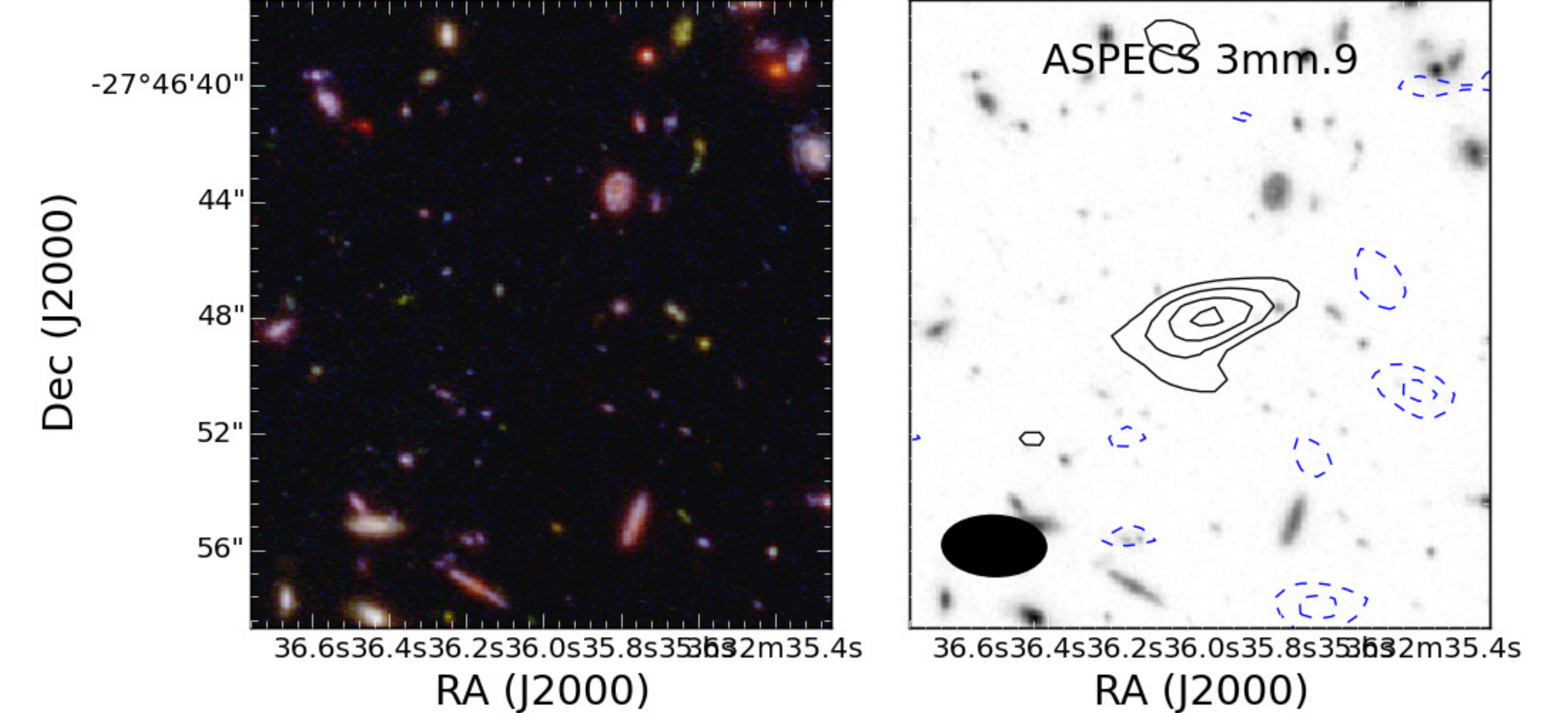}
\includegraphics[width=0.38\columnwidth]{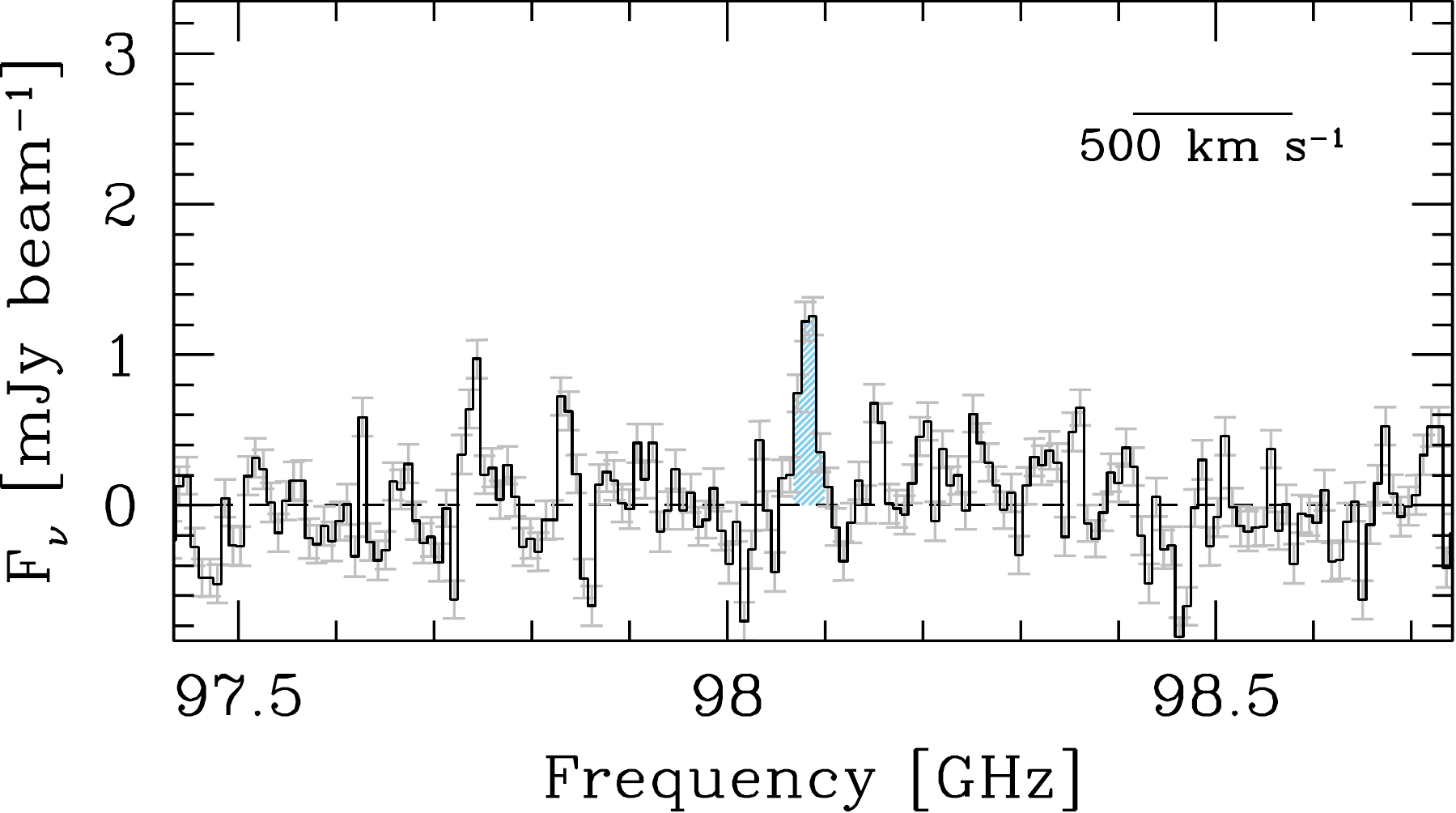}\\
\includegraphics[width=0.48\columnwidth]{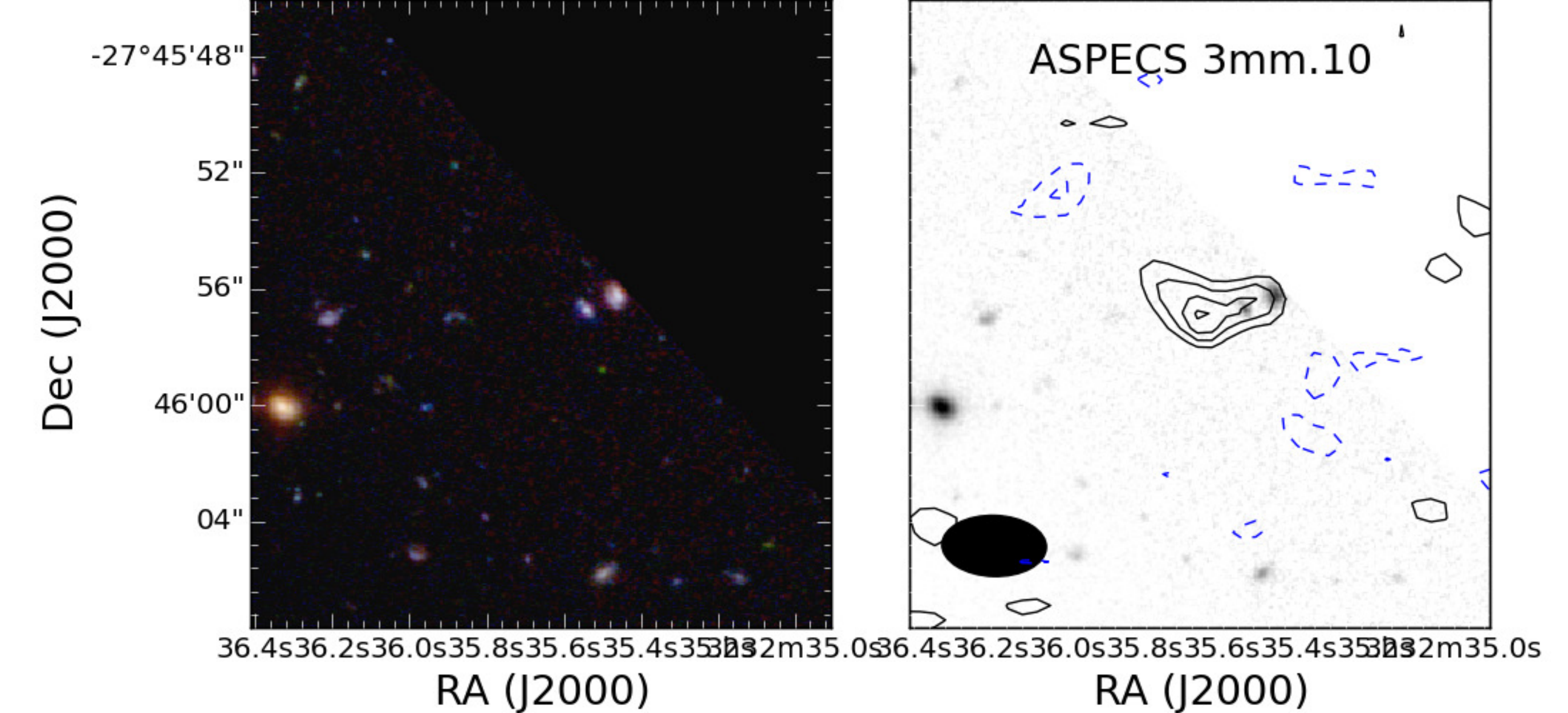}
\includegraphics[width=0.38\columnwidth]{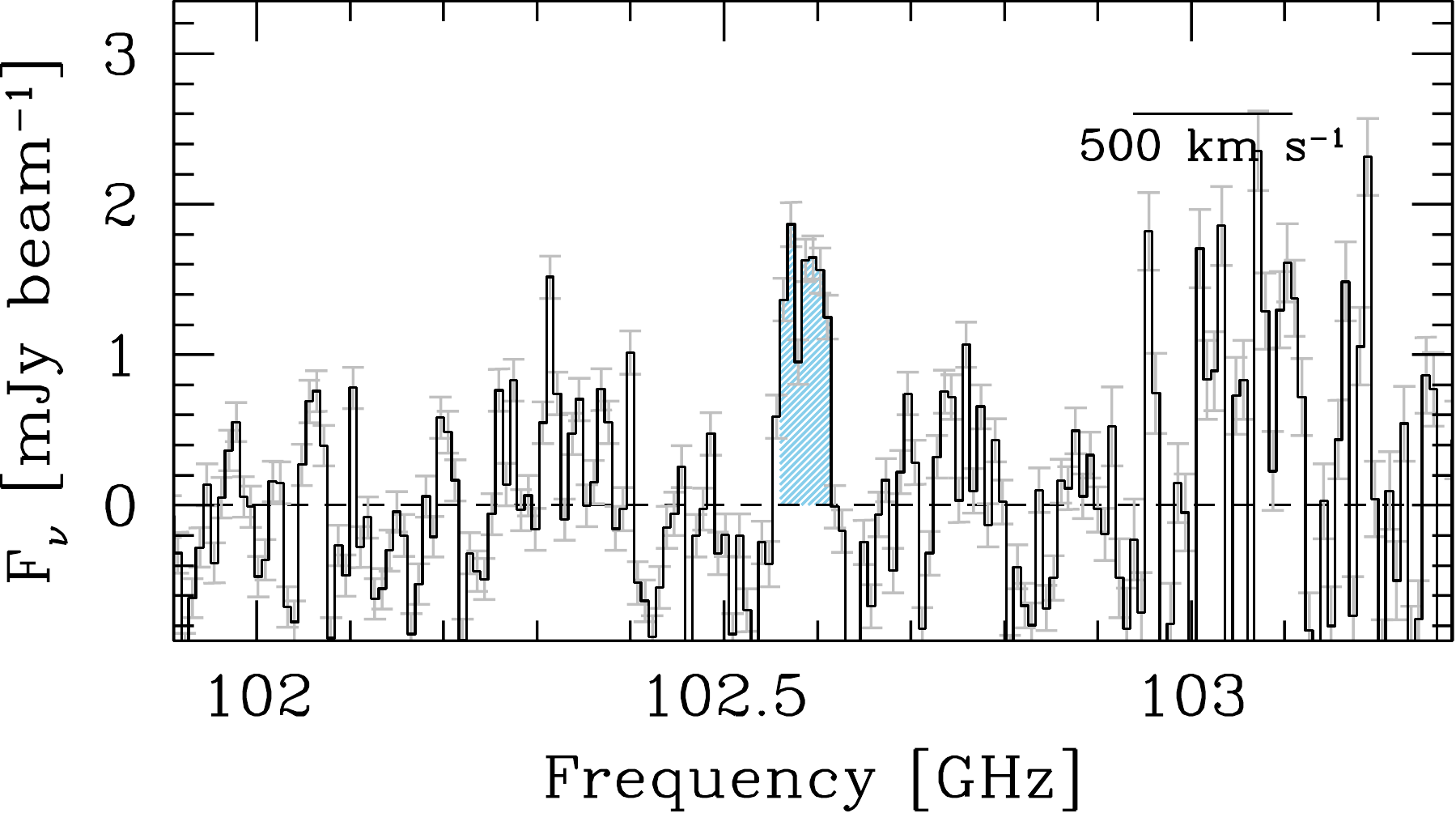}\\
\caption{continued.}
\end{figure*}

\begin{figure*}
\includegraphics[width=0.48\columnwidth]{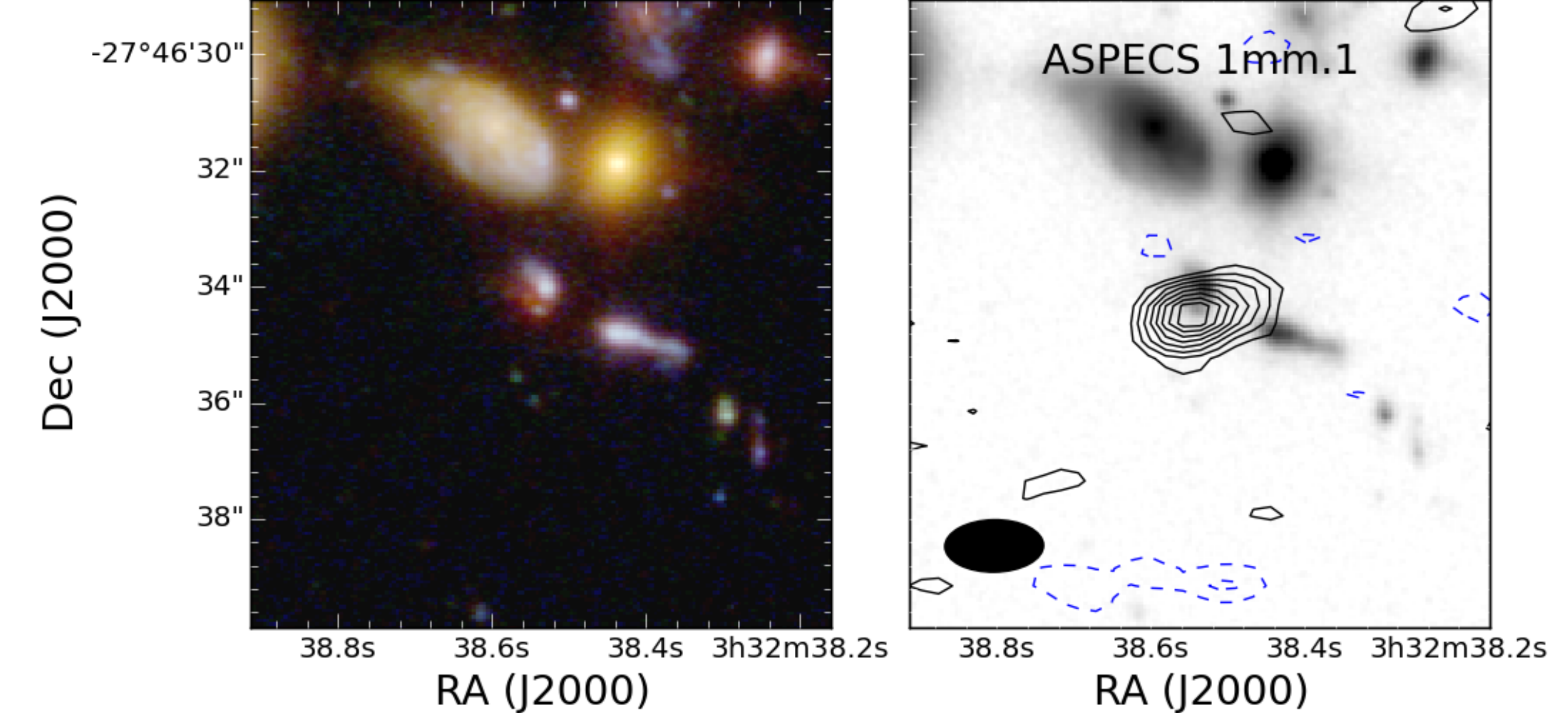}
\includegraphics[width=0.38\columnwidth]{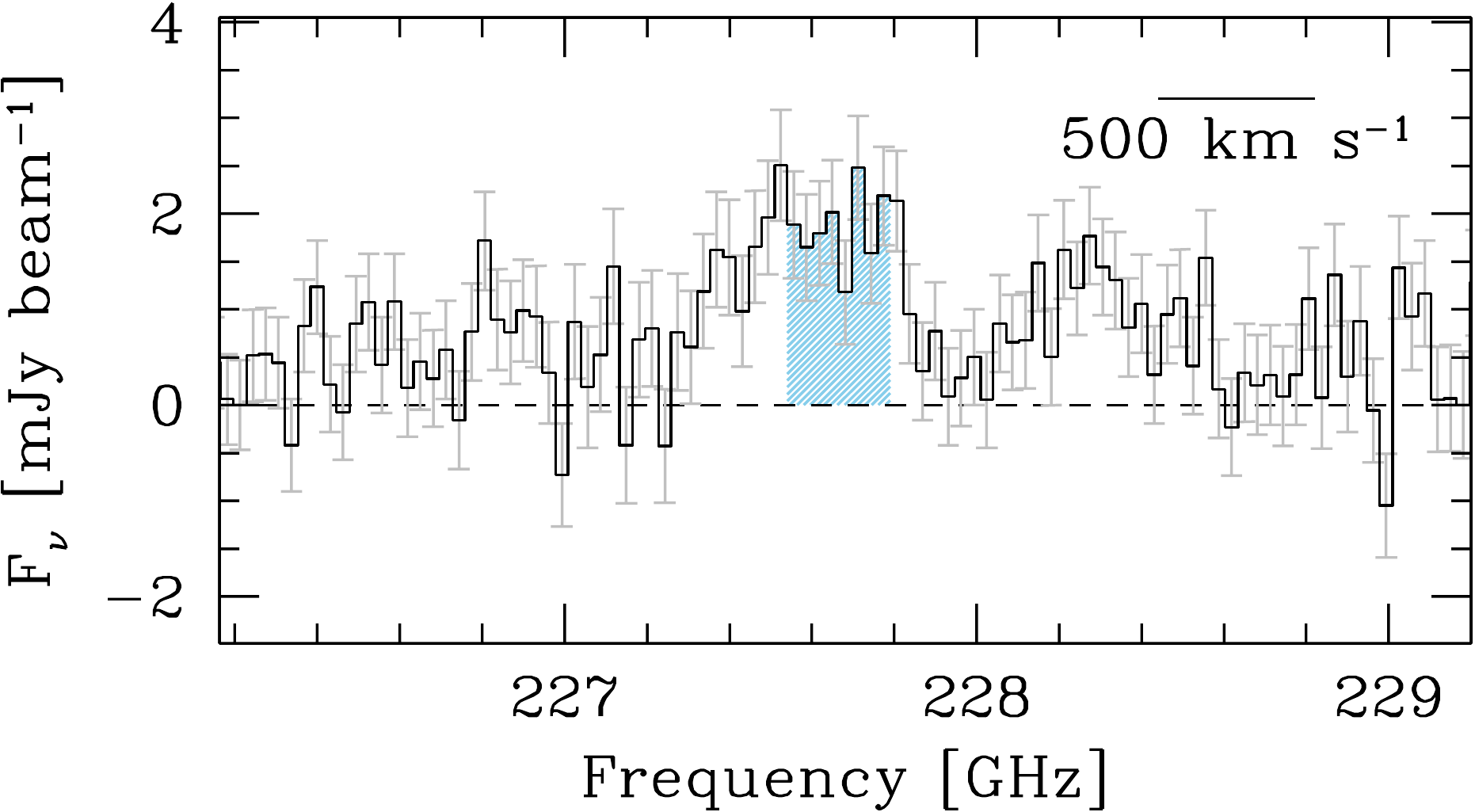}\\
\includegraphics[width=0.48\columnwidth]{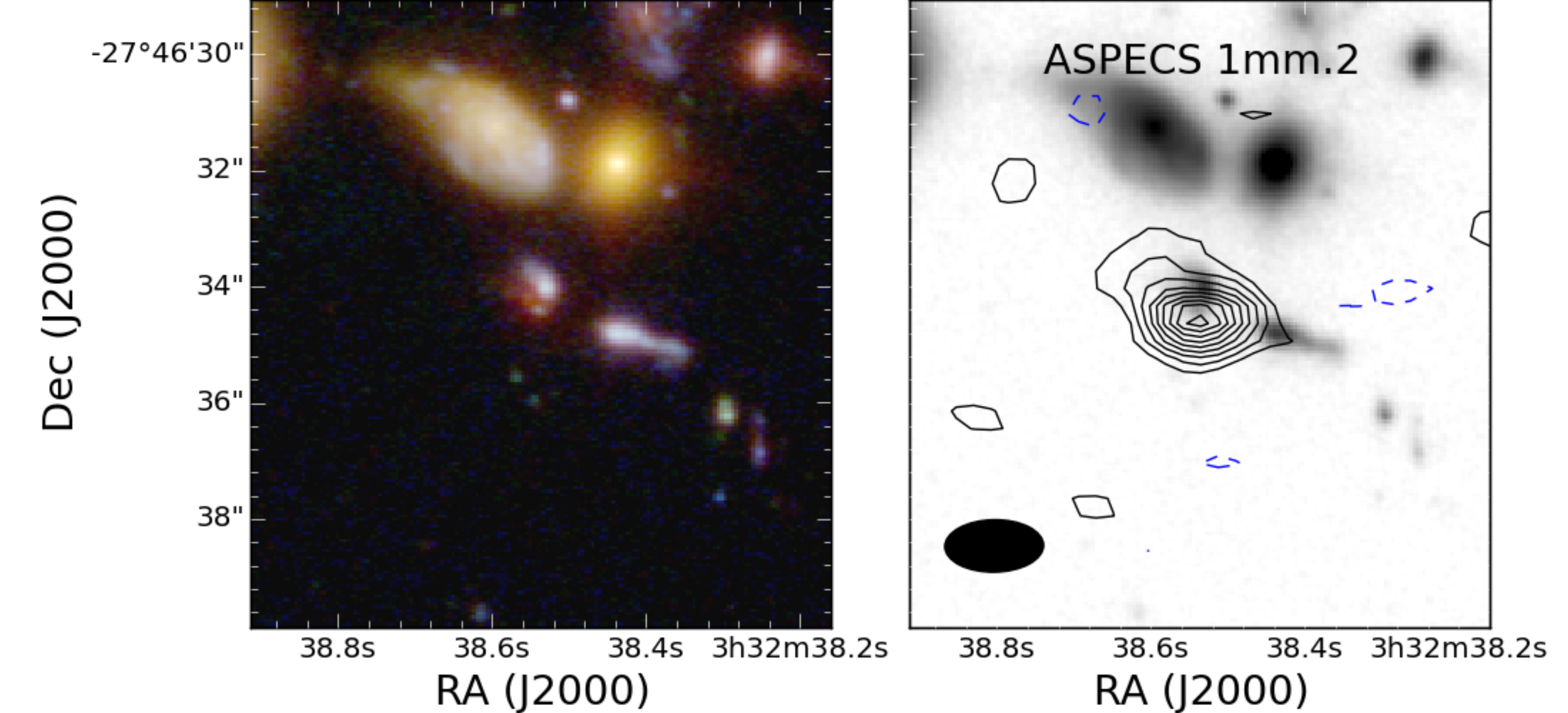}
\includegraphics[width=0.38\columnwidth]{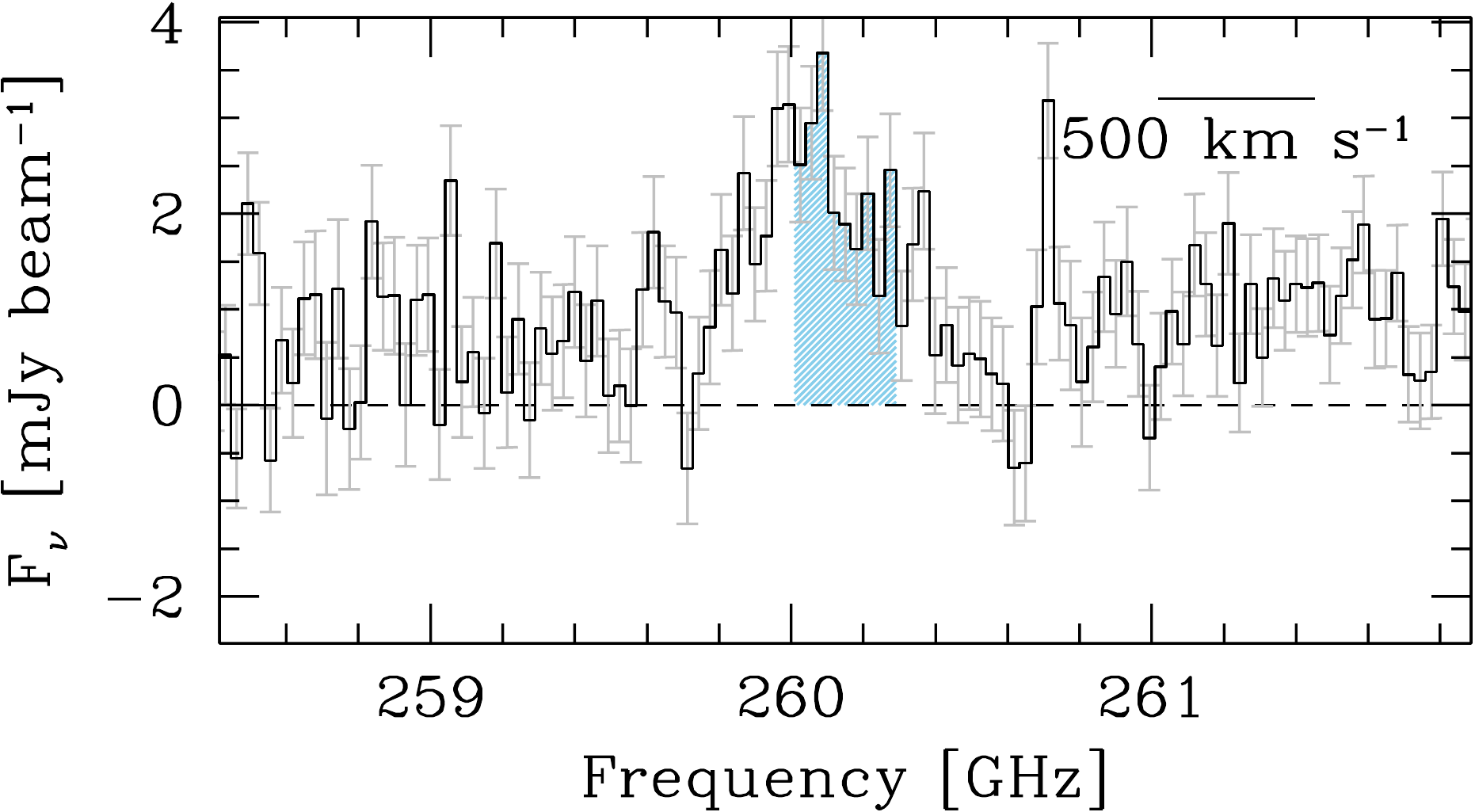}\\
\includegraphics[width=0.48\columnwidth]{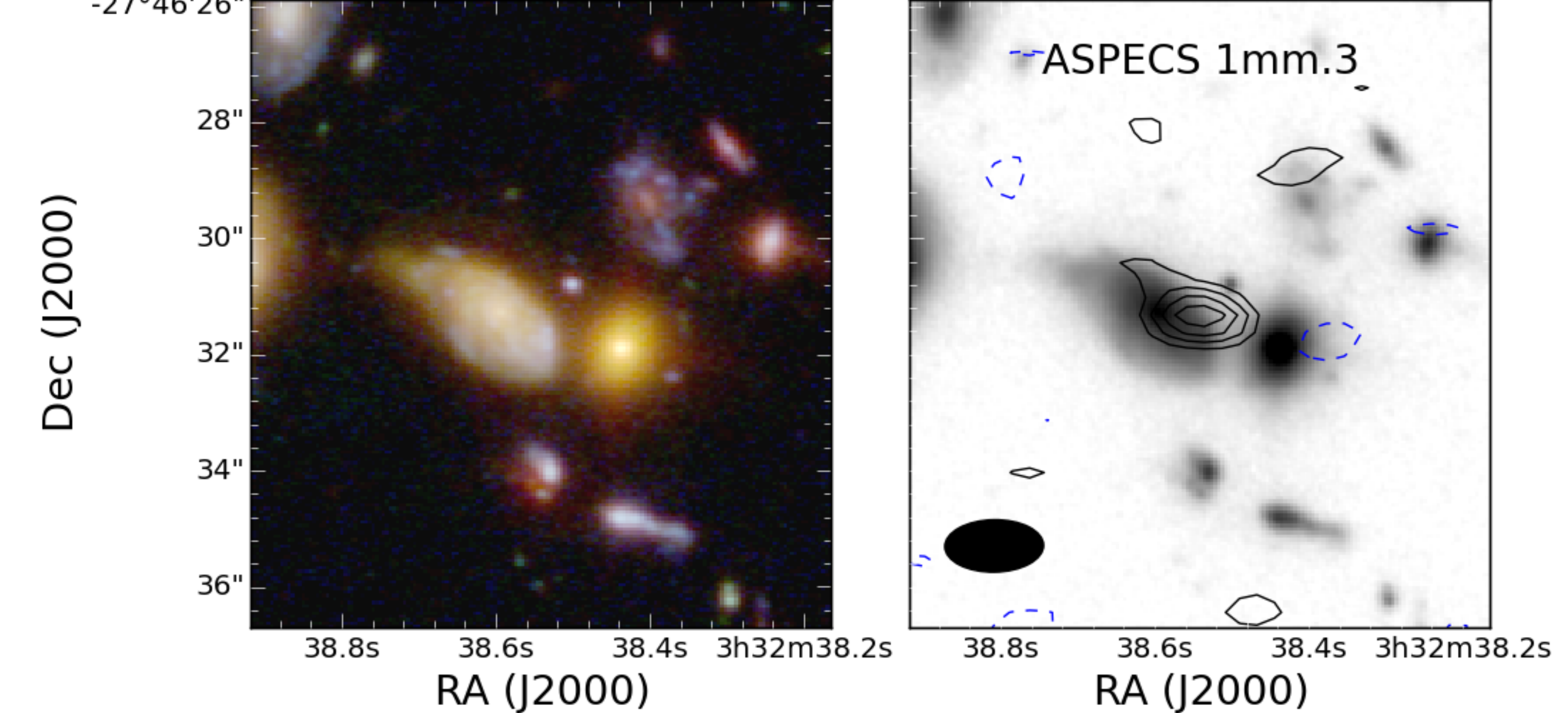}
\includegraphics[width=0.38\columnwidth]{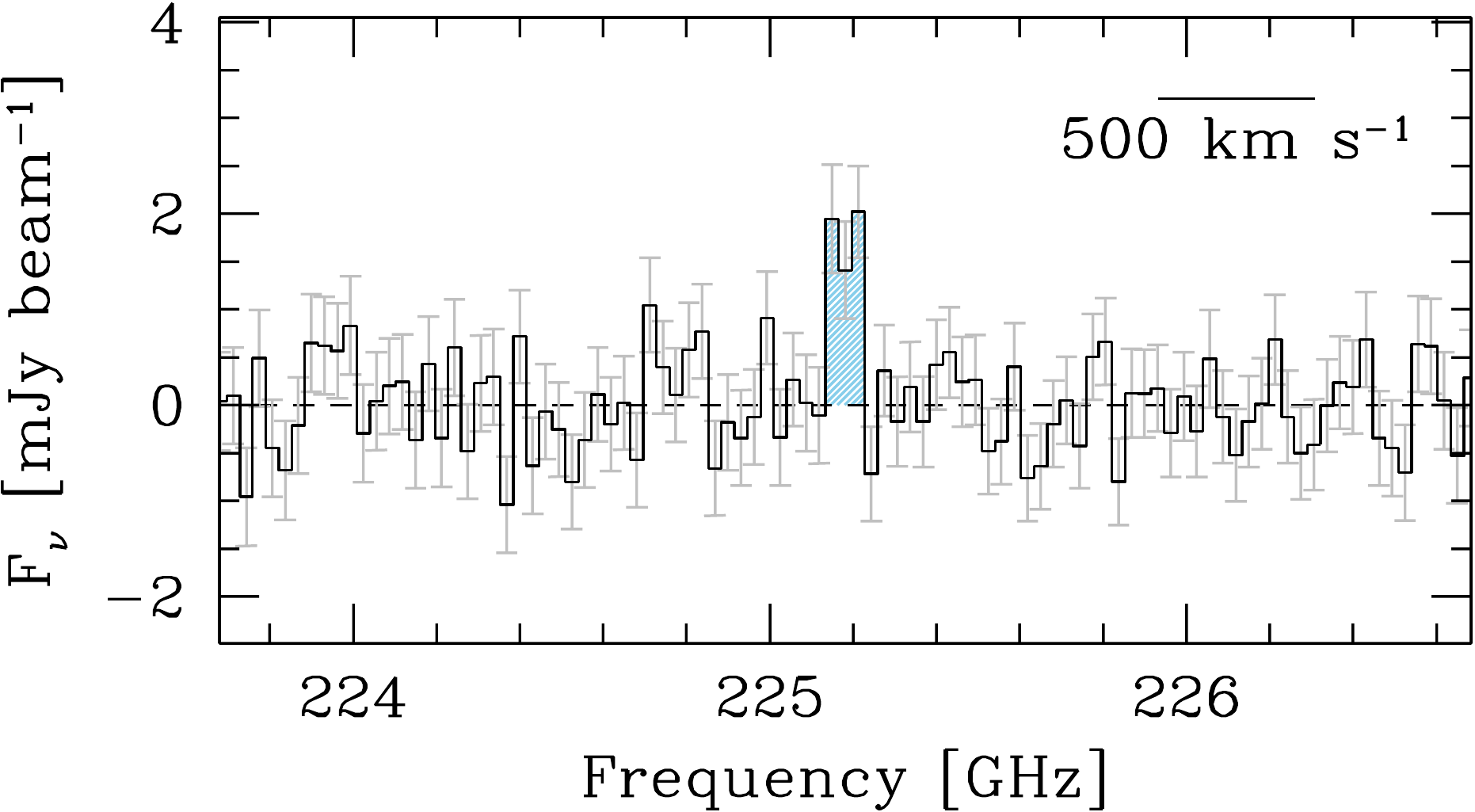}\\
\includegraphics[width=0.48\columnwidth]{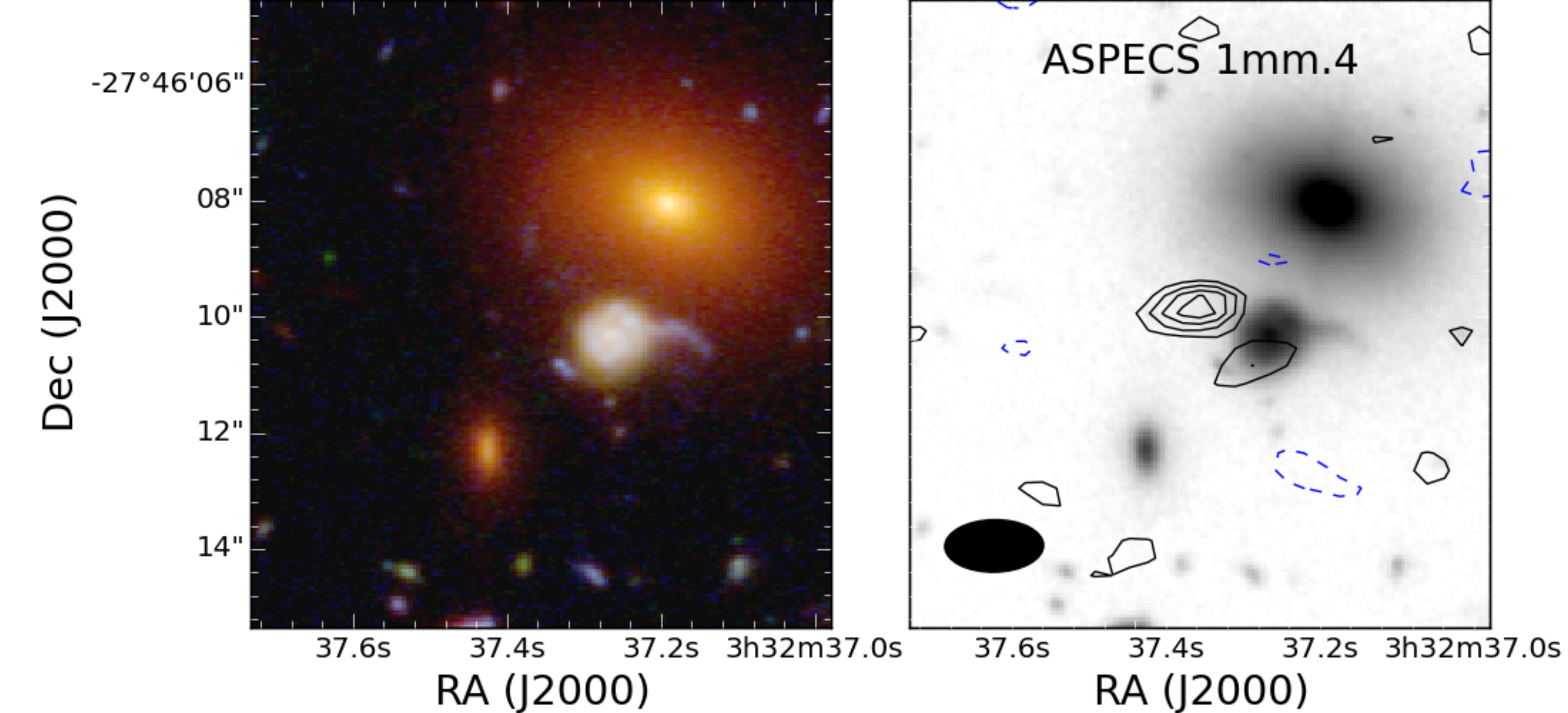}
\includegraphics[width=0.38\columnwidth]{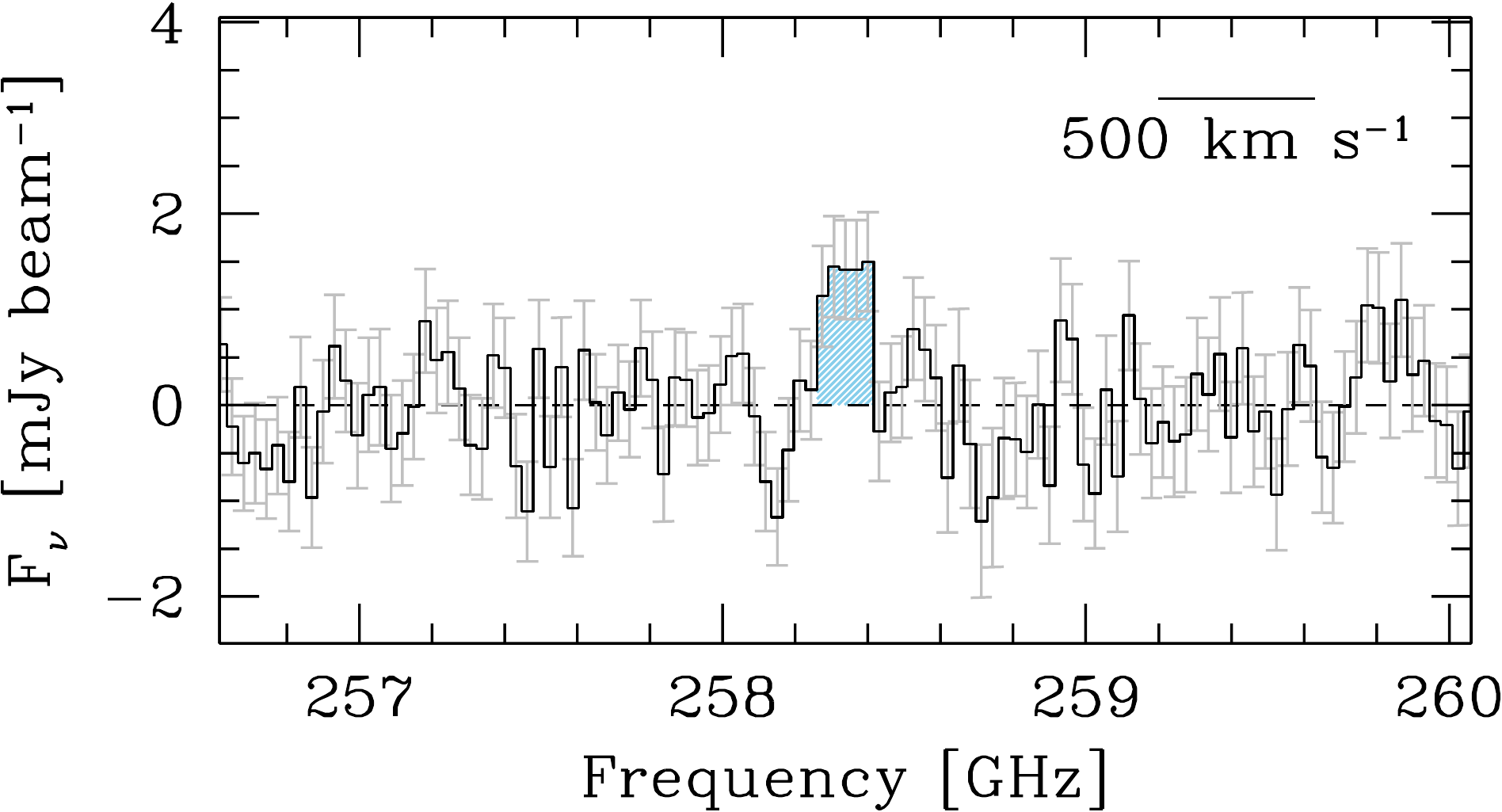}\\
\includegraphics[width=0.48\columnwidth]{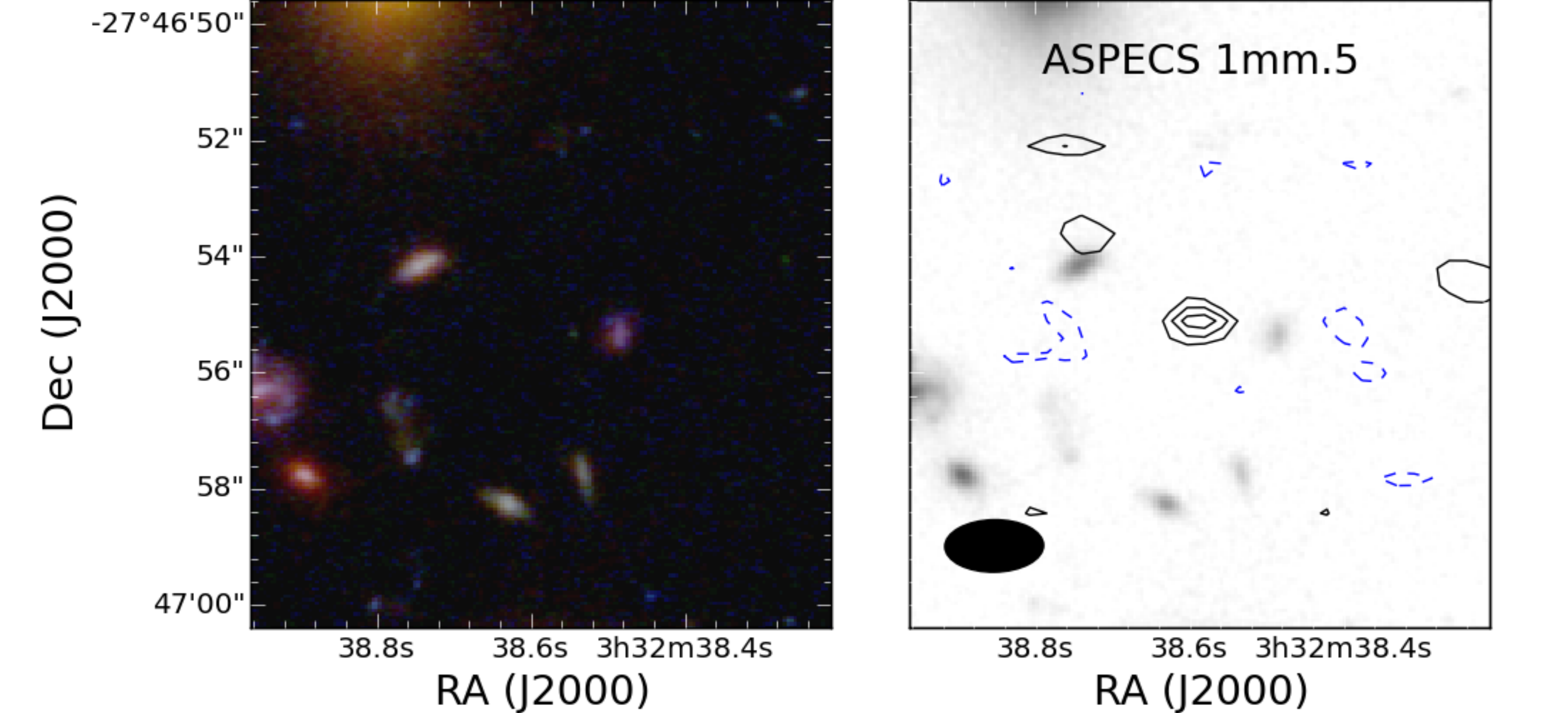}
\includegraphics[width=0.38\columnwidth]{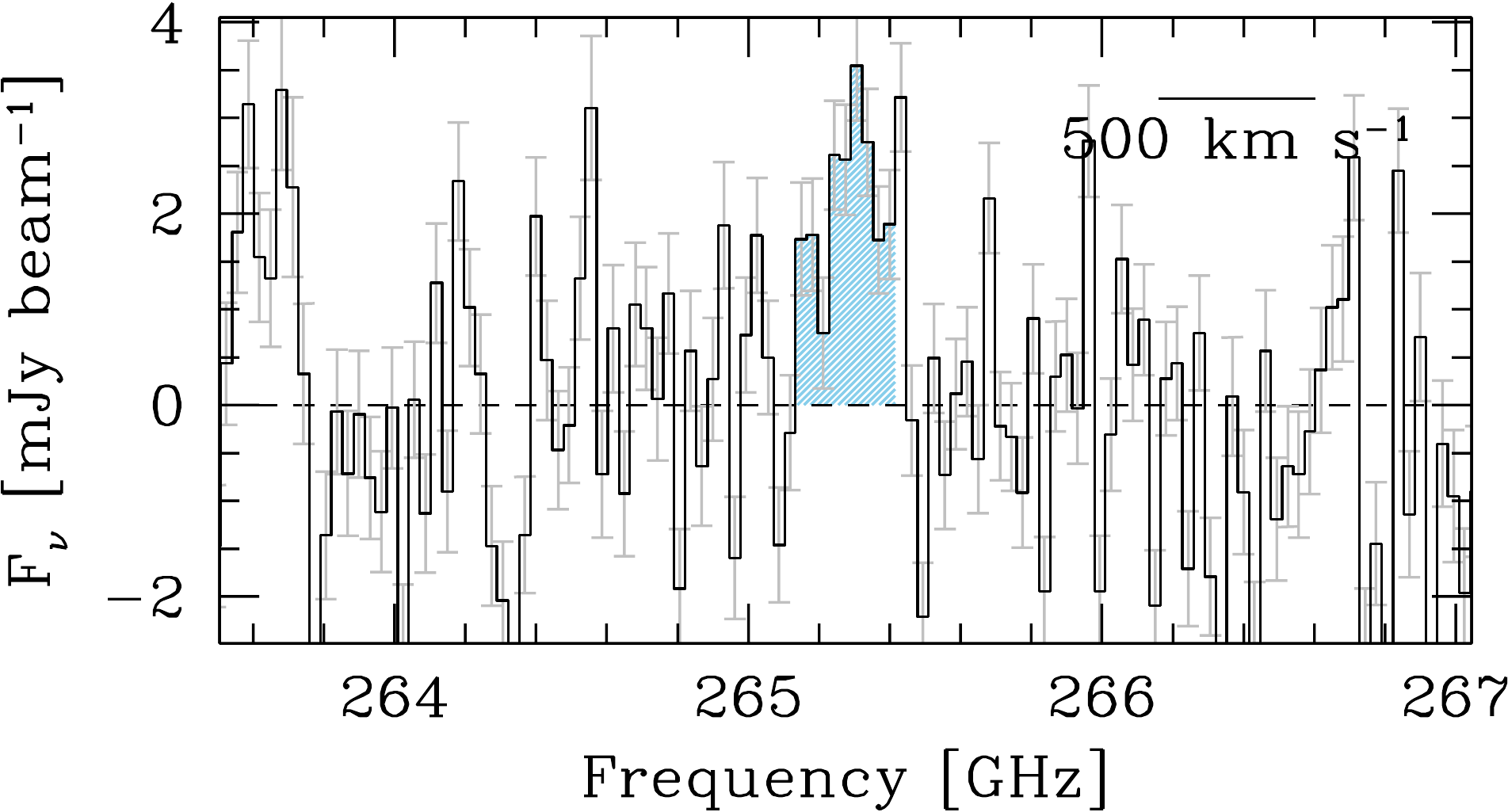}\\
\caption{{\em Left:} Optical/NIR {\it HST} multi--color image centred on the line candidates discovered in the blind search at 1mm (using the F125W, F775W and F435W filters, \citep{illingworth13}). {\em Middle:} Contours of the candidate line maps resulting from our line search described in Sec.~\ref{sec_search}. Positive (negative) contours are plotted in solid black (dashed blue), where the contours mark the $\pm2$,3,4,\ldots-$\sigma$ isophotes ($\sigma$ is derived from the respective line map). Each postage stamp is $20''\times20''$ and the size of the synthesized beam is show in the lower left. {\em Right:} spectrum of the line candidate. The blue shading marks the channels that the line-searching algorithm used to compute the line S/N. All line parameters are summarized in Tab.~\ref{tab_lines}.}
\label{fig_ps_1mm_a}
\end{figure*}

\begin{figure*}
\figurenum{\ref{fig_ps_1mm_a}}
\includegraphics[width=0.48\columnwidth]{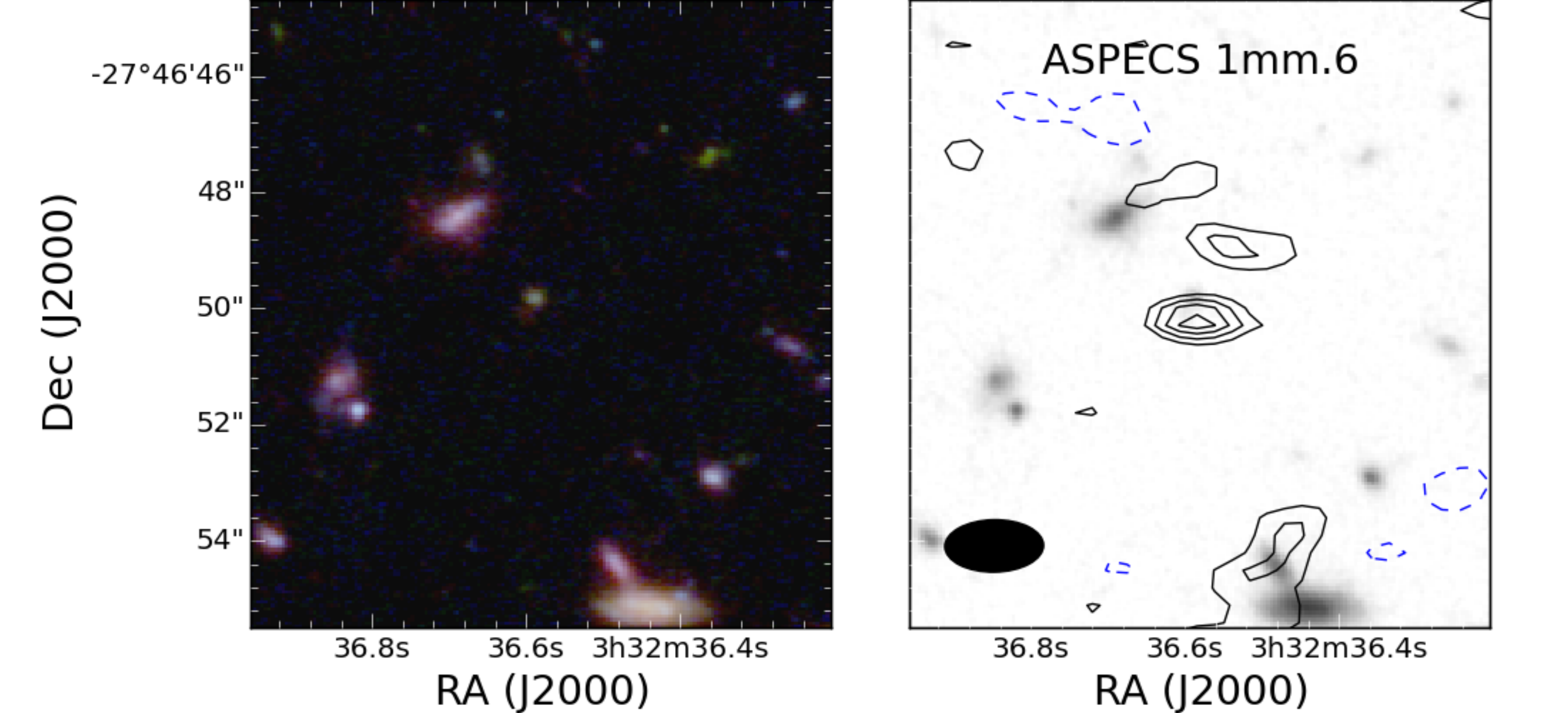}
\includegraphics[width=0.38\columnwidth]{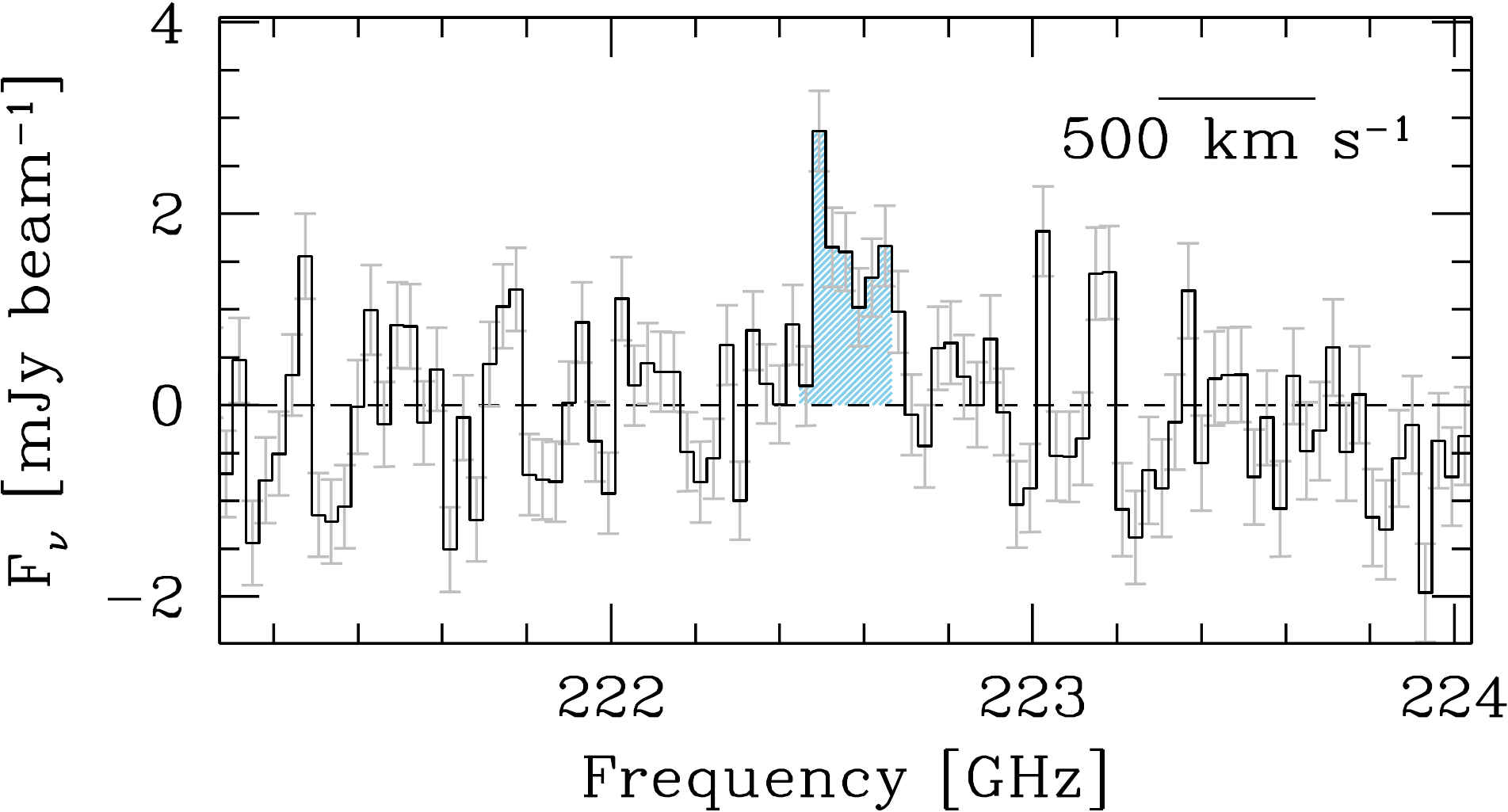}\\
\includegraphics[width=0.48\columnwidth]{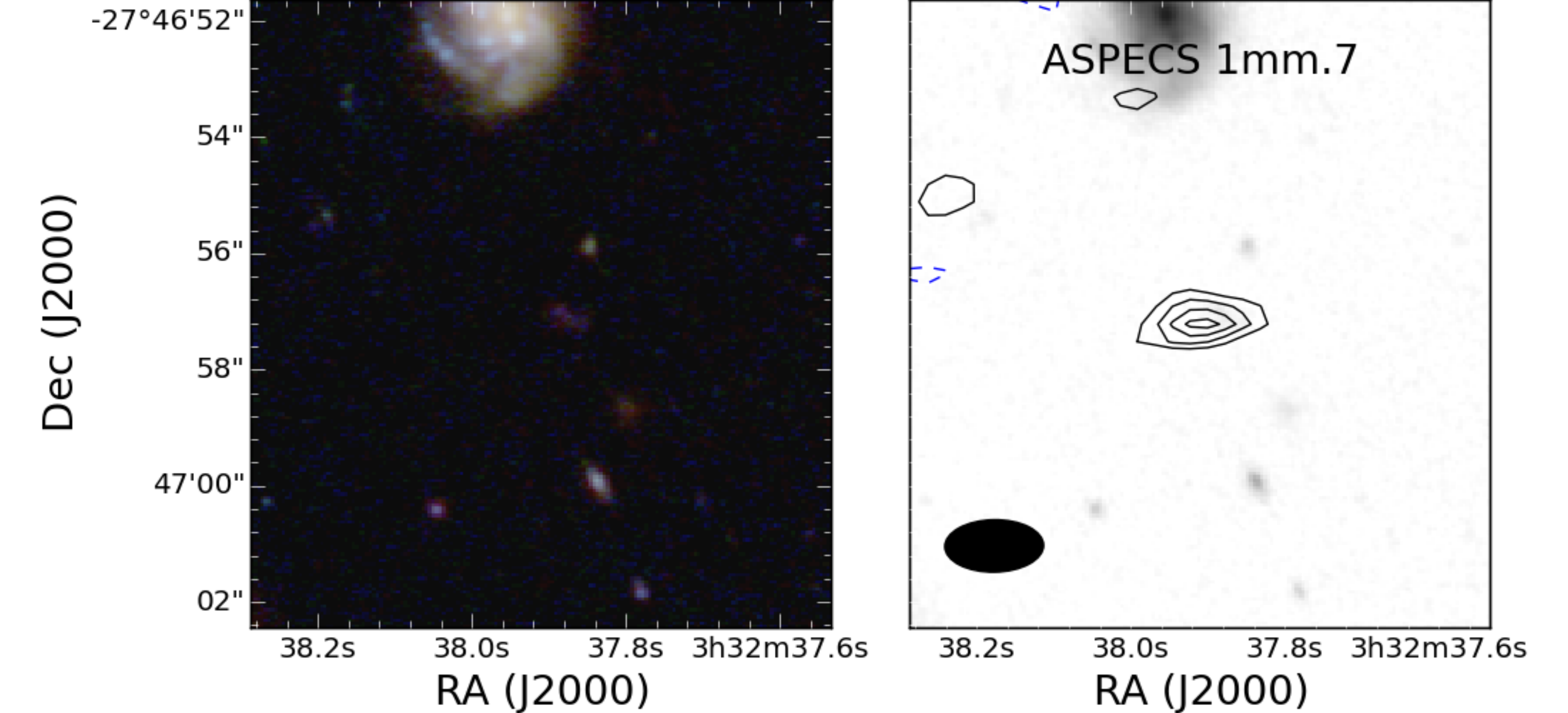}
\includegraphics[width=0.38\columnwidth]{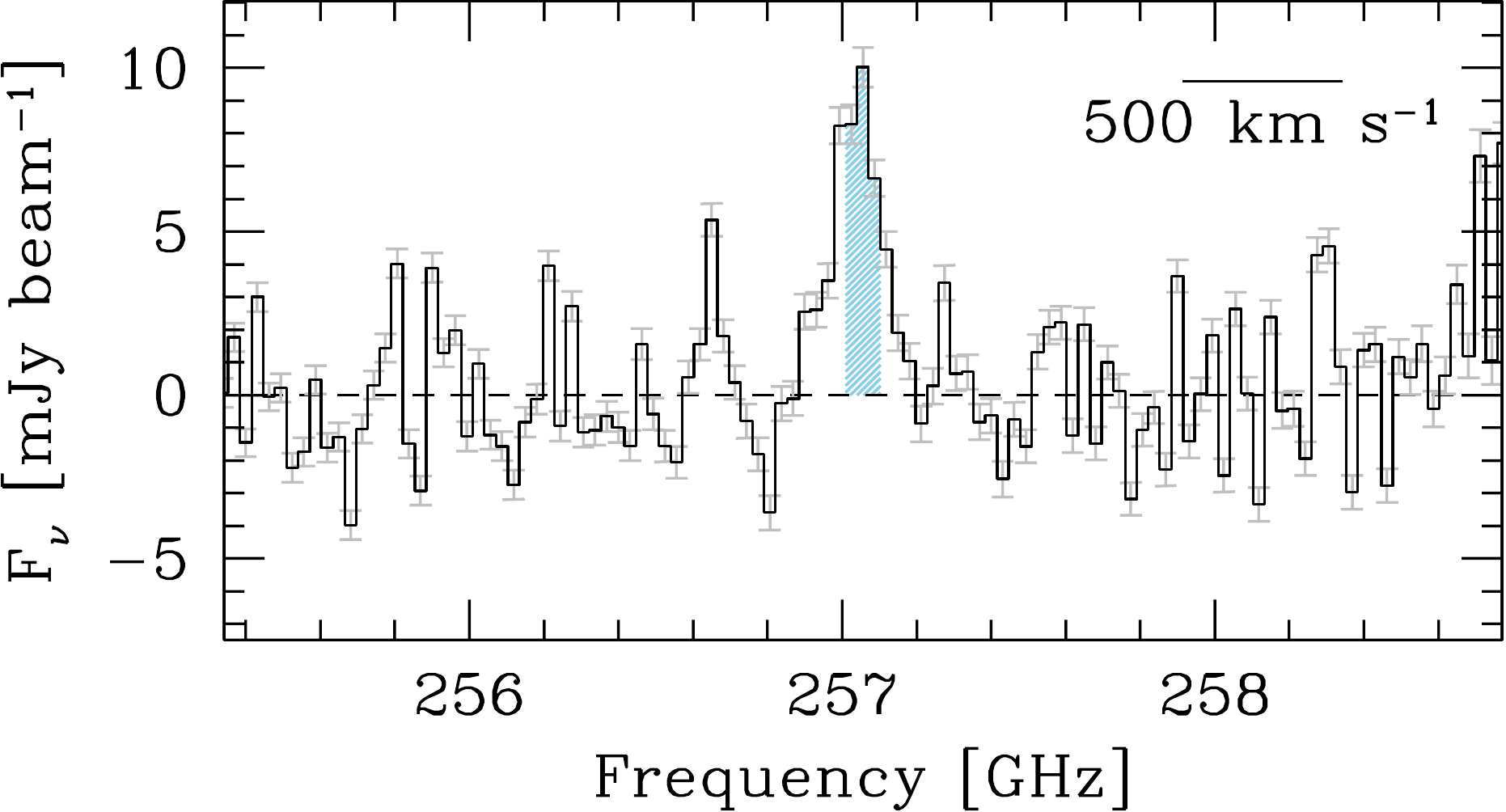}\\
\includegraphics[width=0.48\columnwidth]{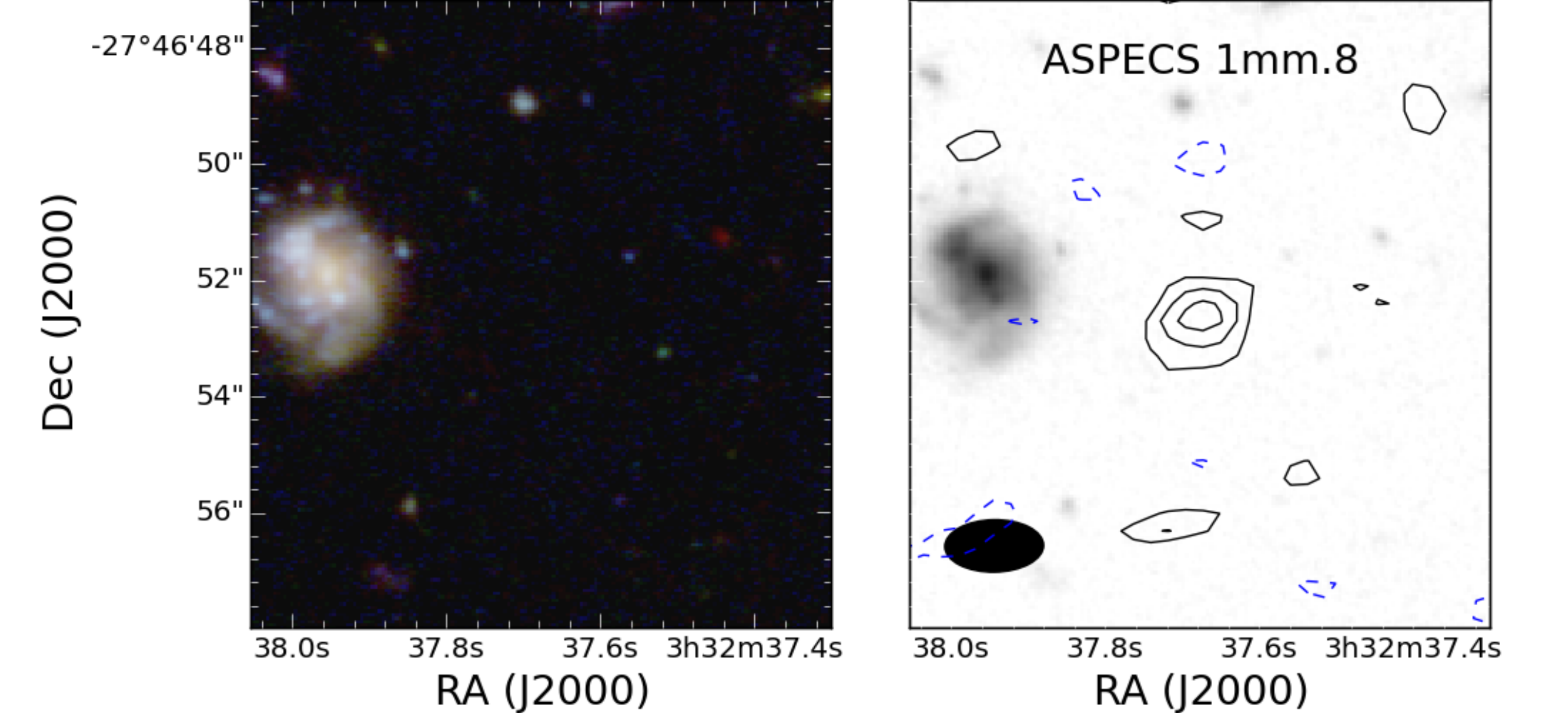}
\includegraphics[width=0.38\columnwidth]{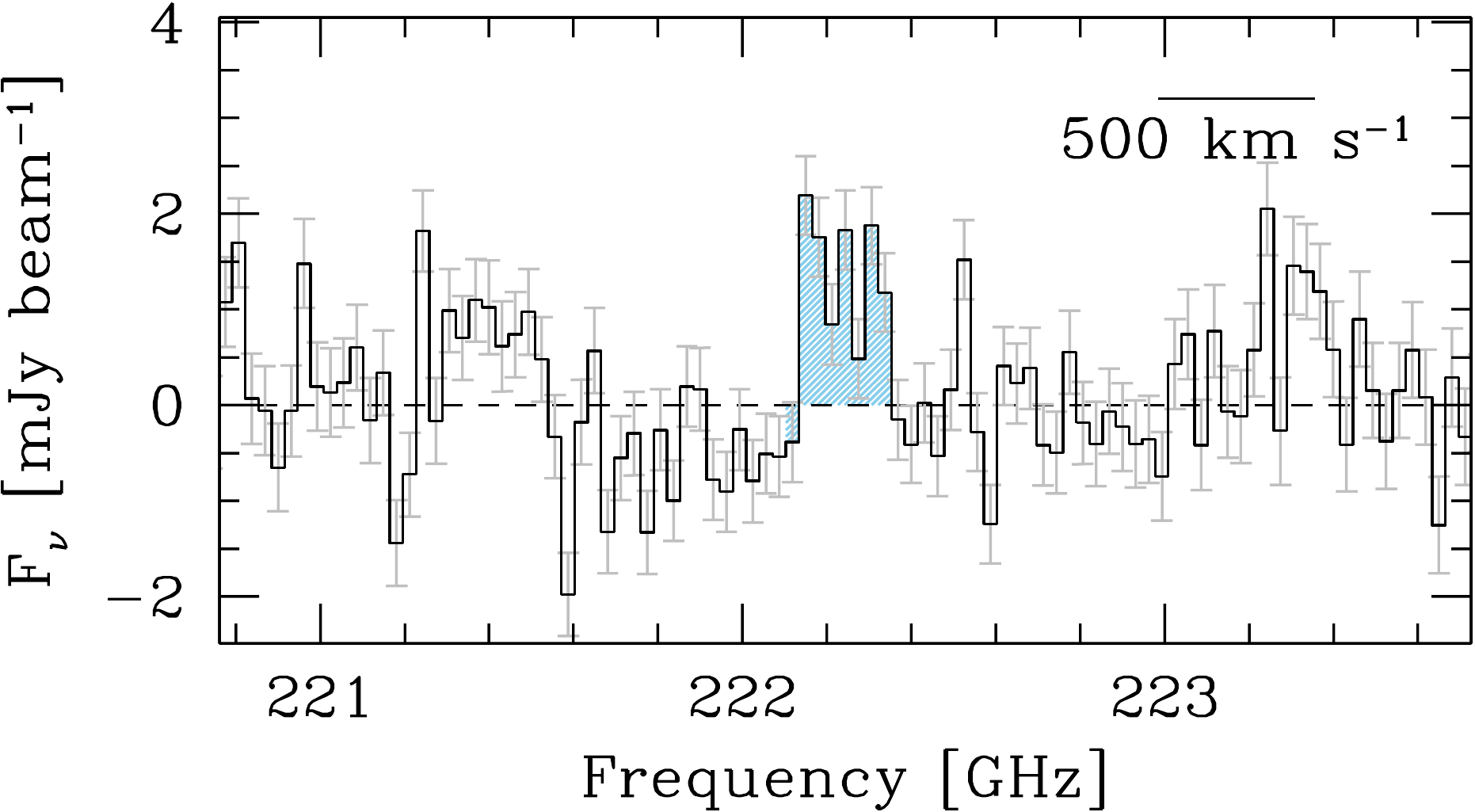}\\
\includegraphics[width=0.48\columnwidth]{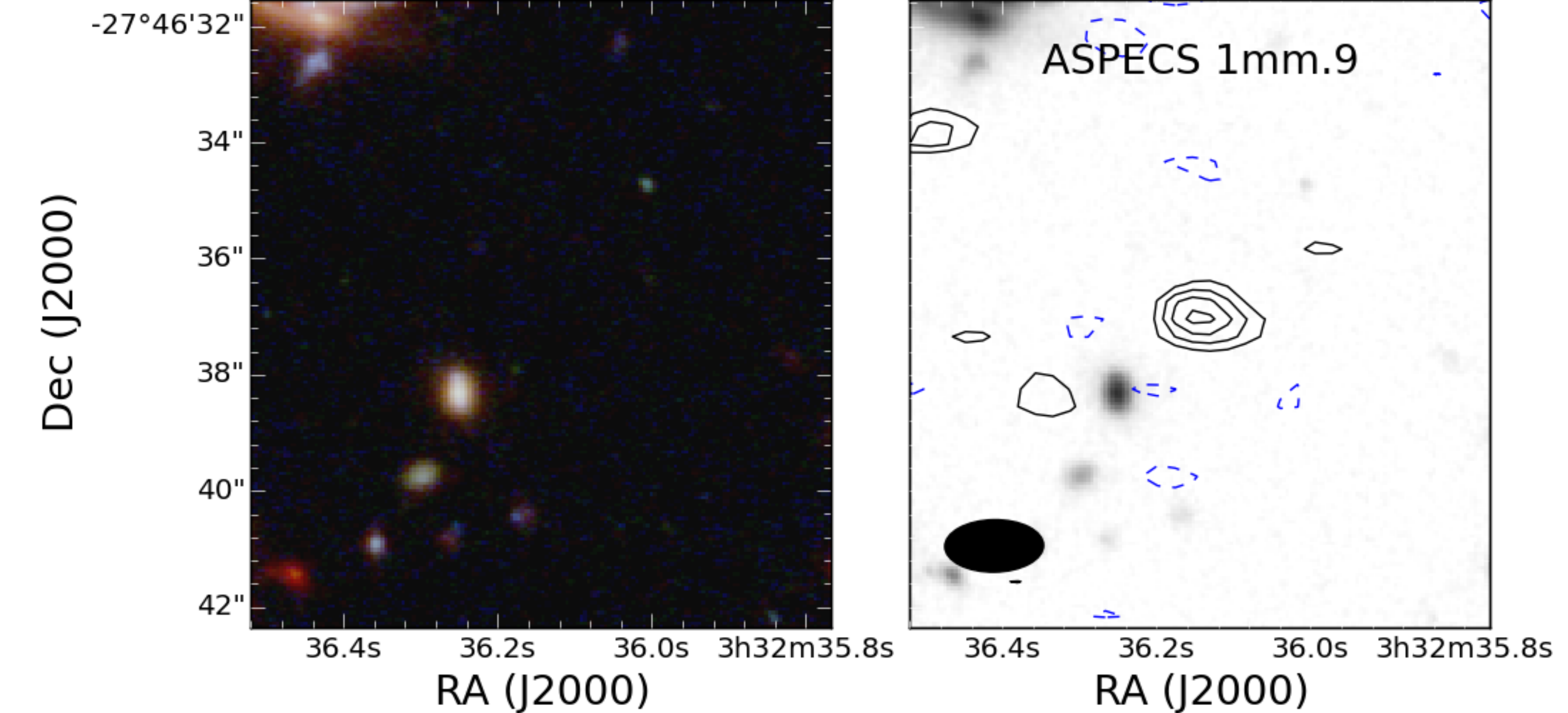}
\includegraphics[width=0.38\columnwidth]{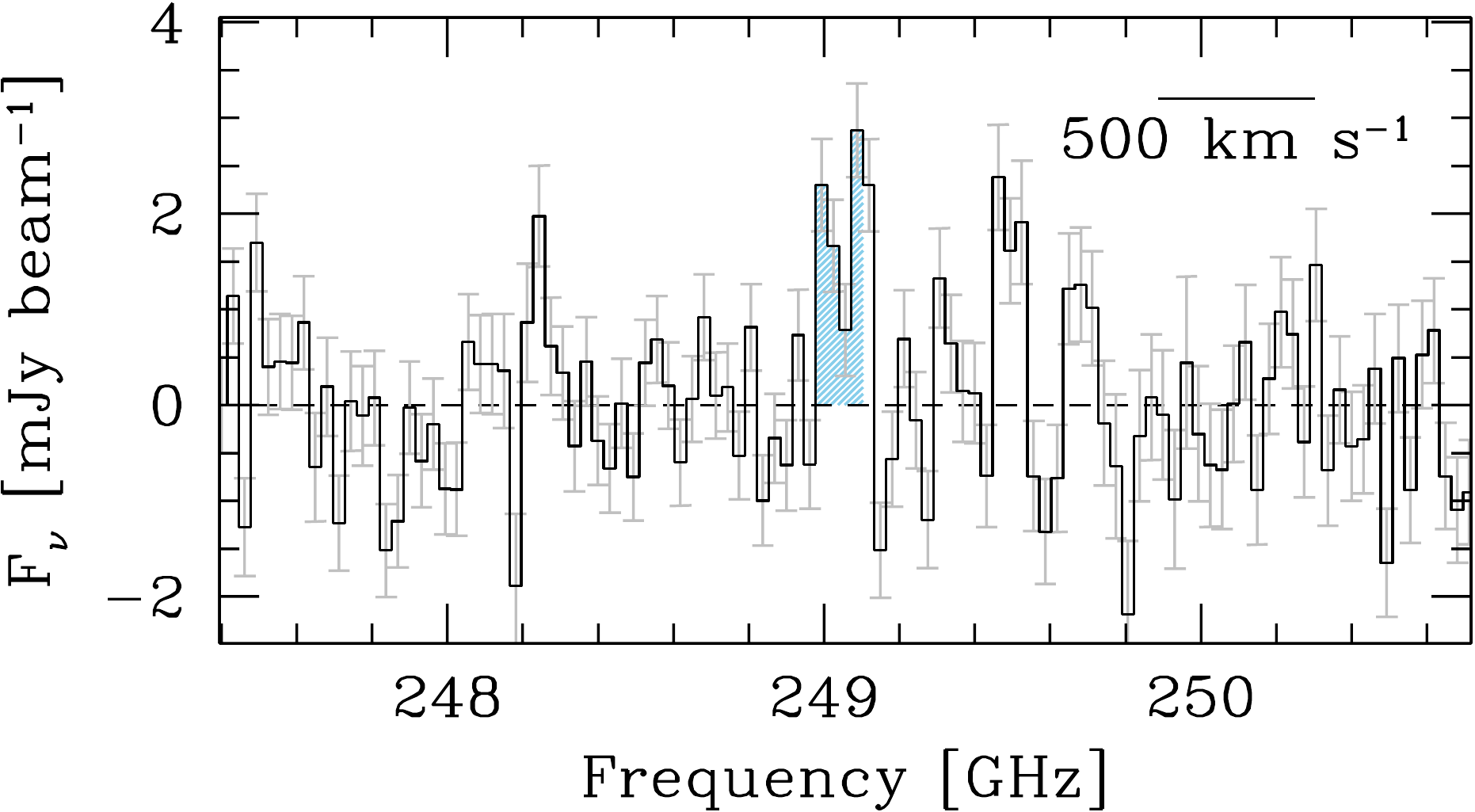}\\
\includegraphics[width=0.48\columnwidth]{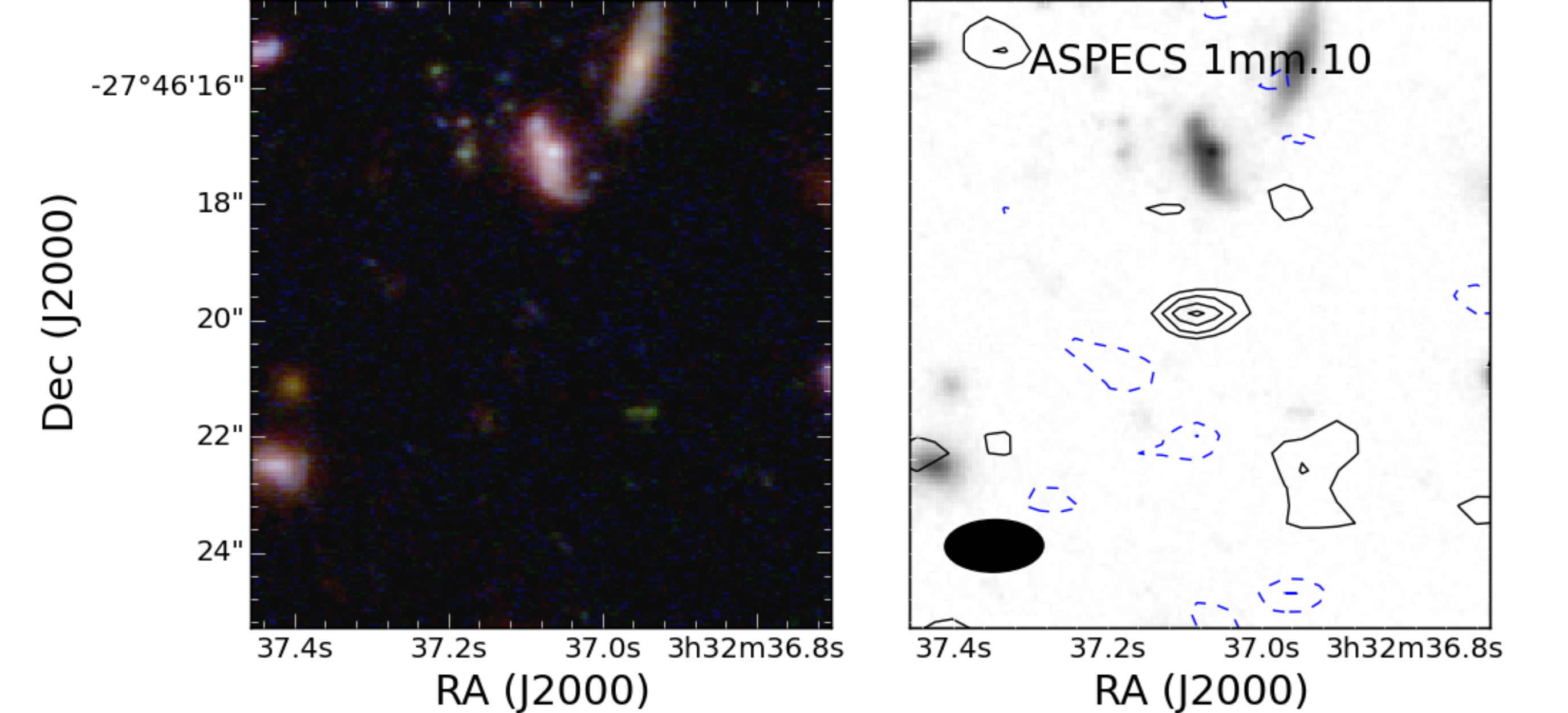}
\includegraphics[width=0.38\columnwidth]{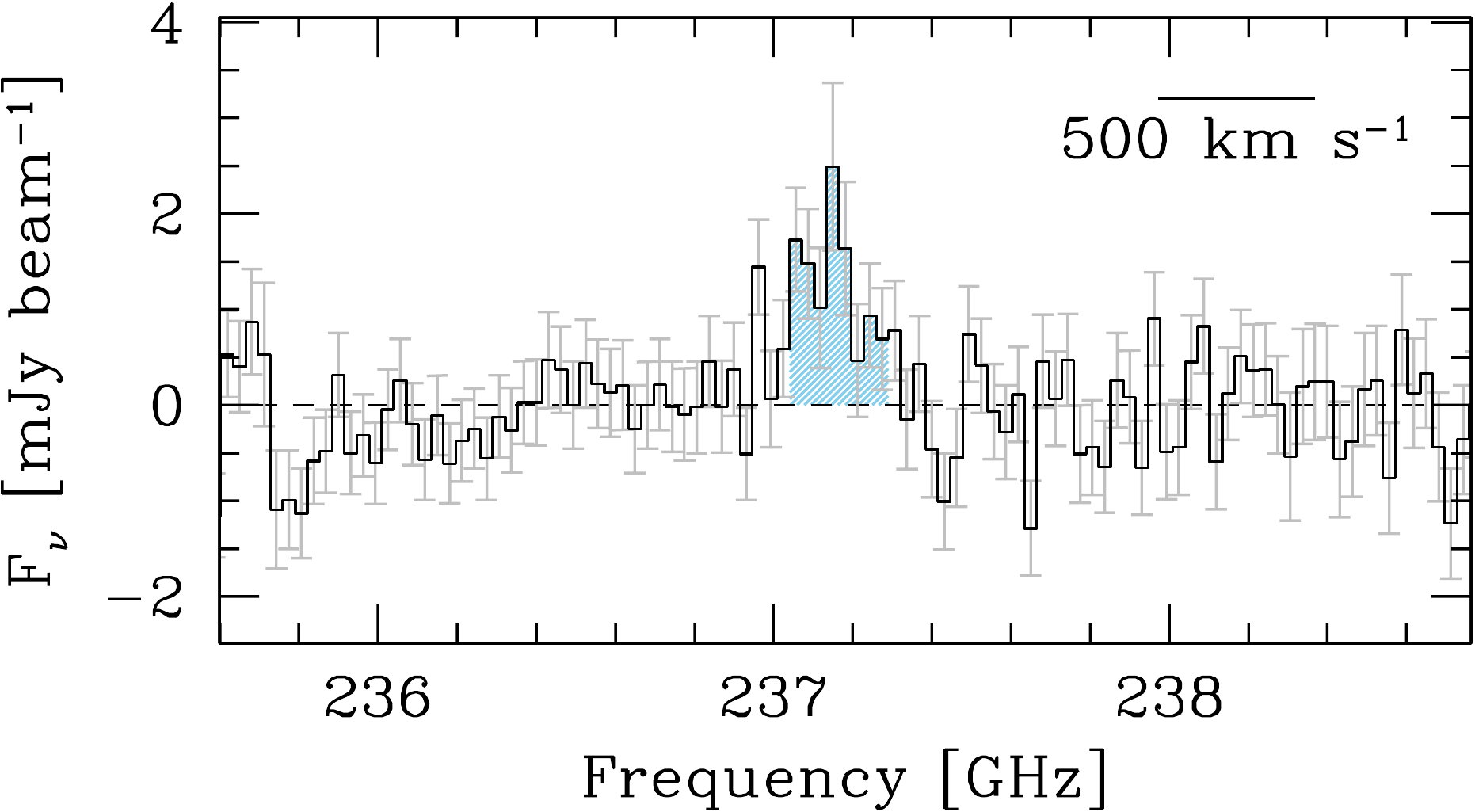}\\
\caption{continued.}
\end{figure*}

\begin{figure*}
\figurenum{\ref{fig_ps_1mm_a}}
\includegraphics[width=0.48\columnwidth]{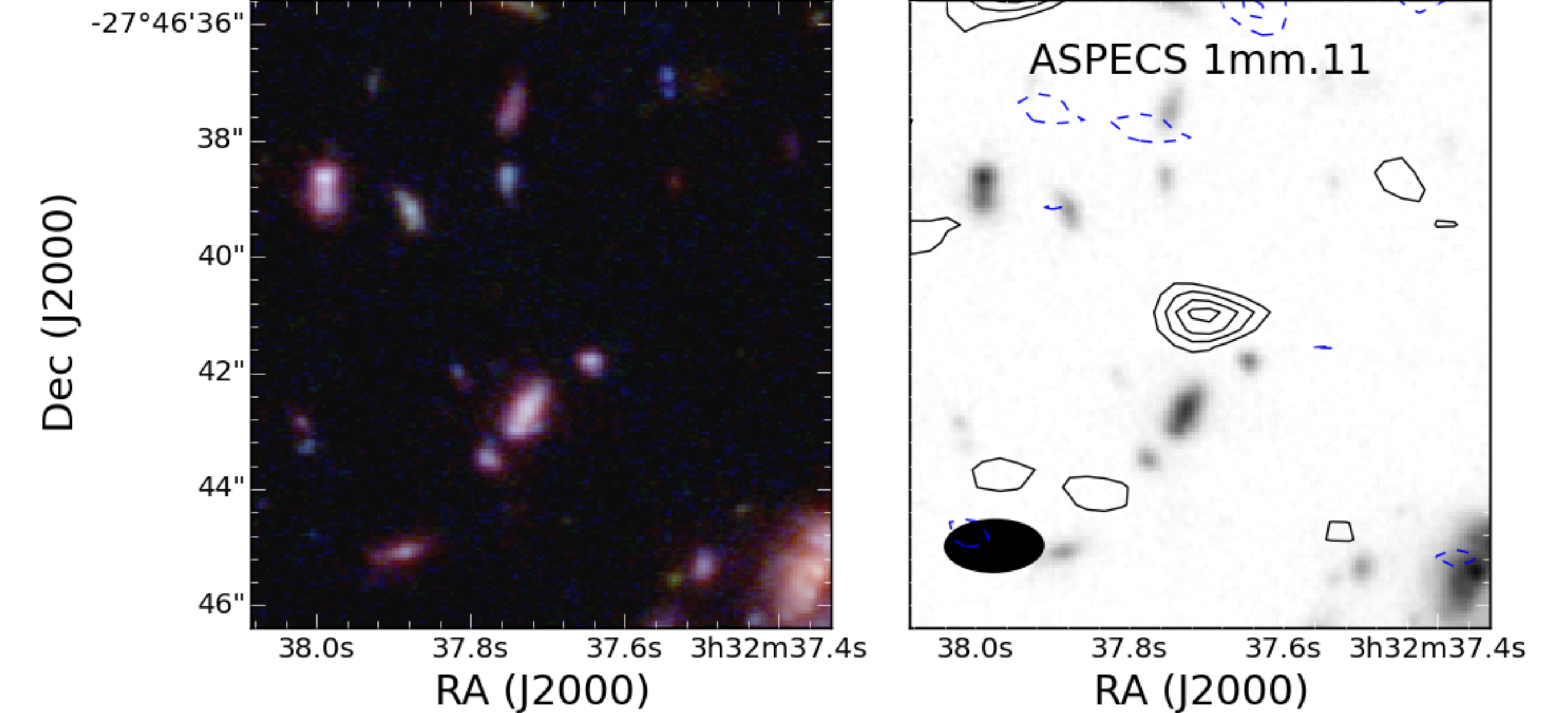}
\includegraphics[width=0.38\columnwidth]{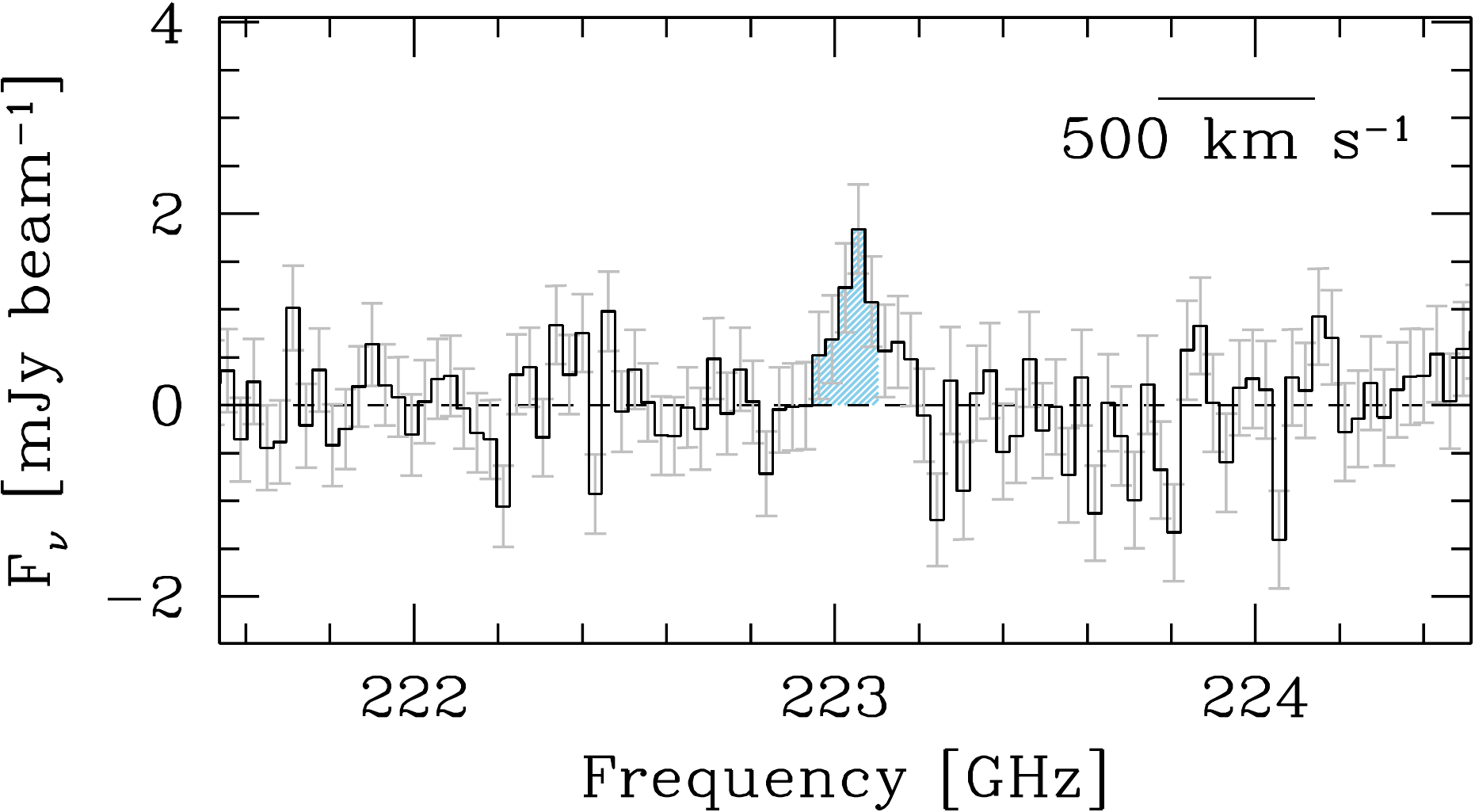}\\
\caption{continued.}
\end{figure*}

\label{lastpage}

\end{document}